\documentclass[nofootinbib,superscriptaddress,prd,aps,floatfix]{revtex4}
\usepackage{amsmath,amssymb,graphicx,amstext,amsfonts,graphics,epstopdf,slashed,color,multirow,adjustbox,lipsum,graphicx,rotating,xcolor,wrapfig}%
\usepackage[unicode=true, bookmarks=false, breaklinks=false,pdfborder={0 0 1},backref=section,colorlinks=false]{hyperref}
\hypersetup{colorlinks,bookmarksopen,bookmarksnumbered,linkcolor=blu,pdfstartview=FitH,urlcolor=rossos,citecolor=rossos}

\definecolor{rosso}{cmyk}{0,1,1,0.4}
\definecolor{rossos}{cmyk}{0,1,1,0.55}
\definecolor{rossoc}{cmyk}{0,1,1,0.2}
\definecolor{blu}{cmyk}{1,1,0,0.3}
\definecolor{blus}{cmyk}{1,1,0,0.6}
\definecolor{bluc}{cmyk}{1,1,0,0.1}
\definecolor{verde}{cmyk}{0.92,0,0.59,0.25}
\definecolor{verdec}{cmyk}{0.92,0,0.59,0.15}
\definecolor{verdes}{cmyk}{0.92,0,0.59,0.4}

\begin{document}

\title{\color{bluc}Identifying the nature of dark matter at $e^{-}e^{+}$ colliders}
\author{Nabil Baouche}
\email{baouche.nabil@gmail.com}
\affiliation{Laboratoire de Physique des Particules et Physique Statistique, Ecole
Normale Supérieure, BP 92 Vieux Kouba, DZ-16050 Algiers, Algeria}
\affiliation{Faculty of Thechnology, University of Dr. Yahia Fares-Medea, DZ-26000
Medea, Algeria}

\author{Amine Ahriche}
\email{aahriche@daad-alumni.de}
\affiliation{Department of Physics, University of Jijel, PB 98 Ouled Aissa, DZ-18000
Jijel, Algeria}
\affiliation{The Abdus Salam International Centre for Theoretical Physics, Strada
Costiera 11, I-34014, Trieste, Italy}
\begin{abstract}
In this work, we consider the process $e^{+}+e^{-}\rightarrow b\bar{b}+\slashed{E}_{T}$,
at the future electron-positron colliders such as the International Linear Collider and Compact Linear Collider, to
look for the dark matter (DM) effect and identify its nature at two
different centre-of-mass energies $E_{c.m.}=500~\mathrm{GeV}~and~1~\mathrm{TeV}$.
For this purpose, we take two extensions of the standard model, in which the DM could be a real scalar or a heavy right-handed neutrino
(RHN) similar to many models motivated by neutrino mass. In the latter
extension, the charged leptons are coupled to the RHNs via a lepton
flavor violating interaction that involves a charged singlet
scalar. After discussing different constraints, we define a set of
kinematical cuts that suppress the background, and generate different
distributions that are useful in identifying the DM nature. The use of
polarized beams (like the polarization $P(e^{-},e^{+})=\left[+0.8,-0.3\right]$
at the International Linear Collider) makes the signal detection easier and the DM identification
more clear, where the statistical significance gets enhanced by twice
(five times) for scalar (RHN) DM. 
\end{abstract}
\maketitle

\section{Introduction}

The standard model (SM) has achieved a great success in describing
the particle physics phenomenology at high energies, especially after
the recent discovery at the LHC of a Higgs
boson with mass around $125~\mathrm{GeV}$~\cite{Aad:2012tfa,Chatrchyan:2012xdj},
which is its most important success until now. Despite its successes,
the SM is unable to explain many questions such as baryon asymmetry
of the Universe, dark matter (DM), and neutrino masses and their mixing.
Indeed, the first strong experimental evidence that the SM is complete
was the neutrino oscillation observation~\cite{Fukuda:2001nj}.

One of the popular mechanisms to explain the smallness of neutrino
masses is the so-called seesaw mechanism~\cite{Gell}. Another approach
is based on getting naturally small neutrino masses radiatively, where
the loop suppression factor, $1/(16~\pi^{2})^{n}$, makes the suppression
natural instead of a suppression by a large scale of new physics (NP)~\cite{Cheng:1980qt,Zee:1980ai,Ma:1998dn,Zee:1985id,Babu:1988ki,Krauss:2002px,Aoki:2008av,Gustafsson:2012vj}
(for a review, see Ref~\cite{Boucenna:2014zba}). Some of these models
address, in addition to neutrino oscillation data, the DM problem
in which a heavy right-handed neutrino (RHN) with a mass range from $\mathrm{GeV}$
to $\mathrm{TeV}$ can play the role of a good DM candidate~\cite{Ma:1998dn,Krauss:2002px,Aoki:2008av,Okada:2015hia,Ahriche:2015loa}.
These models predict an interesting signature at collider experiments~\cite{Aoki:2016wyl,Ahriche:2014xra,Chekkal:2017eka}.
For instance, in Ref.~\cite{Chekkal:2017eka}, the authors have probed
the interactions of RHN with charged leptons via a singlet charged
scalar by considering many final states at $e^{-}e^{+}$ colliders
such as $\ell\ell+\slashed{E}_{T}$, $\ell\ell+\gamma+\slashed{E}_{T}$
and $\gamma+\slashed{E}_{T}$. This analysis was performed by taking
into account all constraints: lepton flavor violating (LFV) processes,
the muon anomalous magnetic moment~\cite{Lindner:2016bgg}, relic density,
and the monophoton negative searches at LEP-II~\cite{Achard:2003tx}.

The International Linear Collider (ILC) and the Compact Linear Collider
(CLIC) were proposed to discover physics beyond the SM, where the
ILC can scan the c.m. energies from 250 to $500~\mathrm{GeV}$,
with a possible expandability to $1~\mathrm{TeV}$~\cite{ILC4,Adolphsen:2013kya,Baer:2013c.m.a}, and the CLIC is subject
to development with c.m. energies from $380~\mathrm{GeV}$
to $3~\mathrm{TeV}$, with luminosity up to $2000~fb^{-1}$~\cite{CLIC:2016zwp}.
The leptonic collider has the option of polarized beams, which may
lead to an increasing signal/background ratio, and therefore enhances
the NP signal strength. This could provide a valuable opportunity
to detect new particles and determine their properties. In Ref.~\cite{Suehara:2015ura},
it has been found that the b-tagging efficiency is about $80\%$ when
the misidentification efficiencies for the c jet and u/d/s jet are below
$10\%$ and $1\%$, respectively. This motivates any analysis that
involves b jets. For instance, in Ref.~\cite{Durig:2014lfa}, it
has been shown that by considering the final state $b\bar{b}+\slashed{E}_{T}$
at the ILC, the $hWW$ coupling can be measured at a precision of 4.8\%
and 1.2\% at 250 and $500~\mathrm{GeV}$, respectively.
This analysis was performed using the beams polarization $P(e^{-},e^{+})=\left[+0.8,-0.3\right]$.

Another approach to dealing with the DM problem is to extend the SM
with singlet scalar(s), which plays a DM candidate role. This scalar
is assigned by a global $Z_{2}$ symmetry in order to ensure the DM
stability~\cite{Mambrini:2011ik,Ahriche:2012ei}. Whatever the DM
nature is, when a DM pair is produced, it does not leave any signature
or trace at the detectors and behaves as missing energy. If one considers
the final state $jj+\slashed{E}_{T}$ at $e^{-}e^{+}$ colliders such
as the ILC or CLIC, where $\slashed{E}_{T}=DM+DM$, then the dijet may
come from the Z/$\gamma^{*}$-gauge boson and/or the Higgs depending
on the model considered: SM, scalar DM, or RHN DM. So, if the dijet
is coming from the Higgs, then it would be suppressed except for the
b jets. Therefore, we will consider here only b-tagged jets that
can come form $Z/\gamma^{*}/Higgs$ according to the model and use
the polarization to identify the DM nature. So, in this work, we will
consider the signal $e^{-}e^{+}\rightarrow b\bar{b}+\slashed{E}_{T}$
and try to propose relevant cuts that reduce the background and identify
the DM nature based on the distributions shape with respect to the background.

This paper is organized as follows. In Sec-\ref{DM}, we describe
the models and different current experimental constraints such as
invisible Higgs decay, the muon anomalous magnetic moment, lepton flavor violation, DM relic
density $\Omega_{DM}h^{2}$, and the LEP-II data. We propose different
values for the model parameters, taking into account different bounds.
In Sec-\ref{PRO}, we describe the investigated process in detail,
and we discuss our results in Sec-\ref{DIS}, where we consider the
cases with polarized and unpolarized beams. Finally, we give our conclusion
in Sec-\ref{CON}.

\section{DM Models and Constraints}

\label{DM}

In this work, we will consider two types of models in which the DM could
be either a real scalar or a heavy RHN. Therefore, we consider for
the case of scalar DM a generic case of the Higgs portal~\cite{Birkedal:2004xn},
and in the case of heavy RHNs, we propose the SM extended with three
heavy RHNs, $N_{i}(i=1,2,3),$ and an electrically charged scalar
field, $S^{\pm}$, which is a singlet under the $SU(2)_{L}$ gauge group.
In addition, to ensure the DM candidate stability, we impose
a global discrete \emph{$\mathbb{Z}_{2}$} symmetry, under which $\{S,N_{i}\}\rightarrow\{-S,-N_{i}\}$
and all other fields are even~\cite{Ahriche:2014xra}.

\subsection{Scalar dark matter}

We consider a very simple extension of the SM by adding a real singlet
scalar defined under $SU(3)_{C}\otimes SU(2)_{L}\otimes U(1)_{Y}$
as $\phi\sim\left(1,1,0\right)$. This scalar field has to obey a
global $\mathbb{Z}_{2}$ symmetry and should not develop a vacuum expectation value (vev), and
therefore it could be a weakly interacting massive particle.
In this setup, the DM candidate can self-annihilate into SM particles
final states via the Higgs mediation. According to the scalar field
mass and its coupling to Higgs, one can get the relic density and
avoid the direct detection cross section bound.\footnote{By adding another scalar to the SM that assists the electroweak symmetry
breaking, one can easily avoid the direct detection bound~\cite{Ahriche:2012ei}.} The Lagrangian reads

\begin{equation}
\mathcal{L}=\mathcal{L}_{SM}+\frac{1}{2}\text{\ensuremath{\partial}}^{\mu}\phi\text{\ensuremath{\partial_{\mu}}}\phi-V\left(\phi,H\right),\label{ch}
\end{equation}
where $H$ is the SM Higgs doublet and $V\left(\phi,H\right)$ is
the scalar potential, which after the electroweak symmetry breaking
reads

\begin{equation}
V\left(\phi,h\right)\supset\frac{1}{2}m_{\phi}^{2}\phi^{2}+\frac{c_{s}\upsilon}{2}h\phi^{2},
\end{equation}
with $m_{\phi}$ the real scalar mass after the symmetry breaking,
$c_{s}$ the quartic coupling constant of the potential term $\phi^{2}\left|H\right|^{2}$,
$\upsilon=246~\mathrm{GeV}$ the SM doublet vev, and $h$ the
usual SM Higgs field. So, the model is defined just by two free parameters:
the coupling constant $c_{s}$ and the scalar mass $m_{\phi}$. At
an electron-positron collider, it is possible to produce the real
singlet scalar $\phi$ via the Z fusion $e^{-}e^{+}\rightarrow e^{-}e^{+}h\rightarrow e^{-}e^{+}\phi\phi$,
or by the associate production $e^{-}e^{+}\rightarrow Zh\rightarrow Z\phi\phi$
, where the $Z$ gauge boson subsequently decays, primarily hadronically~\cite{Olive:2016xmw}.
If the Higgs decay into invisible channel $h\rightarrow\phi\phi$,
the decay width is given by

\begin{equation}
\text{\ensuremath{\Gamma_{inv}\left(h\rightarrow\phi\phi\right)}}=\frac{c_{s}^{2}\upsilon^{2}}{32\pi m_{h}}\left(1-\frac{4m_{\phi}^{2}}{m_{h}^{2}}\right)^{\frac{1}{2}}.\label{Inv}
\end{equation}

The experimental constraint on the invisible Higgs decay reads

\begin{equation}
\mathcal{B}_{inv}\left(h\rightarrow\phi\phi\right)=\frac{\text{\ensuremath{\Gamma_{inv}}}}{\text{\ensuremath{\Gamma_{inv}}}+\text{\ensuremath{\Gamma}}_{SM}^{tot}}\leq0.16,\label{BR}
\end{equation}
where $\text{\ensuremath{\text{\ensuremath{\Gamma}}}}_{SM}^{tot}=4.20~\mathrm{MeV}$ 
is the SM Higgs total width~\cite{Heinemeyer:2013tqa}. This bound
can be translated into a constraint on the couplings $c_{s}$ and
the scalar mass $m_{\phi}$ as 
\begin{equation}
c_{s}\leq1.2882\times10^{-2}\left(1-\left(\frac{m_{\phi}}{62.5~\mathrm{GeV}}\right)^{2}\right)^{-\frac{1}{4}}.
\end{equation}

In our analysis, we focus on the case in which the scalars are pair produced
through an on-shell Higgs decay. This means that we will consider
the light masses range $m_{\phi}\leq m_{h}/2$, and we choose two values
of the model free parameters $\left\{ m_{\phi},c_{s}\right\} $, where
they respect the experimental constraint (\ref{BR}). We call them
model 1 ($M_{1}$) and model 2 ($M_{2}$).

\begin{table}
\centering %
\begin{tabular}{|c|c|}
\hline 
Model & Parameters\tabularnewline
\hline 
\hline 
$M_{1}$ & $\left\{ m_{\phi},c_{s}\right\} =\{10~\mathrm{GeV},\ 1.25\times10^{-2}\},$\tabularnewline
\hline 
$M_{2}$ & $\left\{ m_{\phi},c_{s}\right\} =\{60\mathrm{~\mathrm{GeV}},2.35\times10^{-2}\}.$\tabularnewline
\hline 
\end{tabular}\caption{The parameters values for model 1 and model 2.}

\label{Model12} 
\end{table}

\subsection{Fermionic dark matter}

In this case, the SM was extended with an electrically charged singlet
scalar field $S^{+}\sim\left(1,1,2\right)$ and three RHNs, $N_{i}\sim\left(1,1,0\right)$~\cite{Krauss:2002px}.
The Lagrangian has the form~\cite{Ahriche:2013zwa} 
\begin{equation}
\mathcal{L}=\mathcal{L}_{SM}+\{g_{i\alpha}N_{i}^{C}\ell_{\alpha R}S^{+}+\frac{1}{2}m_{N_{i}}N_{i}^{C}N_{i}+h.c\}-V,\label{L}
\end{equation}
where $\ell_{\alpha R}$ is the right-handed charged lepton, $m_{N_{i}}$ are
the heavy RHN's masses, $C$ denotes the charge conjugation operator,
and $g_{i\alpha}$ are the new Yukawa couplings. Here, V is the scalar
potential. The Greek letters denote $\alpha=\mu,e,\tau$, and the fermion
generations are labelled by $i=1,2,3$. When the \emph{$\mathbb{Z}_{2}$
}symmetry is imposed, the lightest RHNs becomes stable and could
be a good DM candidate~\cite{Krauss:2002px,Ahriche:2015lqa}. These
couplings as well the RHNs and the charged scalar masses enter the
expression of the neutrino mass matrix elements depending on the model
details.

The interactions (\ref{L}) induce a new contribution to the muon's
anomalous magnetic moment and LFV processes such as $\ell_{\alpha}\rightarrow\ell_{\beta}+\gamma$
and $\ell_{\alpha}\rightarrow\ell_{\beta}+\bar{\ell_{\beta}}+\ell_{\beta}$,
and all are generated at one loop via the exchange of the charged scalar
$S^{\pm}$, where the branching ratios are given in Refs~\cite{Toma:2013zsa,Hisano:1995cp,Ahriche:2015loa}.
Unlike other models~\cite{Chiang:2017tai}, the contribution to the
muon anomalous magnetic moments in this model is negative~\cite{Ahriche:2015loa},
and therefore does not to close the gap between the experimental measurement
and the SM prediction $\delta a_{\mu}=a_{\mu}^{exp}-a_{\mu}^{SM}=288(63)(43)\times10^{-11}$~\cite{Olive:2016xmw}.
In Table \ref{Bounds}, we present the current bounds on different
LFV observables.

\begin{table}
\centering %
\begin{tabular}{|c|c|}
\hline 
LFV process & Current bound\tabularnewline
\hline 
\hline 
$\mathcal{B}\left(\mu\rightarrow e+\gamma\right)$ & $4.2\times10^{-13}$~\cite{TheMEG:2016wtm}\tabularnewline
\hline 
$\mathcal{B}\left(\tau\rightarrow\mu+\gamma\right)$ & $4.4\times10^{-8}$~\cite{Olive:2016xmw}\tabularnewline
\hline 
$\mathcal{B}\left(\tau\rightarrow e+\gamma\right)$ & $3.3\times10^{-8}$~\cite{Aubert:2009ag}\tabularnewline
\hline 
$\mathcal{B}\left(\tau\rightarrow e^{-}+e^{+}+e^{-}\right)$ & $2.7\times10^{-8}$~\cite{Hayasaka:2010np}\tabularnewline
\hline 
$\mathcal{B}\left(\mu\rightarrow e^{-}+e^{+}+e^{-}\right)$ & $1.0\times10^{-12}$~\cite{Bellgardt:1987du}\tabularnewline
\hline 
$\mathcal{B}\left(\tau\rightarrow\mu^{-}+\mu^{+}+\mu^{-}\right)$ & $2.1\times10^{-8}$~\cite{Hayasaka:2010np}\tabularnewline
\hline 
\end{tabular}\caption{The current bounds for different LFV observables.}
\label{Bounds} 
\end{table}

The current experimental bounds in Table \ref{Bounds} must be fulfilled
by the interactions (\ref{L}), as well other bounds such as DM relic
density if $N_{1}$ is considered as a DM candidate. If this is the
case, the main annihilation channel would be the $S^{\pm}$-mediated
process $N_{1}N_{1}\rightarrow\ell_{\alpha}\ell_{\beta}$. In case
in which there exist other annihilation channels,\footnote{Similar to the cases in Ref.~\cite{Ahriche:2015loa}.}
an extra contribution to the total annihilation cross section will
affect the relic density value. Therefore, to take into account
this case, one has to adjust the charged scalar mass and the new Yukawa
couplings in order to ensure $\Omega_{DM}h^{2}$ $=0.1186\pm0.0020$~\cite{Ade:2015xua}.

In this work, we will focus on the process $e^{-}e^{+}\rightarrow b\bar{b}+\slashed{E}_{T}$
considering unpolarized beams for two c.m. energies: $E_{c.m.}=500~\mathrm{GeV}$
and $1~\mathrm{TeV}$. Then, we expand our analysis and discussion
by using the different beam polarizations at the electron-positron
linear colliders that can be available at the ILC and CLIC, where
in SM, the process mentioned above has three subprocesses, in which
the missing energy is the light SM neutrinos $\slashed{E}_{T}^{(SM)}\equiv\nu_{\alpha}\bar{\nu}_{\alpha}$,
where $\alpha=\mu,e,\tau$. In the case of RHN DM, the heavier RHNs, $N_{2,3}$, are pair produced at the
collider, and decay into pairs of charged leptons $\ell_{\alpha R}$ $\ell_{\beta R}$ ($\alpha,\beta=\mu,e,\tau$)
and a pair of DM $N_{1}N_{1}$ via $S^{\pm}$-mediated processes.

If $m_{N_{1,2,3}}<m_{S}$, in this case $N_{2,3}$ has a three-body
decay, and therefore may decay outside of the detector. In the inverse
case $m_{N_{1}}<m_{S}<m_{N_{2,3}}$, $N_{2,3}$ has a two-body decay
with a larger decay width and a smaller distance, which should be inside
the detector. Then, the missing energy in the process $e^{-}e^{+}\rightarrow b\bar{b}+\slashed{E}_{T}$
could be defined in the three cases as:

1. $\slashed{E}_{T}=N_{1}N_{1}$, if $N_{2}$ and $N_{3}$ decays
inside the detector,

2. $\slashed{E}_{T}=N_{1}N_{1},N_{1}N_{2},N_{2}N_{2}$, if only $N_{2}$
decays outside the detector,

3. $\slashed{E}_{T}=N_{1}N_{1},N_{1}N_{2},N_{1}N_{3},N_{2}N_{2},N_{2}N_{3},N_{3}N_{3}$,
if all $N_{2,3}$ decay outside the detector.

To check whether these three cases correspond to $m_{N_{1,2,3}}<m_{S}$,
$m_{N_{1,2}}<m_{S}<m_{N_{3}}$, and $m_{N_{1}}<m_{S}<m_{N_{2,3}}$,
respectively, one should estimate the distance travelled by the heavier
RHNs $N_{2,3}$.

The distance $D_{i}$ travelled by the heavier RHNs $N_{i}$ can be
defined by

\begin{equation}
\frac{D_{i}}{1~c.m.}=1.98\times10^{-4}\left(\frac{\Gamma_{i}}{10^{-7}~MeV}\right)^{-1}\left(\frac{E_{i}^{2}}{m_{N_{i}}^{2}}-1\right)^{1/2},
\end{equation}
where $\Gamma_{i}$, $m_{N_{i}}$, and $E_{i}$ are the heavy RHN's decays
widths, masses, and energies respectively, and $i=2,3$. Here, the total
decay width of $N_{i}$, $\Gamma_{i}$, is estimated using LANHEP/CALCHEP~\cite{Semenov:2008jy,Belyaev:2012qa}.

\begin{figure}[ht]
\includegraphics[width=0.33\textwidth]{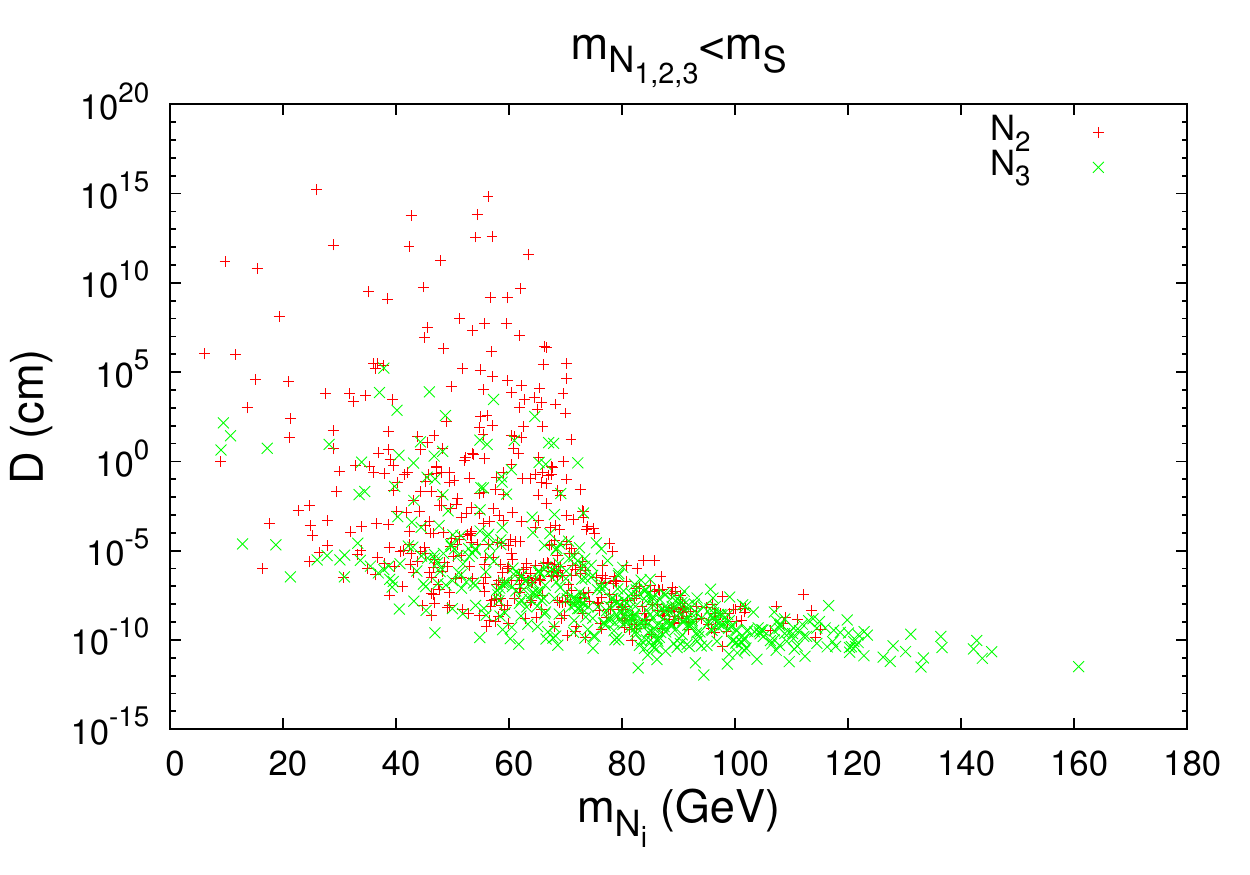}~\includegraphics[width=0.33\textwidth]{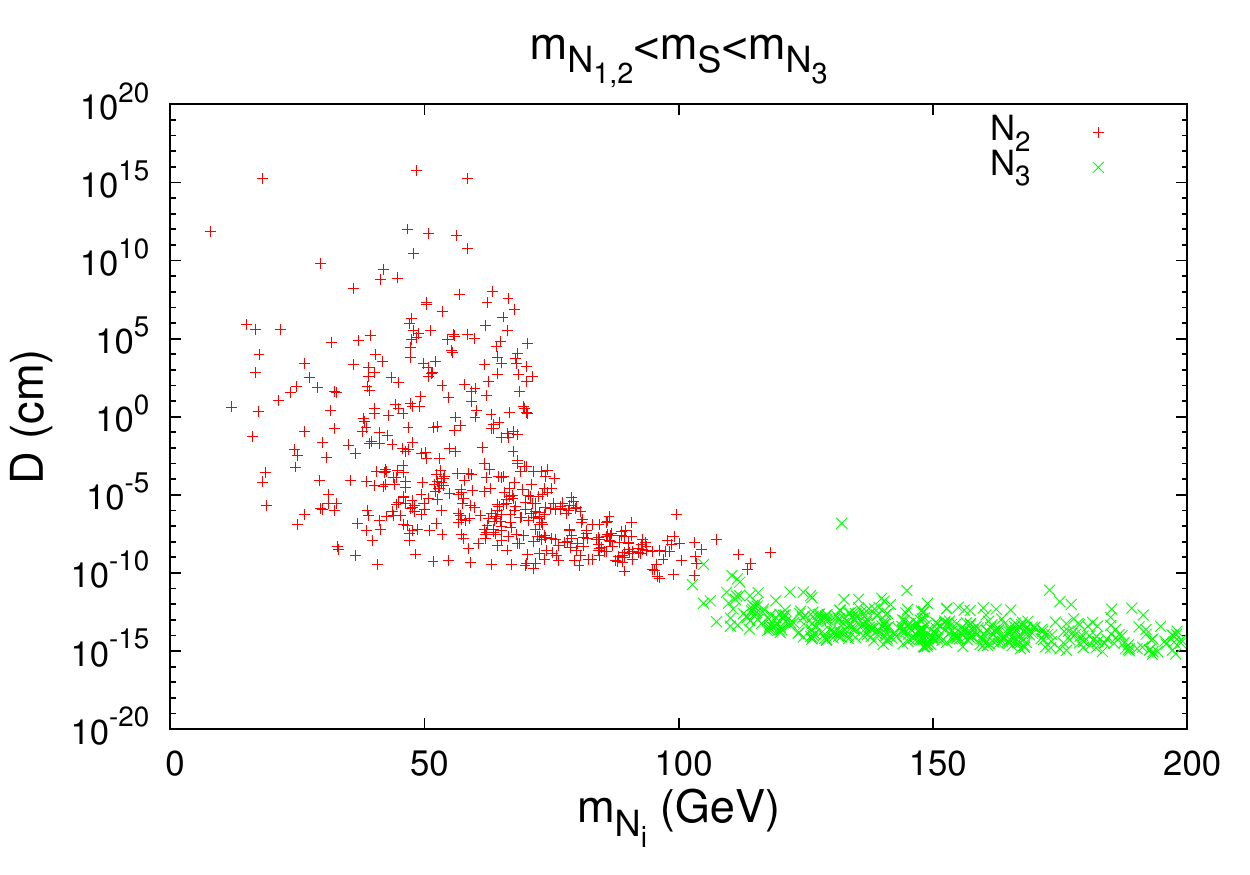}~\includegraphics[width=0.33\textwidth]{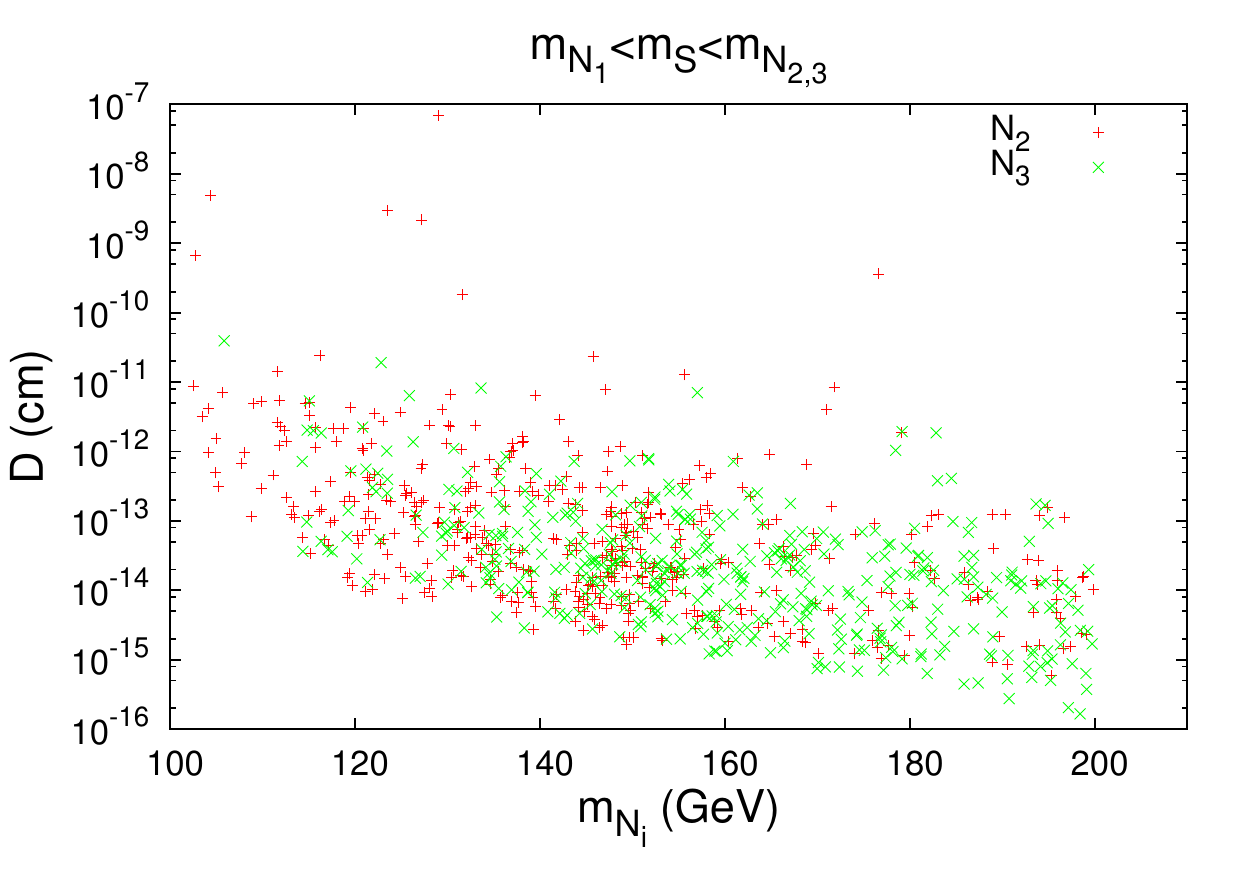}
\caption{The distance $D_{i}$ travelled by the heavier RHNs as a function
of their masses for the three cases: $m_{N_{1,2,3}}<m_{S}$ (left),
$m_{N_{1,2}}<m_{S}<m_{N_{3}}$ (middle) and $m_{N_{1}}<m_{S}<m_{N_{2,3}}$
(right). Here, we consider the typical energy for $N_{2,3}$ to be
$200\,GeV$. }
\label{Disp} 
\end{figure}

In Fig.~\ref{Disp}, we show the travelled distance $D_{i}$ as a function
of $m_{N_{2,3}}$ for the three aforementioned cases for 500 benchmark
points that fulfil the muon anomalous magnetic moment and the LFV
bounds on $\ell_{\alpha}\rightarrow\ell_{\beta}+\gamma$ and $\ell_{\alpha}\rightarrow\ell_{\beta}+\bar{\ell_{\beta}}+\ell_{\beta}$.

It is clear from Fig.~\ref{Disp} that $N_{3}$ decays mostly inside
the detector except for a few benchmark points in the case in which the
charged scalar is the heaviest. For the RHN $N_{2}$, it decays inside
the detector in the case in which it is heavier than the charged scalar.
In the inverse case, it could decay either inside or outside the
detector depending on the couplings.

A search with a negative result had been performed by the L3 Collaboration
at LEP-II about a single photon with a missing energy signal at c.m. energies
189 and $209~\mathrm{GeV}$ with the corresponding
luminosity values 176 and $130.2~\mathrm{pb^{-1}}$,
respectively~\cite{Achard:2003tx}. Based on this negative search,
we will constrain our parameters space.

Using LANHEP~\cite{Semenov:2008jy} to implement the model (\ref{L})
and CALCHEP~\cite{Belyaev:2012qa} to compute the cross sections
for the background $e^{-}e^{+}\rightarrow\nu\bar{\nu}\gamma$ and
the signal $e^{-}e^{+}\rightarrow\gamma+\slashed{E}_{T}$, taking
into account the same cuts used by LEP-II to search for the single
photon events~\cite{Achard:2003tx}, we have the following:

$\bullet$ The polar angle of the photon is $|cos\theta^{\gamma}|<0.97$.

$\bullet$ The transverse momentum of photon must satisfy $p_{T}^{\gamma}>0.02~E_{c.m.}~(\mathrm{GeV})$.

$\bullet$ The energy of the photon must satisfy $E^{\gamma}>1~\mathrm{GeV}$.

\begin{figure}[ht]
\includegraphics[width=0.48\textwidth]{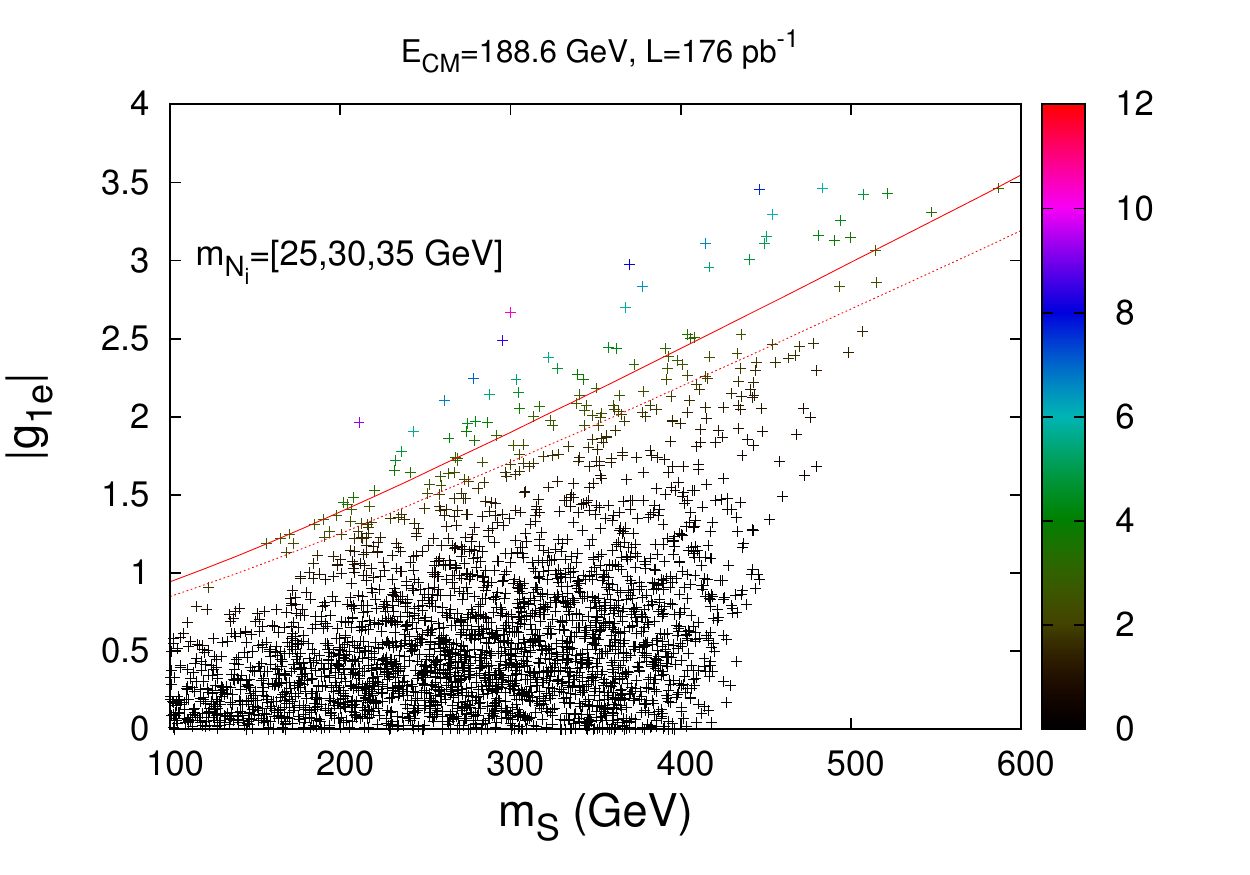}~\includegraphics[width=0.48\textwidth]{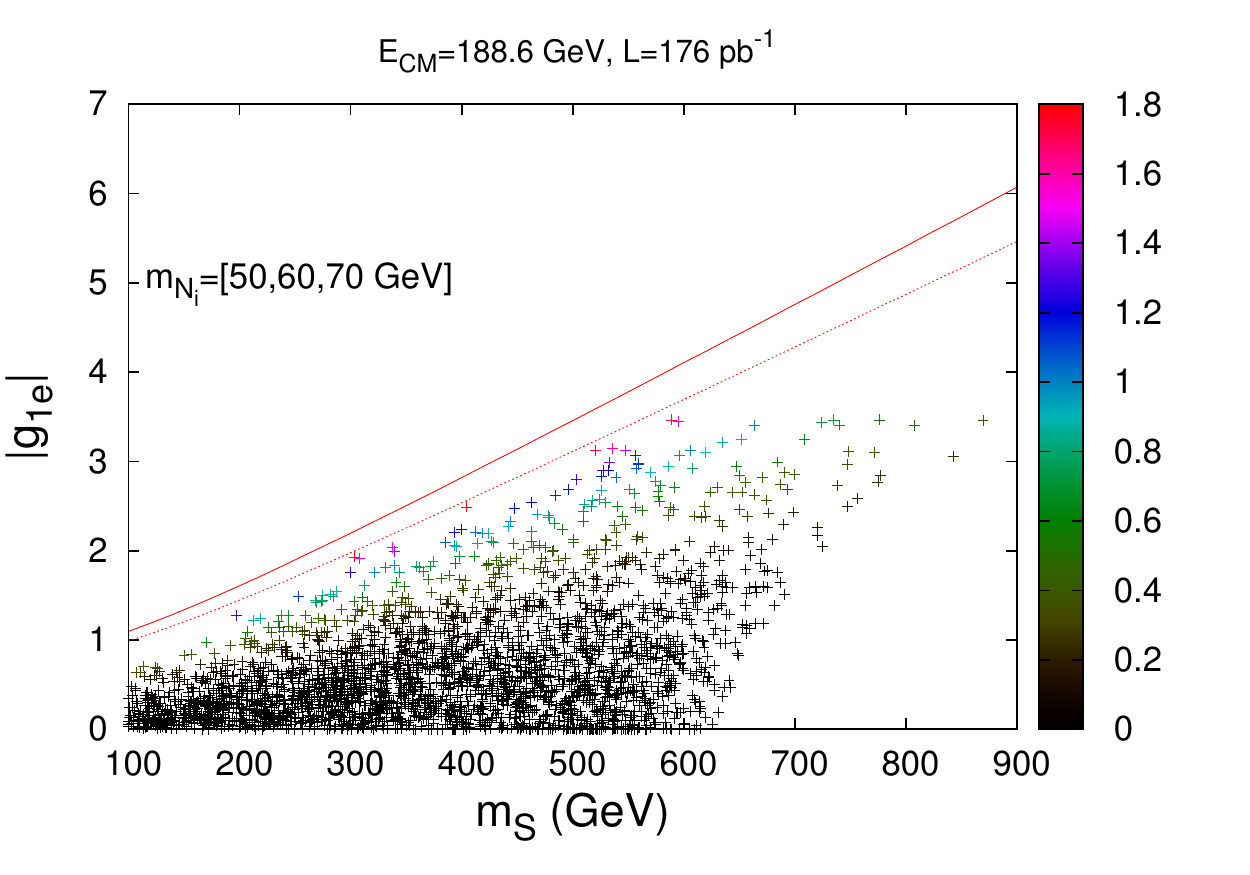}\\
 \includegraphics[width=0.48\textwidth]{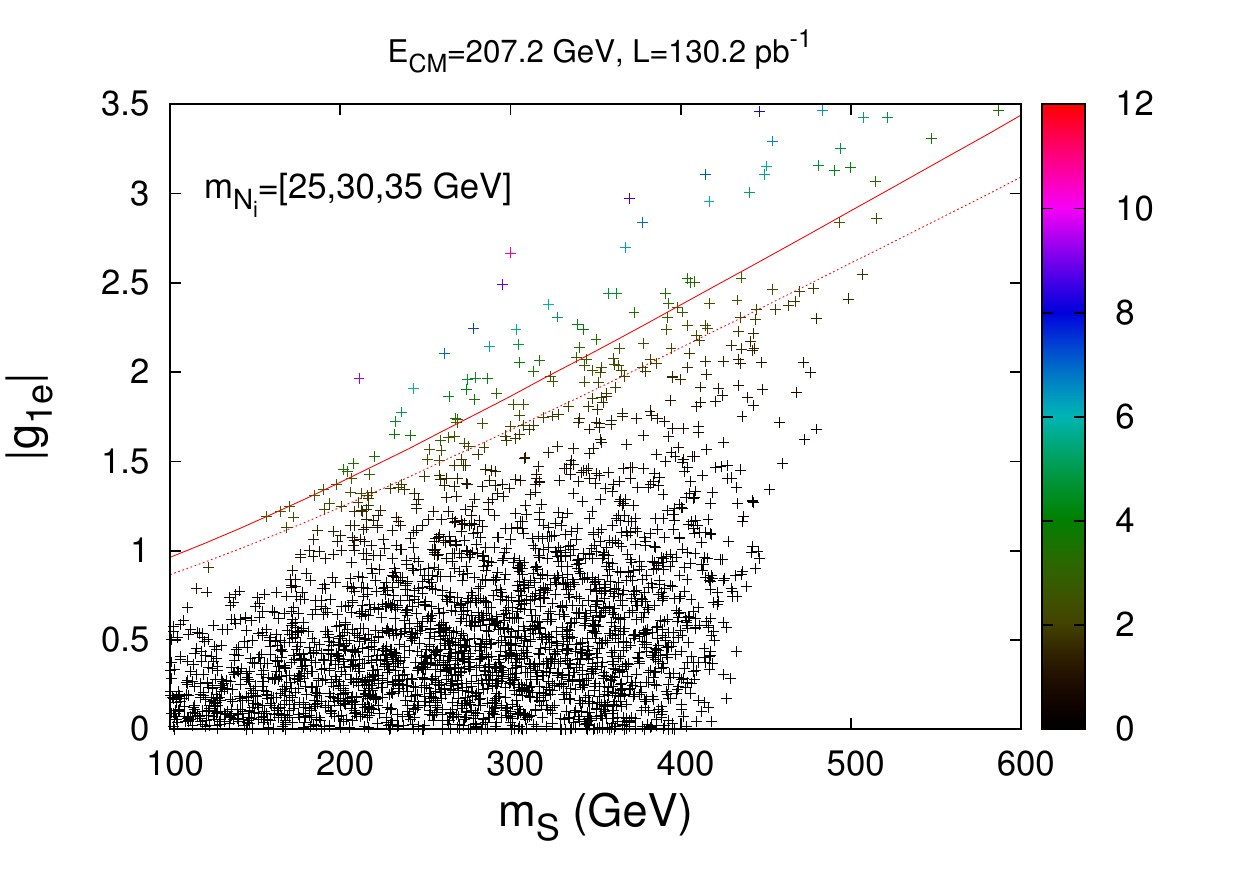}~\includegraphics[width=0.48\textwidth]{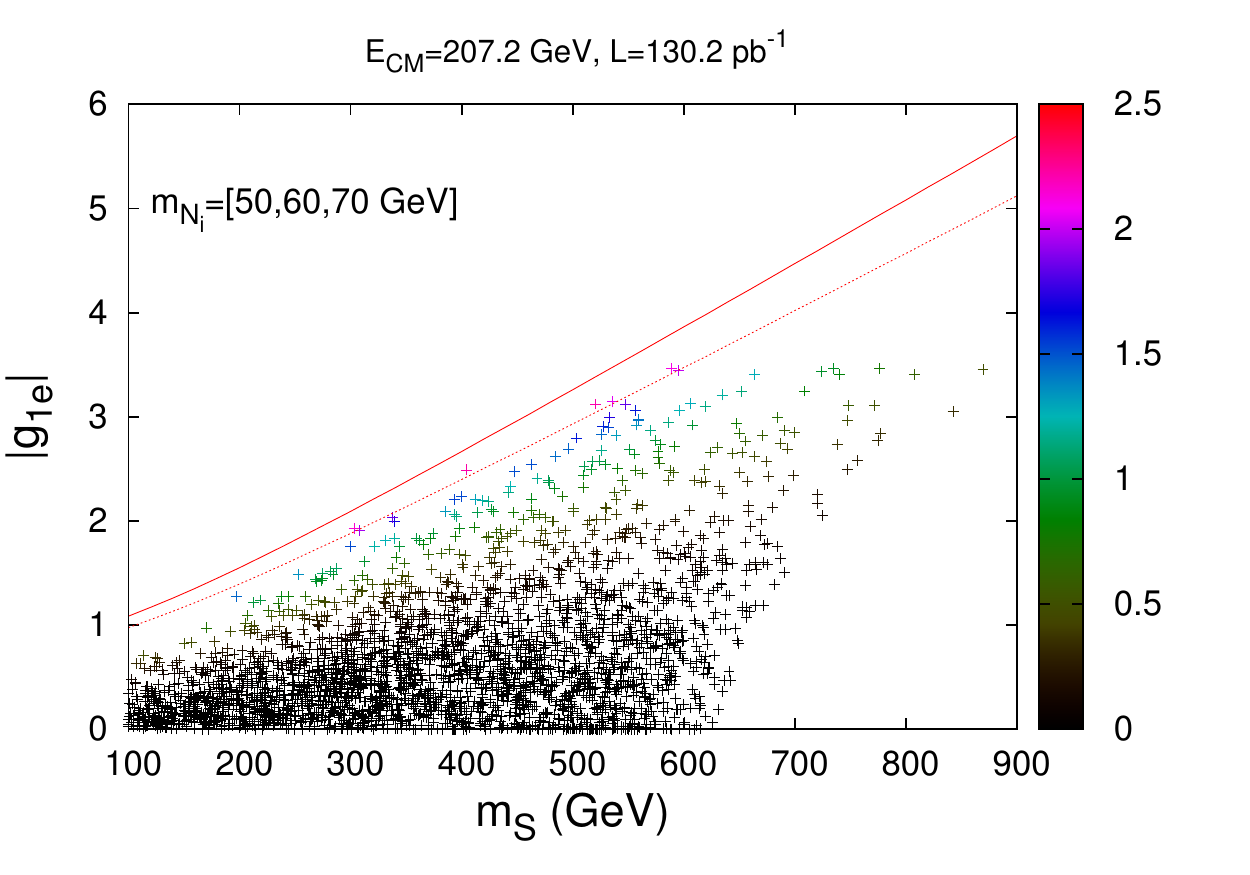}\\
 \caption{The coupling $\left|g_{1e}\right|$ as a function of $m_{S}$ for
3000 benchmark points with the two cases $m_{N_{i}}=\{25,30,35~\mathrm{GeV}\}$
(left) and $m_{N_{i}}=\{50,60,70~\mathrm{GeV}\}$ (right). The palette
represents the signal significance $S$ of the process $e^{-}e^{+}\rightarrow\gamma+\slashed{E}_{T}$
for the c.m. energy values $E_{c.m.}=188.6$ (up) and $E_{c.m.}=207.2$
(down) with the integrated luminosity $L=176~pb^{-1}$ and $L=130.2~pb^{-1}$,
respectively. The solid (dashed) line corresponds to the new constraint
at LEP II, which makes the signal significance smaller than $\mathcal{S}<3$
$(\mathcal{S}<2)$. For most of the benchmark points used here, the
missing energy is identified as $\slashed{E}_{T}=N_{1}N_{1}$, which
justifies the choice of $\left|g_{1e}\right|$ in the y axes.}

\label{exclud} 
\end{figure}

At the end, we generate 3000 benchmark points that are in agreement
with the bounds from the muon anomalous magnetic moment and the LFV
processes $\ell_{\alpha}\rightarrow\ell_{\beta}+\gamma$ and $\ell_{\alpha}\rightarrow\ell_{\beta}+\bar{\ell_{\beta}}+\ell_{\beta}$.
We distinguish two cases with $m_{N_{i}}=\{25,30,35~\mathrm{GeV}\}$
and $m_{N_{i}}=\{50,60,70~\mathrm{GeV}\}$. In Fig.~\ref{exclud},
we display the significance of the signal $e^{-}e^{+}\rightarrow\gamma+\slashed{E}_{T}$
(in the palette) for different values of the coupling $\left|g_{1e}\right|$
and the charged scalar mass for the two cases mentioned previously.

From Fig.~\ref{exclud}, one can remark that once the LFV bounds are
fulfilled the bound from LEP-II is also satisfied for $N_{1}$ heavier
than $50~\mathrm{GeV}$, whereas LEP-II could exclude some benchmark
points, especially using the analysis with $E_{c.m.}=207.2$~$\mathrm{GeV}$.
For our analysis, we consider the following numerical values shown
in Table \ref{Model34}, which we call model 3 ($M_{3}$) and model
4 ($M_{4}$). These values respect the muon anomalous magnetic moment
and LFV bounds in Table \ref{Bounds}.

\begin{table}[ht]
\centering \begin{adjustbox}{max width=\textwidth} %
\begin{tabular}{|c|c|c|}
\hline 
Models & \multicolumn{1}{c|}{} & Parameters\tabularnewline
\hline 
\hline 
 & $m_{N_{i}}\left(\mathrm{GeV}\right)$ & $25.788,~28.885,~36.274$,\tabularnewline
$M_{3}$ & $m_{S}$ $\left(\mathrm{GeV}\right)$ & $196.75$,\tabularnewline
 & $g_{i\alpha}/10^{-2}$ & $\left(\begin{array}{ccc}
75.063-i0.14367 & 0.0026819-i0.015758 & -136.03i-70.675\\
-3.6203-i35.9460 & -0.0035368+i0.041316 & 120.47-i286.100\\
-3.0602-i0.49553 & 0.057628-i0.2462700 & -235.27+i33.529
\end{array}\right)$,\tabularnewline
\hline 
 & $m_{N_{i}}\left(\mathrm{GeV}\right)$ & $62.184,~76.275,~95.736$,\tabularnewline
$M_{4}$ & $m_{S}$ $\left(\mathrm{GeV}\right)$ & $126.78$,\tabularnewline
 & $g_{i\alpha}/10^{-2}$ & $\left(\begin{array}{ccc}
-60.008+i2.4015 & -0.55187-i1.1133 & -32.641+i41.313\\
5.0213+i22.533 & 3.5209-i2.2480 & -112.35-i32.473\\
4.2829+i3.7764 & -2.2562+i2.3886 & -171.25-i94.890
\end{array}\right)$.\tabularnewline
\hline 
\end{tabular}\end{adjustbox} \caption{The parameters values for model 3 and model 4.}

\label{Model34} 
\end{table}

\begin{figure}[ht]
\centering \includegraphics[width=0.48\textwidth]{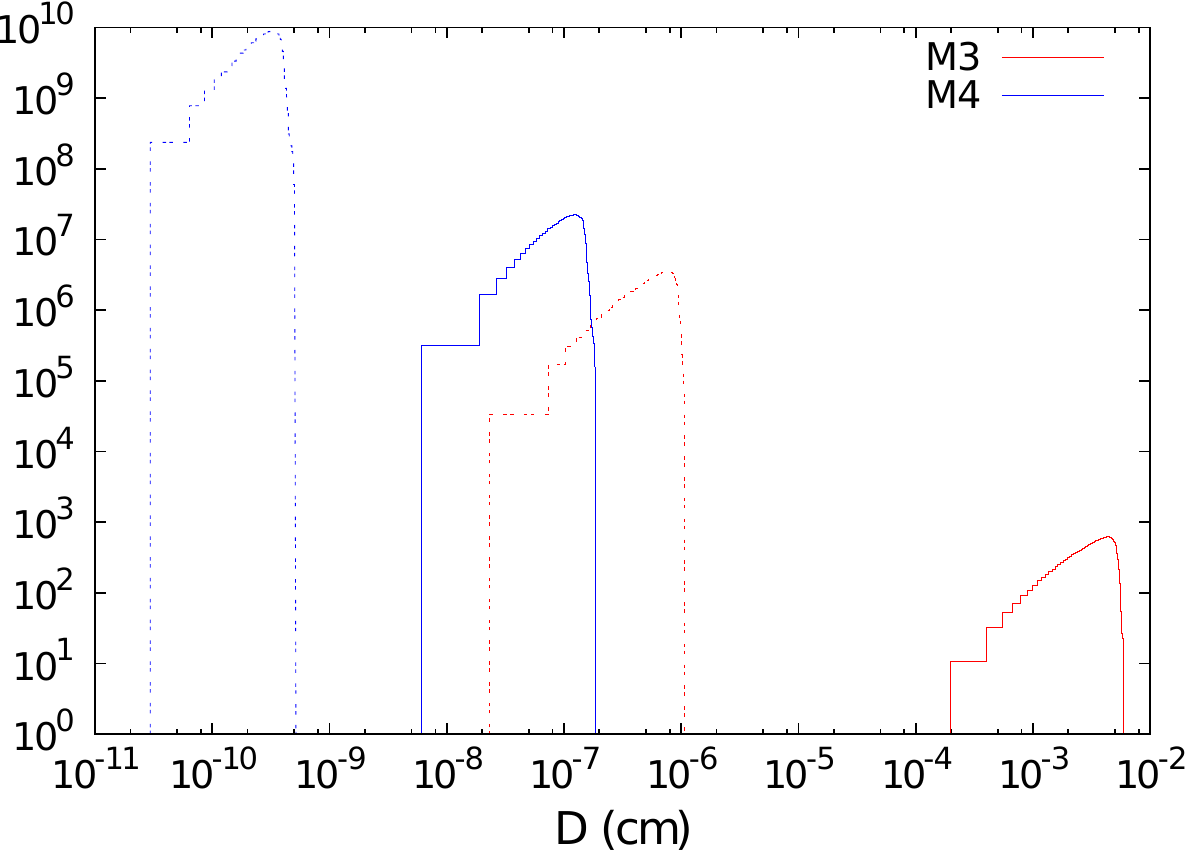}
\caption{The normalized distribution of the travelled distance $D_{i}$ (in cm)
by the heavier RHNs $N_{2}$ (solid) and $N_{3}$ (dashed) at $E_{c.m.}=500~\mathrm{GeV}$.}

\label{Di} 
\end{figure}

In Fig.~\ref{Di}, the normalized distribution of the travelled distance
$D_{i}$ for the heavier RHNs $N_{2,3}$ is shown for $M_{3,4}$.

It is very clear that the travelled distance $D_{i}$ is very small
for both heavy RHNs $N_{2,3}$ since their decay via a three-body
process $N_{2,3}\rightarrow N_{1}+\ell_{\alpha}+\ell_{\beta}$ for
both $M_{3}$ and $M_{4}$. This means that they both decay inside
the detector and can be accounted for missing energy, i.e., $\slashed{E}_{T}=N_{1}N_{1}$.

\section{FINAL STATE $b\bar{b}+\slashed{E}_{T}$ AT $e^{-}e^{+}$ COLLIDERS}\label{PRO}

\begin{figure}[ht]
\centering \includegraphics[width=0.8\textwidth]{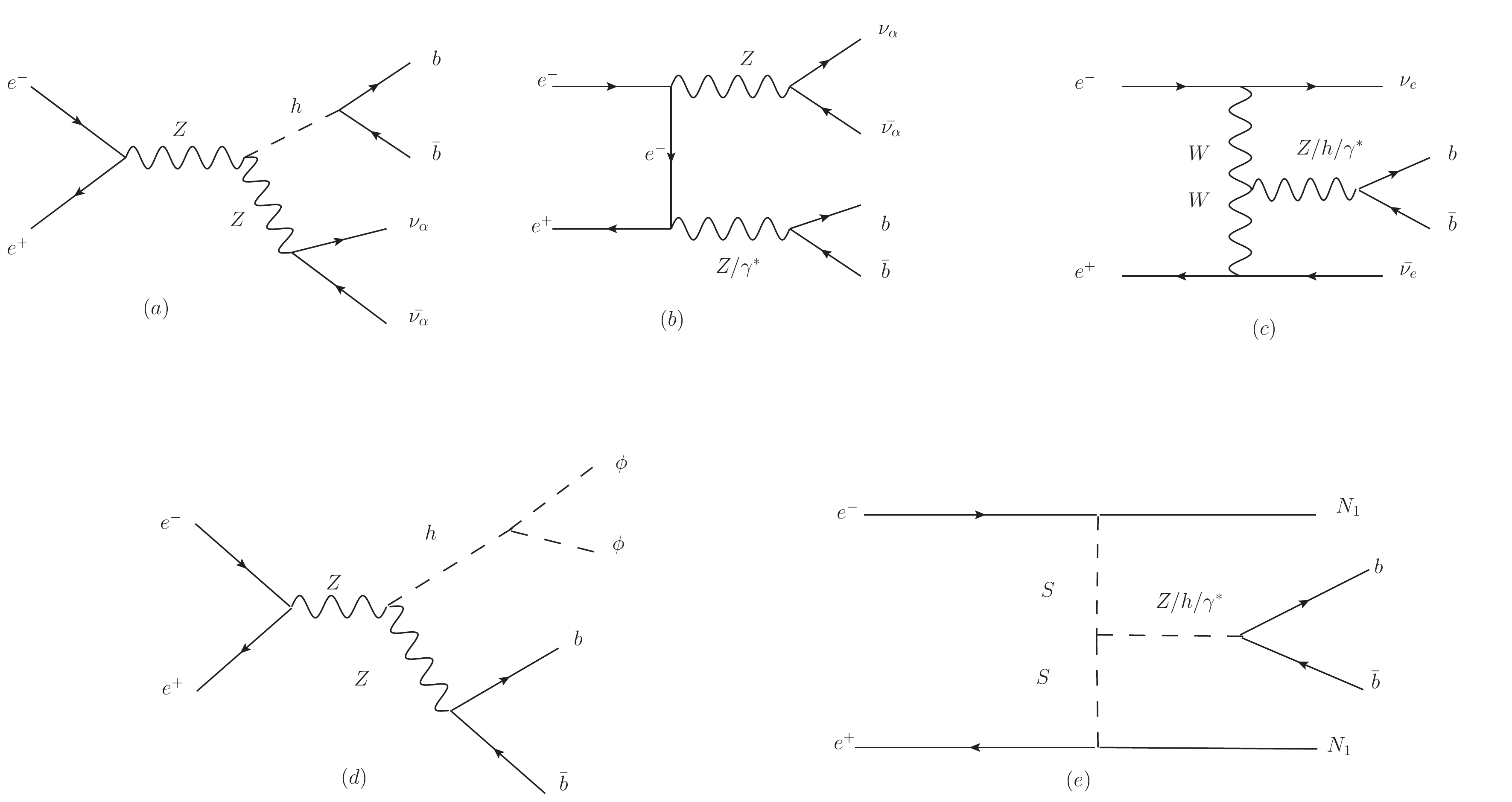}
\caption{The most important Feynman diagrams that contribute to background
(a), (b), and (c) [(d) and (e)] [the signal] for the process
$e^{-}e^{+}\rightarrow b\bar{b}+\slashed{E}_{T}$.}
\label{Diag} 
\end{figure}

The $m_{h}=125.09~\mathrm{GeV}$ Higgs has the dominant decay mode
${\cal B}\left(h\rightarrow b\bar{b}\right)=57.7\%$~\cite{Olive:2016xmw},
while the Z branching ratio ${\cal B}\left(Z\rightarrow b\bar{b}\right)=15.12\%$~\cite{Olive:2016xmw}
is also significant. Then, the choice of the channel $b\bar{b}+\slashed{E}_{T}$
is interesting since the b-tagging efficiency is shown to be about
$80\%$ when the misidentification efficiencies for c jet and u/d/s jet
are below $10\%$ and $1\%$, respectively, at both the ILC and CLIC~\cite{Suehara:2015ura}.
This is encouraging in considering the $b\bar{b}$ final state for our
studied models $M_{i}$ due to a possible clear signal. In this work,
we want to probe the interactions (\ref{ch}) and (\ref{L}) through
the final state $b\bar{b}+\slashed{E}_{T}$ at a leptonic collider.
This signal [Figs.~\ref{Diag}(d) and ~\ref{Diag}(e)] has the
background contributions $e^{-}e^{+}\rightarrow Z(Z,h,\gamma^{*})\rightarrow b\bar{b}+\slashed{E}_{T}$
[Fig.~\ref{Diag}(a) and (b)] in addition to the W-fusion diagrams
[Fig.~\ref{Diag}(c)].

\begin{figure}[ht]
\includegraphics[width=0.33\textwidth]{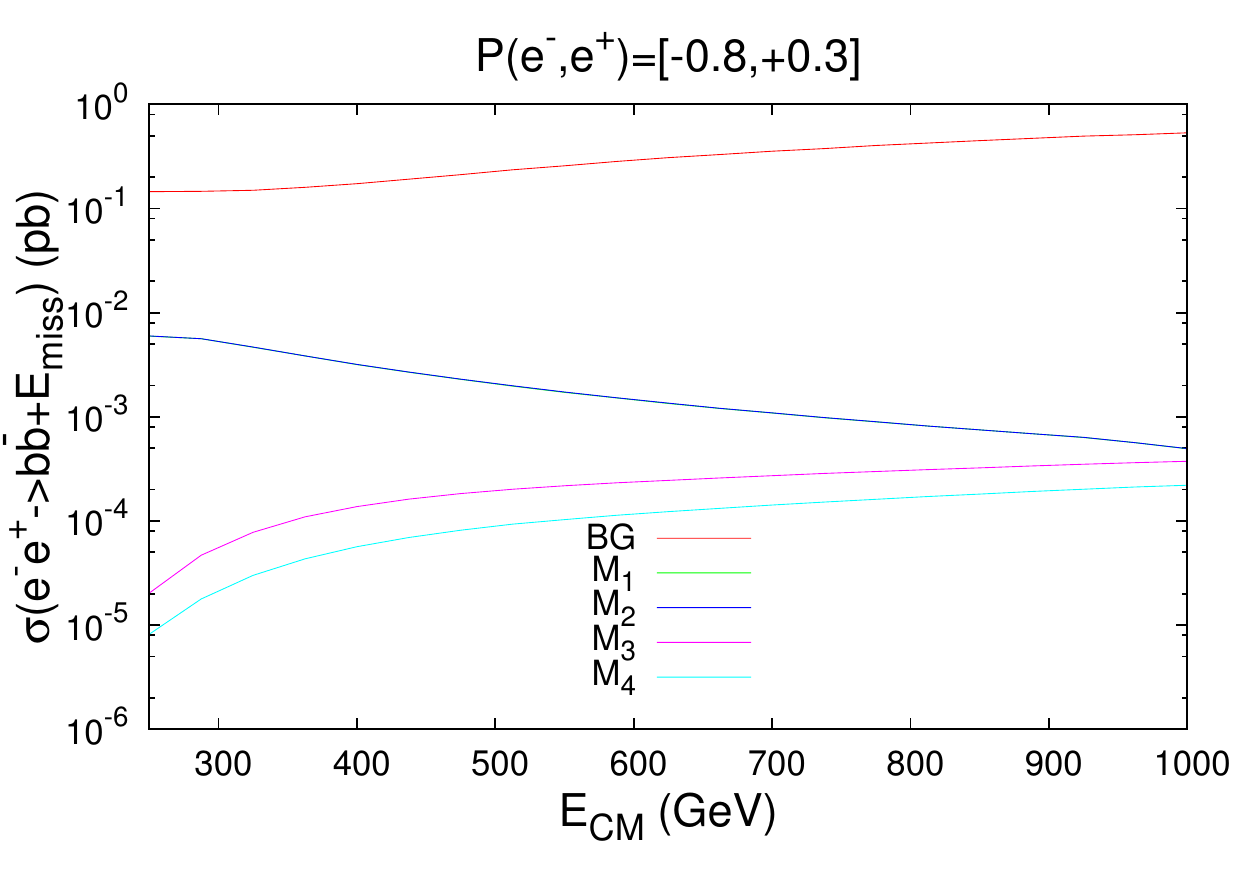}~\includegraphics[width=0.33\textwidth]{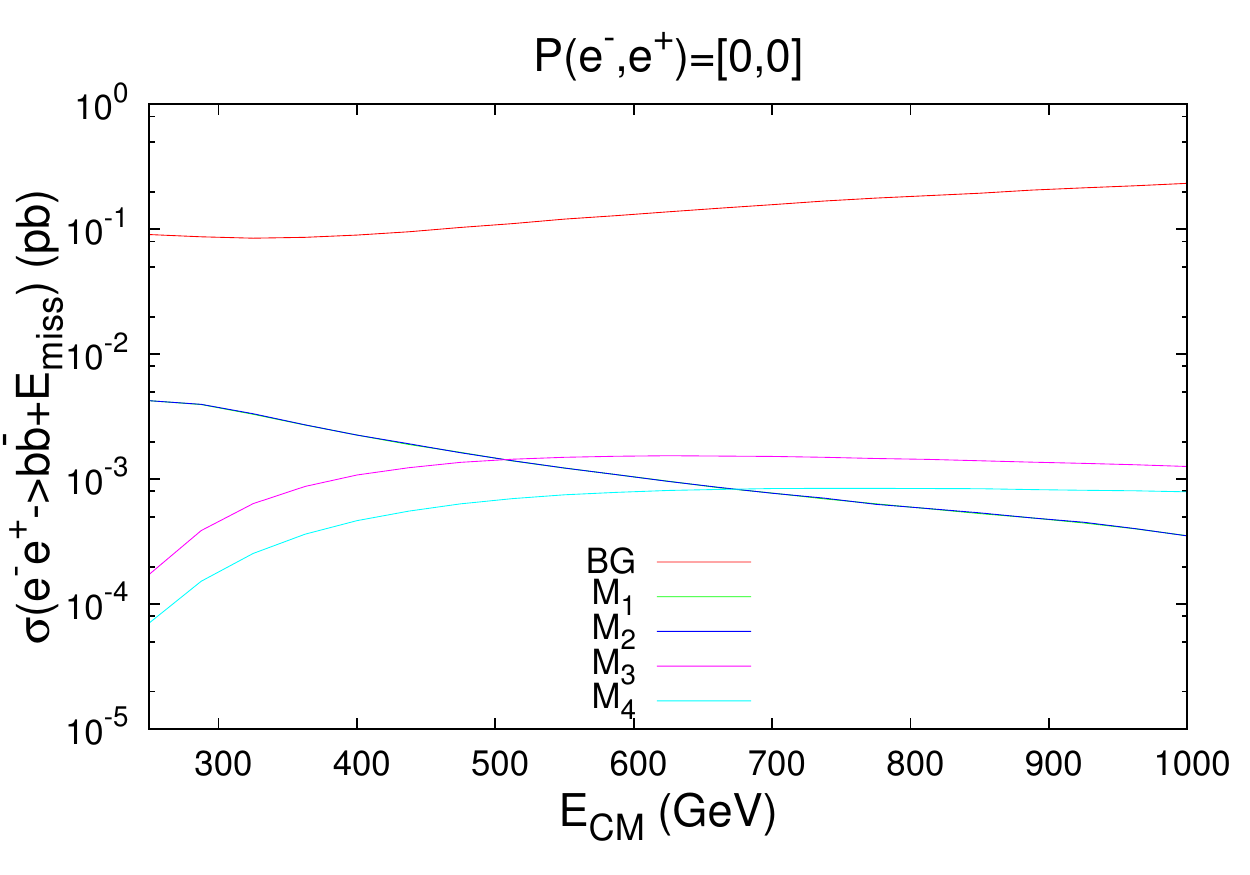}~\includegraphics[width=0.33\textwidth]{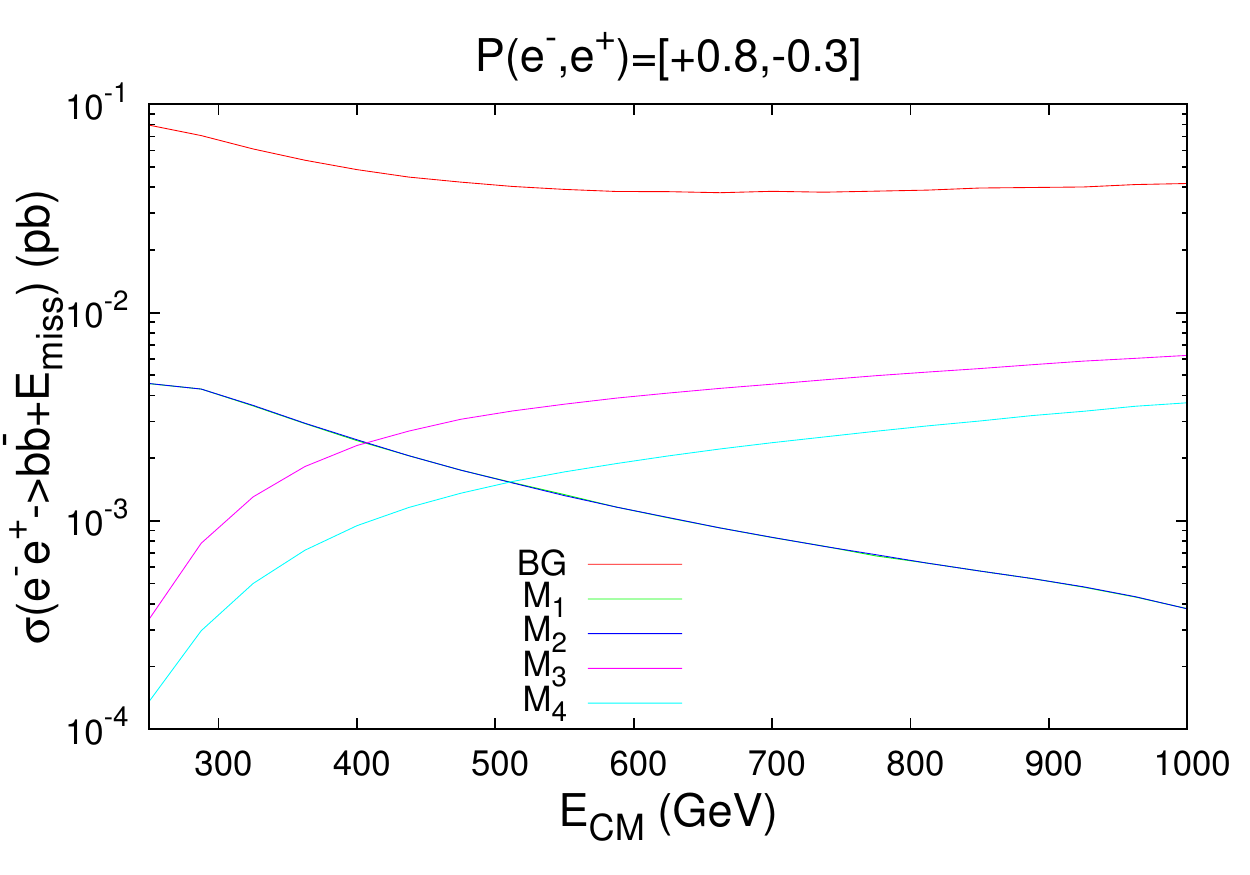}
\caption{The cross section of different models and background as a function
of $E_{c.m.}$ with the polarization $P\left(e^{-},e^{+}\right)=[-0.8,+0.3]$
(left) and $P\left(e^{-},e^{+}\right)=[+0.8,-0.3]$ (right) and
without polarization (middle). These figures are produced using the
packages LANHEP/CALCHEP~\cite{Semenov:2008jy,Belyaev:2012qa}.}
\label{Cross} 
\end{figure}

Future experiments such as the ILC~\cite{ILC4,Adolphsen:2013kya}
and CILC~\cite{Battaglia:2004mw,CLICP2} may use polarized beams
of electrons and positrons. This feature could help us to identify the
DM nature whether it is fermionic, vectorial, or scalar. Here, we will
consider both cases with and without polarized beams. By varying the
c.m. energy in the range $250~\mathrm{GeV}<E_{c.m.}<1~\mathrm{TeV}$,
we get in Fig.~\ref{Cross} the cross section of different models
and the background as a function of $E_{c.m.}$ with and without polarized
beams.

One can see from Fig.~\ref{Cross} that the cross sections of $M_{1}$
and $M_{2}$ are identical for the cases with polarized and unpolarized
beams. This feature is a numerical accident since the cross section
is proportional to the Higgs invisible branching ratio ${\cal B}_{inv}\left(h\rightarrow\phi\phi\right)$,
which has the same numerical value for $M_{1}$ and $M_{2}$, so the
aim of the choice in Table~\ref{Model12} is to find out the effect
of the scalar mass $m_{\phi}$ and the coupling $c_{s}$. One notices
also that the background cross section is increasing (decreasing)
for the cases with the polarizations $P\left(e^{-},e^{+}\right)=[0,0]$ and $[-0.8,+0.3]$
($P\left(e^{-},e^{+}\right)=[+0.8,-0.3]$) as a function of the c.m.
energy. This guides us to not consider the polarization $P\left(e^{-},e^{+}\right)=[-0.8,+0.3]$
in our analysis. For $M_{3}$ and $M_{4}$, the cross section is increasing
with respect $E_{c.m.}$ especially within the polarization $P\left(e^{-},e^{+}\right)=[+0.8,-0.3]$.
Then, we will consider the c.m. energy values $E_{c.m.}=500~\mathrm{GeV}$
and $1~\mathrm{TeV}$ in the rest of our work.

The general signal significance definition is given by\footnote{In Ref.~\cite{Cowan:2010js}, the authors used the notation $\mathcal{Z}_{0}$
for the significance, and here we use $\mathcal{S}$ instead.}~\cite{Cowan:2010js} 
\begin{equation}
\mathcal{S}=\sqrt{2\times\left[(N_{S}+N_{BG})\times log(1+N_{S}/N_{BG})-N_{S}\right]},\label{S}
\end{equation}
where $N_{S}$ and $N_{B}$ are the signal and background events numbers,
respectively. Here, $N_{S}$ is given by 
\begin{equation}
N_{S,BG}=\epsilon_{b}^{2}\times\mathcal{L}_{int}\times\sigma_{S,BG},
\end{equation}
with $\epsilon_{b}=0.8$ being the b-tagging efficiency factor,
$\mathcal{L}_{int}$ being the integrated luminosity, and $\sigma_{S,BG}$
being the signal or background cross section value.

\section{Analysis and Discussion}\label{DIS}

In this work, we used LANHEP packages~\cite{Semenov:2008jy} to implement
the models and generate their Feynman rules, and then we used CALCHep~\cite{Belyaev:2012qa}
to estimate the cross section and produce the differential cross section
for the background and signal at both $E_{c.m.}=500~\mathrm{GeV}$ and
$1~\mathrm{TeV}$. To define the cuts on the kinematic variables
that maximize the significance, we produced different distributions
and looked for ranges in which the background is reduced while keeping
the signal value. Therefore, we generated different distributions, taking
into account the following pre-cuts:

$\bullet$ The transverse momentum of the bottom quark $\left(b\right)$
and the bottom antiquark $\left(\bar{b}\right)$ must satisfy $p_{T}>15$
$\ \mathrm{GeV}$.

$\bullet$ The missing energy $\slashed{E}_{T}>30\ \mathrm{GeV}$.

$\bullet$ The invariant mass of the bottom quark $\left(b\right)$
and the bottom antiquark $\left(\bar{b}\right)$ must be in the range
$71~\mathrm{GeV}<M^{b,\bar{b}}<145~\mathrm{GeV}$.

$\bullet$ The jet separation radius must satisfy $\triangle R_{b,\bar{b}}>0.4$,
where $\triangle R$ is given by

\begin{equation}
\triangle R=\sqrt{\triangle\phi^{2}+\triangle\eta^{2}},
\end{equation}
where $\phi$ is the azimuthal angle and $\eta$ is the pseudorapidity.

The first two cuts helped too much to reduce the contamination in the
signal region. To ensure that the $b\bar{b}$ pair was produced
through a $Z$-gauge boson and/or the Higgs as shown in Figs.~\ref{Diag}(d)
and ~\ref{Diag} (e), we considered the third cut.

In the first step, we considered unpolarized beams of electrons and positrons
to generate the differential cross section for the background (SM)
and the signal (the models $M_{i}$) at both c.m. energies $E_{c.m.}=500~\mathrm{GeV}$
and $1~\mathrm{TeV}$. Then, we looked for kinematical variables regions
where the background was reduced and the signal was as maintained as possible.
Then, the full set of cuts is given in Table \ref{cut}.

\begin{table}[ht]
\centering \begin{adjustbox}{max width=\textwidth} %
\begin{tabular}{|c|c|}
\hline 
$E_{c.m.}$ & Selection cuts\tabularnewline
\hline 
\hline 
$500$ & $15~<p_{T}^{b}~$,$~30~<\slashed{E}_{T}~$,$~71~<M^{b,\bar{b}}<145~$,$~0.4<\triangle R_{b,\bar{b}},~90~\leq E_{T}^{b,\bar{b}}\leq230~,~210~\leq M_{T}^{b,\slashed{E}_{T}},$ \tabularnewline
\hline 
$1000$ & $15~<p_{T}^{b}~$,$~30~<\slashed{E}_{T}~$,$~71~<M^{b,\bar{b}}<145~$,$~0.4<\triangle R_{b,\bar{b}}$,$125~\leq E_{T}^{b,\bar{b}}~,~240~\leq M_{T}^{b,\slashed{E}_{T}}$.\tabularnewline
\hline 
\end{tabular}\end{adjustbox} \caption{The full set of cuts for the process $e^{-}e^{+}\rightarrow b\bar{b}+\slashed{E}_{T}$
at both c.m. energies $E_{c.m.}=500~\mathrm{GeV}$ and $1~\mathrm{TeV}$.
Here, $p_{T}^{b}$ is the transverse momentum of the bottom quark $\left(b\right)$,
$\slashed{E}_{T}$ is the missing energy, $M^{b,\bar{b}}$ is the
invariant mass of the bottom quark $\left(b\right)$ and the bottom
antiquark $\left(\bar{b}\right)$, $\triangle R_{b,\bar{b}}$ is the
jet cone angle, $E_{T}^{b,\bar{b}}$ is the transverse energy for
the bottom quark $\left(b\right)$ and bottom antiquark $\left(\bar{b}\right)$, and
$M_{T}^{b,\slashed{E}_{T}}$ is the transverse mass of bottom-missing 
energy. All masses and energies are given in $\mathrm{GeV}$.}

\label{cut} 
\end{table}

\subsection{Analysis using unpolarized beams}

By imposing the full set of cuts in Table \ref{cut} at both c.m. energies
$E_{c.m.}=500~\mathrm{GeV}$, and $1~\mathrm{TeV}$, using unpolarized
beams, we get the results shown in Table \ref{sini}.

\begin{table}[ht]
\centering \begin{adjustbox}{max width=\textwidth} %
\begin{tabular}{|c|c|c|c|c|c|c|c|}
\hline 
$E_{c.m.}~(\mathrm{GeV})$ & $\sigma^{BG}~(fb)$ & Models & $\sigma^{S}~(fb)$ & $\sigma'^{BG}~(fb)$ & $\sigma'^{S}~(fb)$ & $\mathcal{S}_{100}$ & $\mathcal{S}_{500}$ \tabularnewline
\hline 
\hline 
 & & $M_{1}$ & $1.475$ & & $0.520$ & $0.9808$ & $2.1936$ \tabularnewline
\cline{3-4} \cline{6-8} 
$500$ & $108.19$ & $M_{2}$ & $1.479$ & $17.804$ & $0.638$ & $1.2024$ & $2.6888$ \tabularnewline
\cline{3-4} \cline{6-8} 
 & & $M_{3}$ & $1.425$ & & $0.956$ & $1.7960$ & $4.0168$ \tabularnewline
\cline{3-4} \cline{6-8} 
 & & $M_{4}$ & $1.338$ & & $1.070$ & $2.0088$ & $4.4912$ \tabularnewline
\hline 
 & & $M_{1}$ & $0.352$ & & $0.282$ & $0.3216$ & $0.7192$ \tabularnewline
\cline{3-4} \cline{6-8} 
 & & $M_{2}$ & $0.353$ & & $0.292$ & $0.3328$ & $0.7448$\tabularnewline
\cline{3-4} \cline{6-8} 
$1000$ & $233.27$ & $M_{3}$ & $1.265$ & $49.072$ & $0.942$ & $1.0720$ & $2.3976$ \tabularnewline
\cline{3-4} \cline{6-8} 
 & & $M_{4}$ & $0.954$ & & $0.760$ & $0.8656$ & $1.9352$ \tabularnewline
\hline 
\end{tabular}\end{adjustbox} \caption{The cross section values of the background and the signal for each
model within the pre-cuts $\sigma^{BG}$, $\sigma^{S}$ and after
applying the full cuts set given in Table \ref{cut} $\sigma'^{S}$,$\sigma'^{BG}$
at both c.m. energies $E_{c.m.}=500~\mathrm{GeV}$ and $1~\mathrm{TeV}$.
The corresponding signal significance is shown for the luminosity values
$L=100~,~500~fb^{-1}$. }

\label{sini} 
\end{table}

Through the results presented in Table \ref{sini}, one notices that
the signal cross section within the full set of cuts gets reduced
a bit with respect to the case within the pre-cuts for all models
at both $E_{c.m.}=500~\mathrm{GeV}$ and $1~\mathrm{TeV}$, whereas,
the background cross section gets reduced by about $83.5\%$ ($79\%$)
at $E_{c.m.}=500~\mathrm{GeV}$ ($E_{c.m.}=~1~\mathrm{TeV}$). For luminosity
$L=100~fb^{-1}$, we do not see any deviation from the SM at both $E_{c.m.}=500~\mathrm{GeV}$
and $1~\mathrm{TeV}$. However, for $L=500~fb^{-1}$, one could notice
a deviation from the SM at $E_{c.m.}=500~\mathrm{GeV}$ for $M_{3,4}$.
At $E_{c.m.}=1~\mathrm{TeV}$, within the same luminosity value, we could
not even see a deviation from the SM for all models. Therefore, for
this c.m. energy, we require a large luminosity value (1 $ab^{-1}$ or
more) in order to see such a signal.

\begin{figure}[ht]
\includegraphics[width=0.33\textwidth]{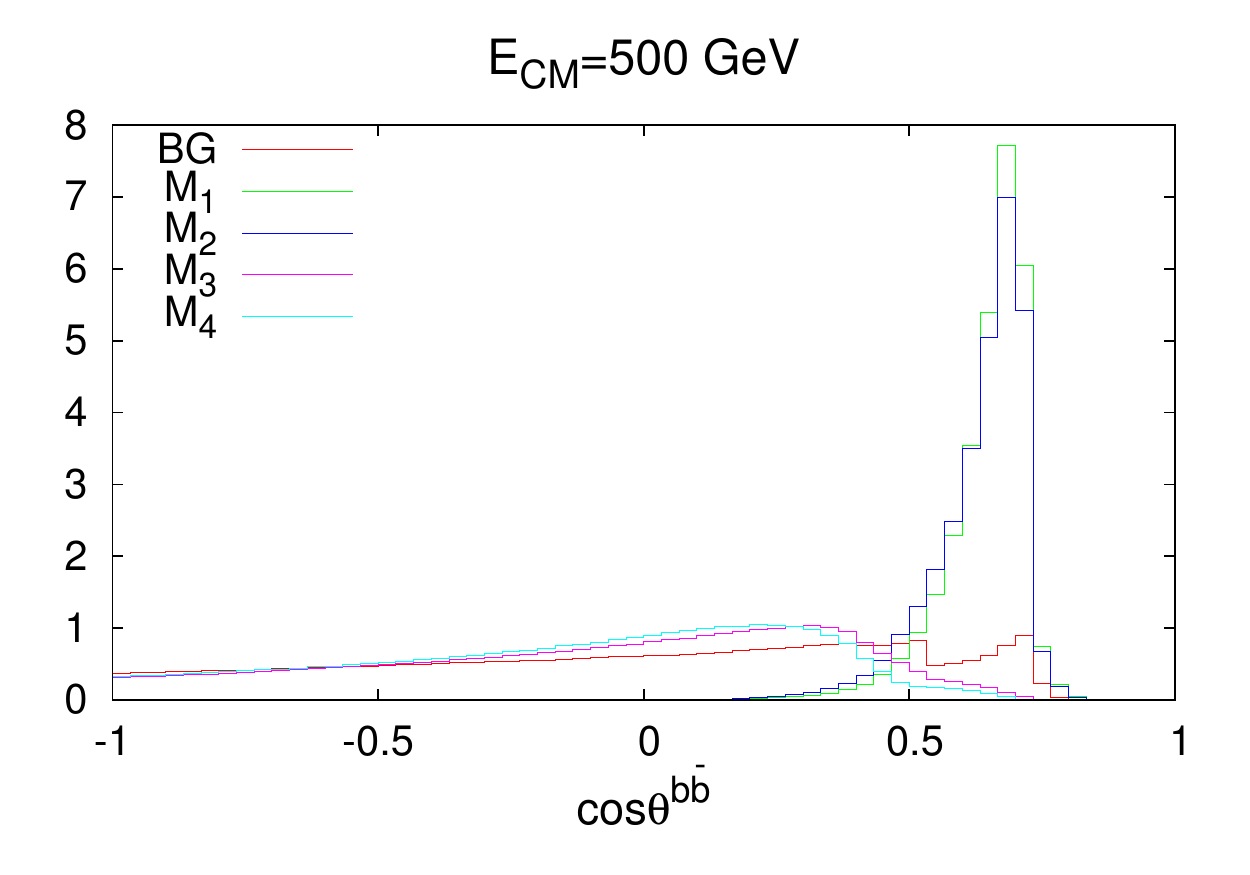}~\includegraphics[width=0.33\textwidth]{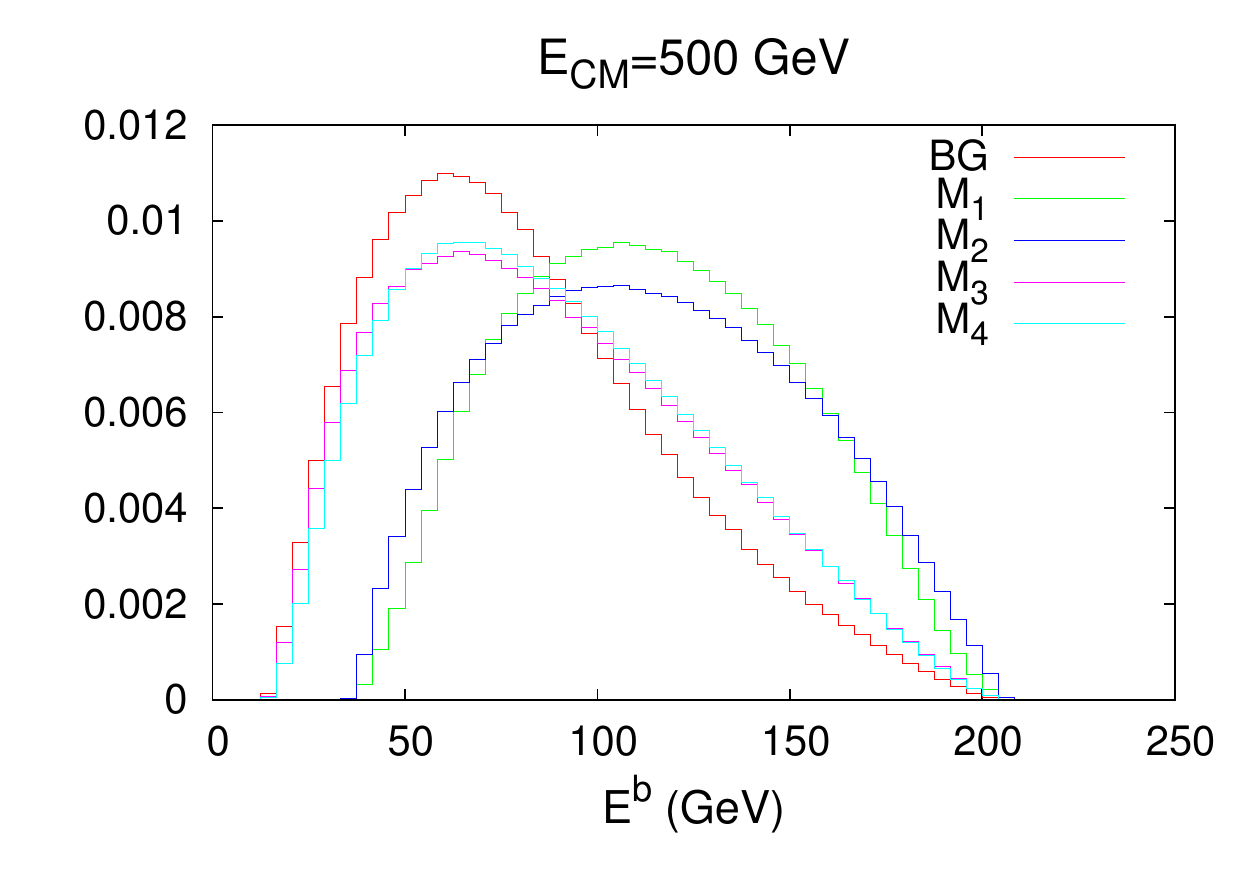}~\includegraphics[width=0.33\textwidth]{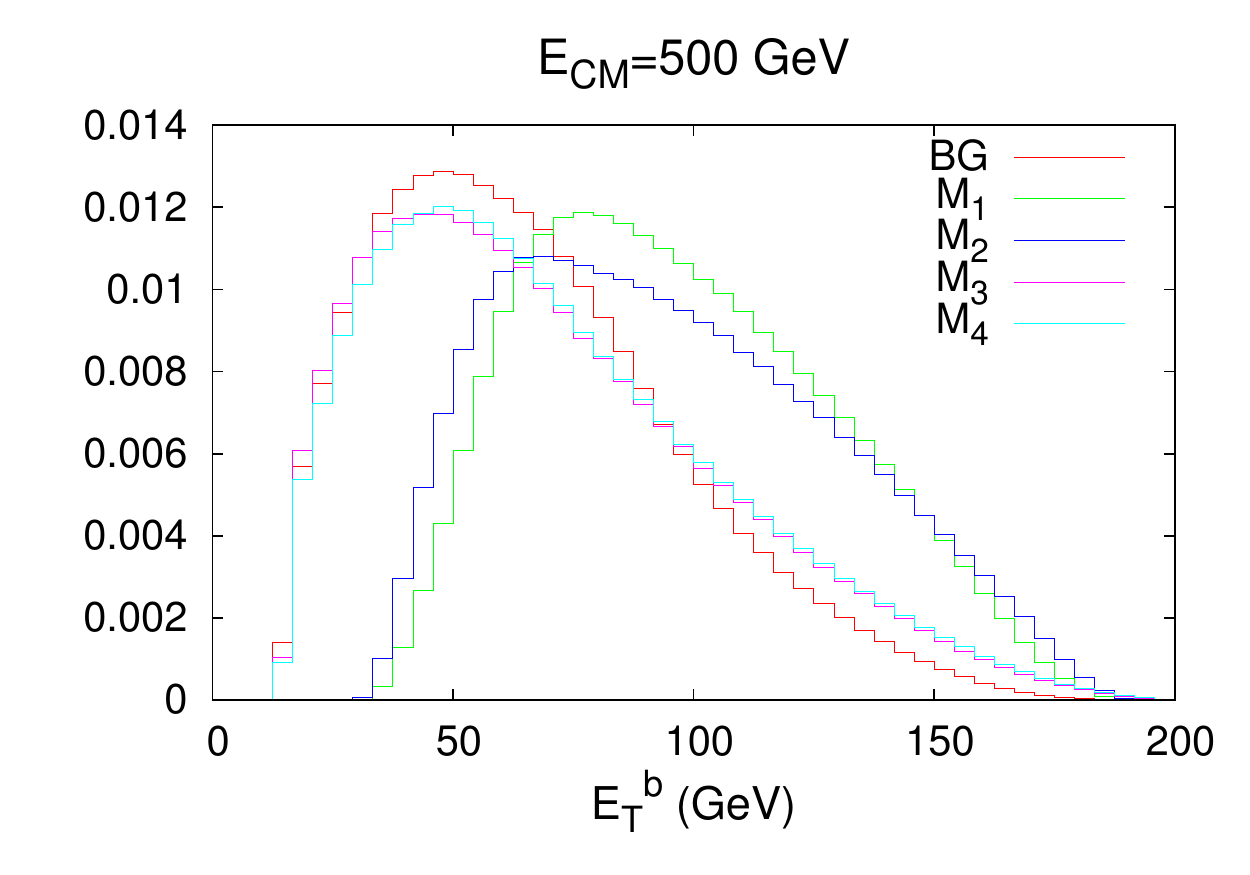}\\
 \includegraphics[width=0.33\textwidth]{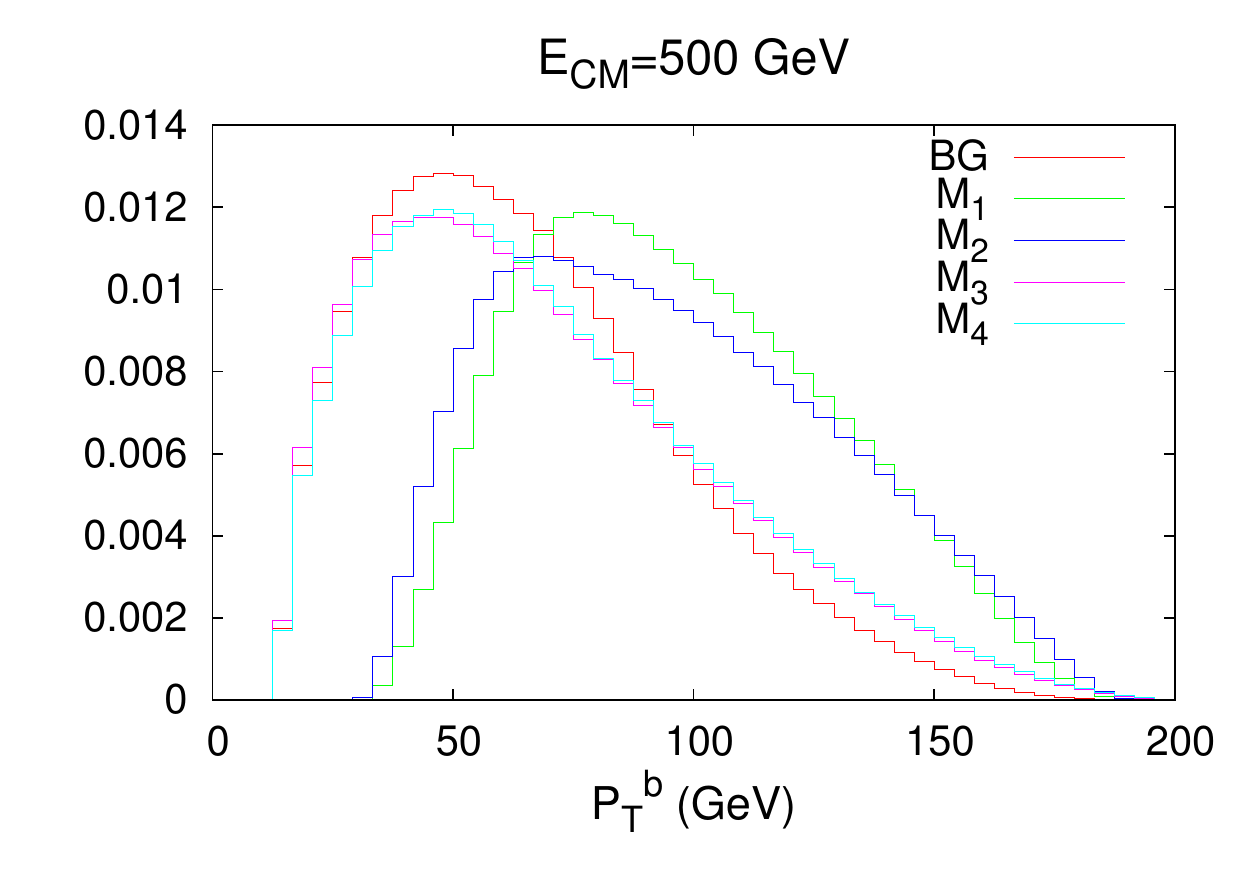}~\includegraphics[width=0.33\textwidth]{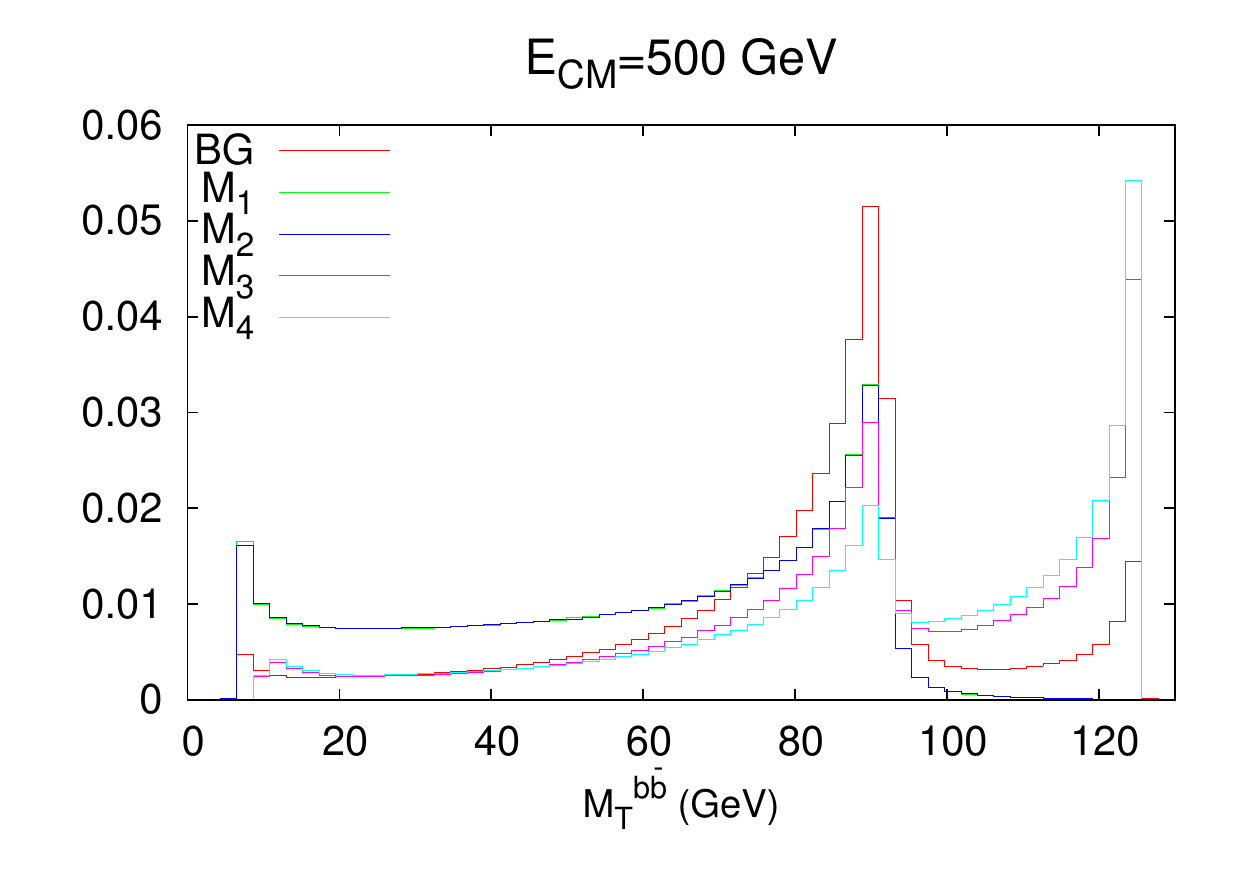}~\includegraphics[width=0.33\textwidth]{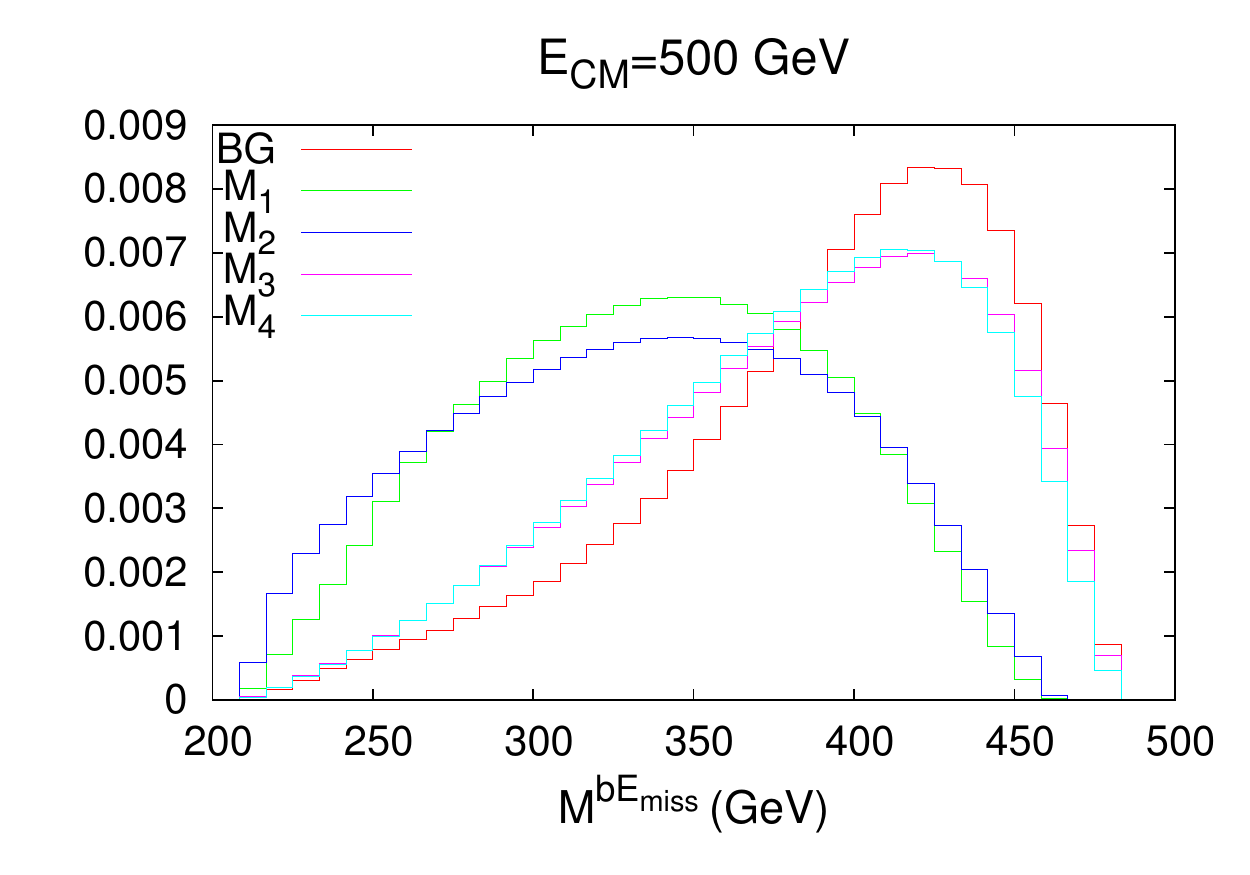}\\
 \includegraphics[width=0.33\textwidth]{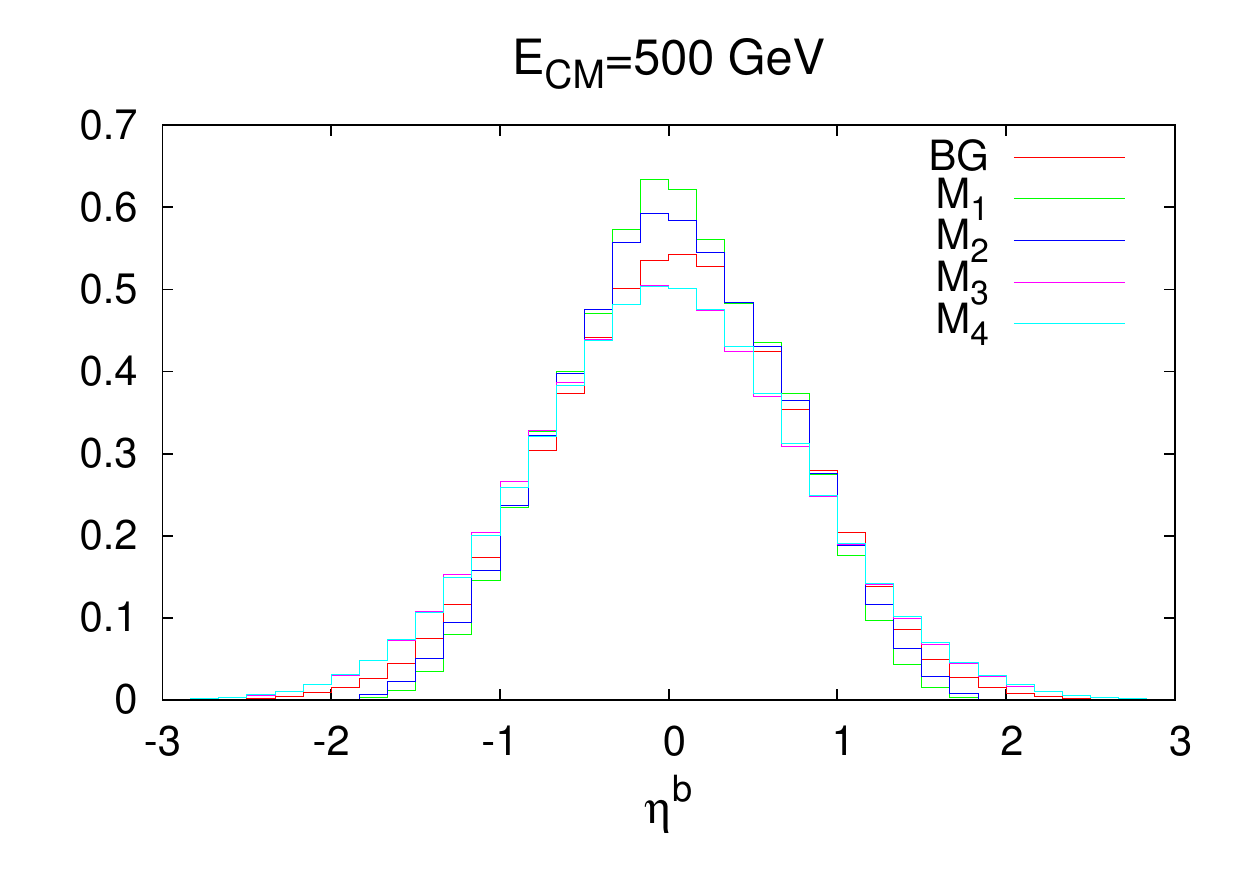}~\includegraphics[width=0.33\textwidth]{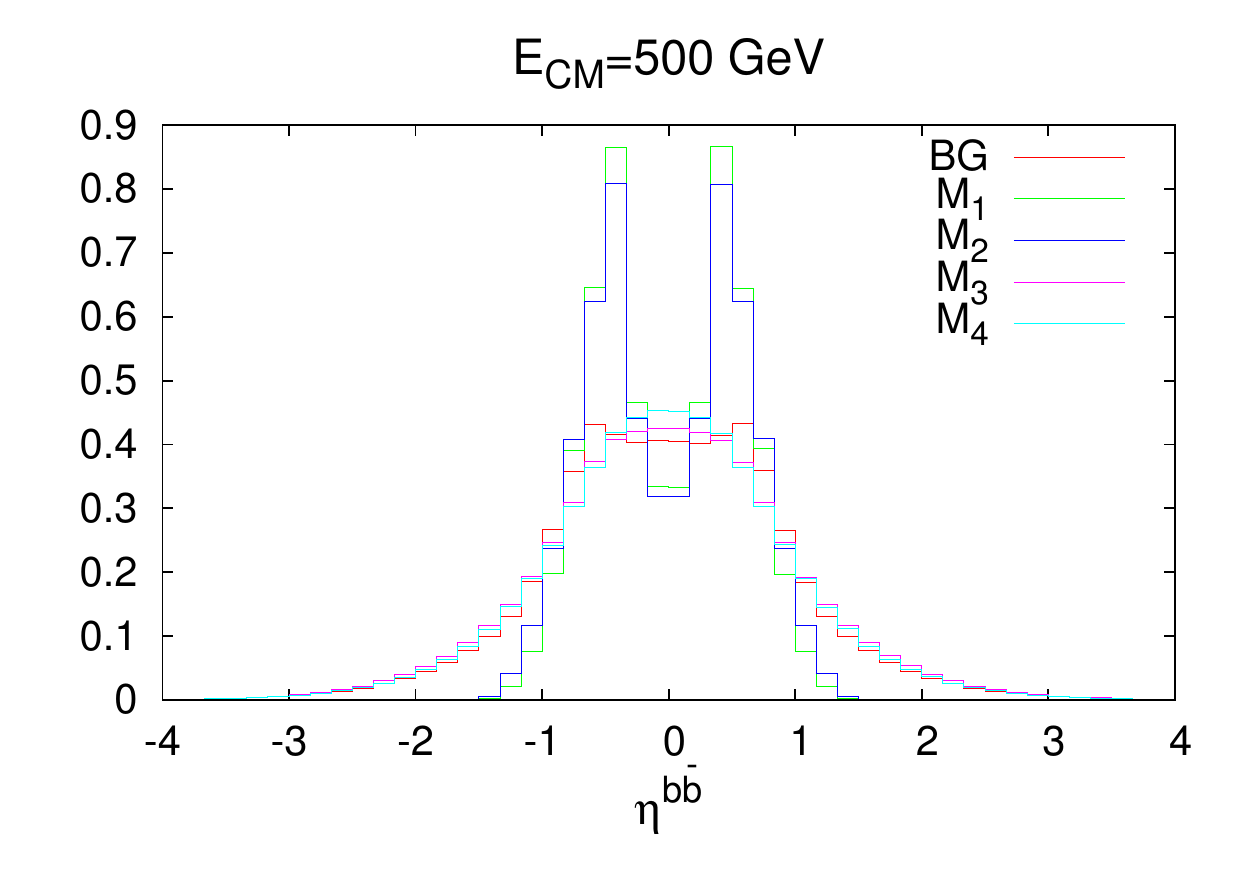}~\includegraphics[width=0.33\textwidth]{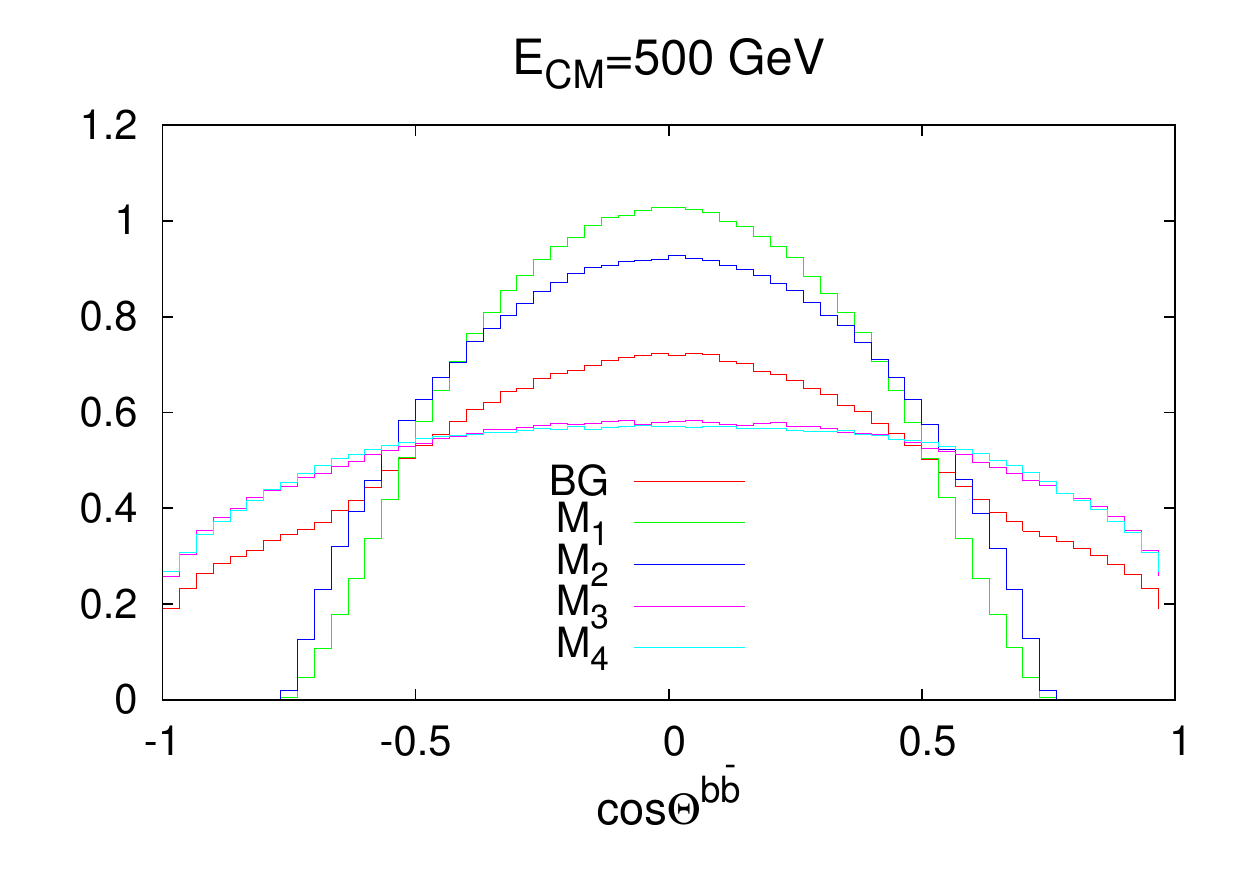}\\
 \caption{The relevant normalized distributions of the process $e^{-}e^{+}\rightarrow b\bar{b}+\slashed{E}_{T}$
at $E_{c.m.}=500~\mathrm{GeV}$.}

\label{dis.cuts500GeV} 
\end{figure}

\begin{figure}[ht]
\includegraphics[width=0.33\textwidth]{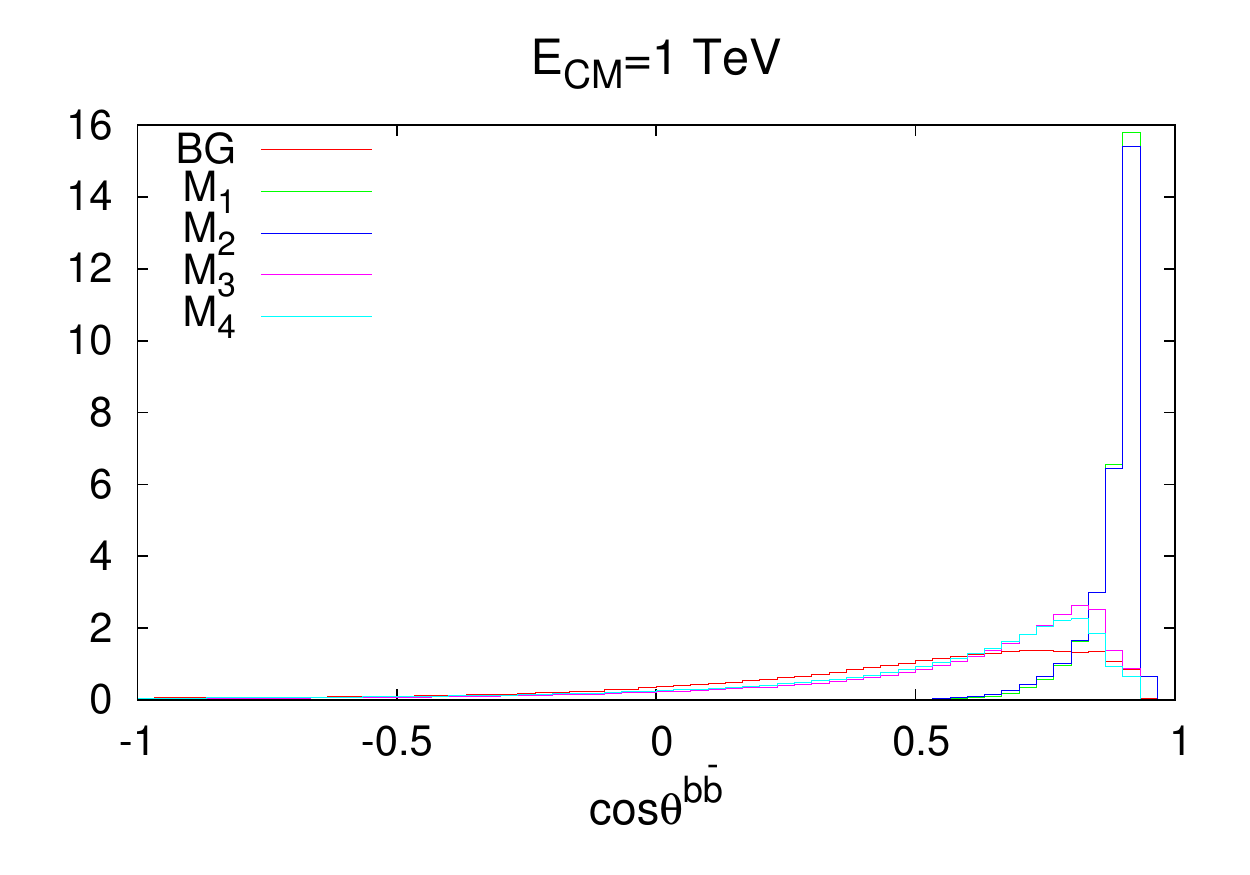}~\includegraphics[width=0.33\textwidth]{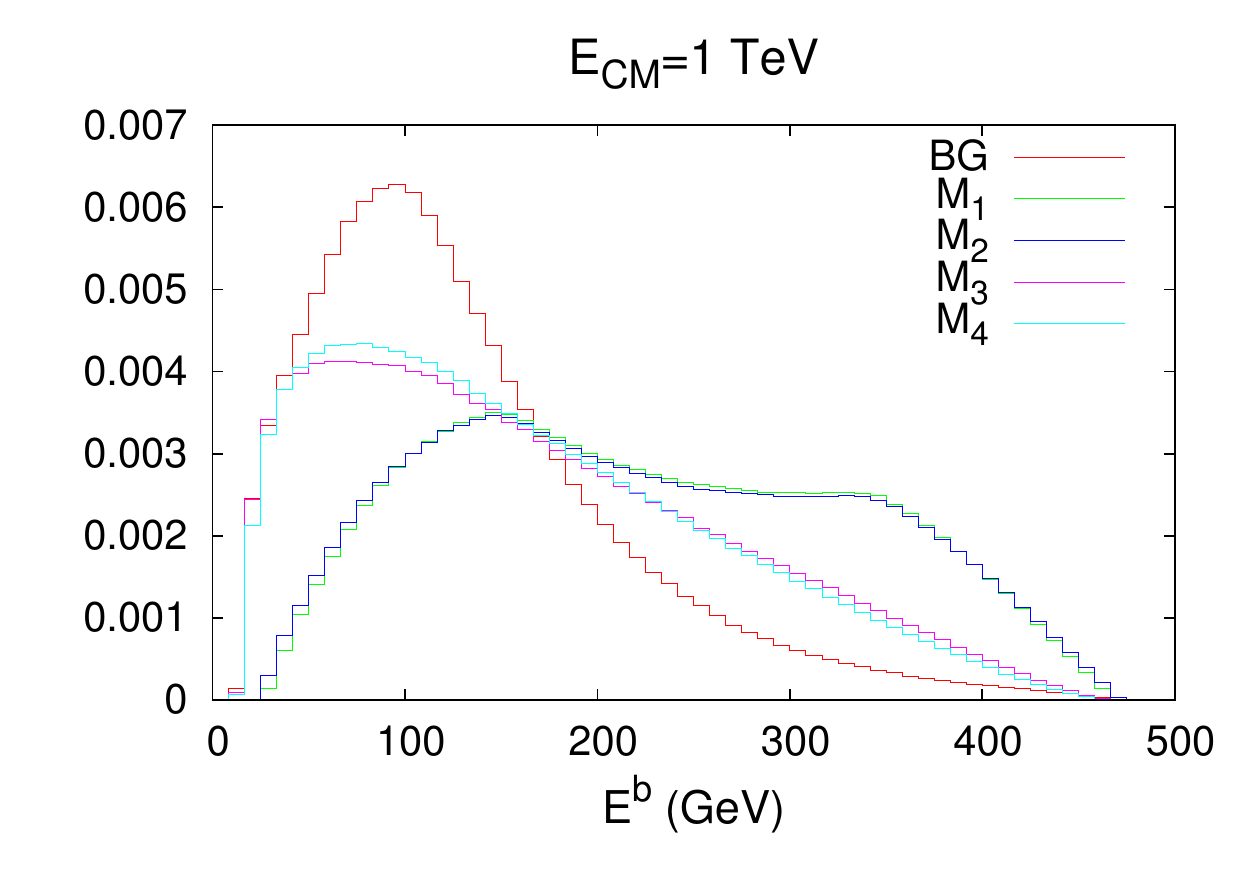}~\includegraphics[width=0.33\textwidth]{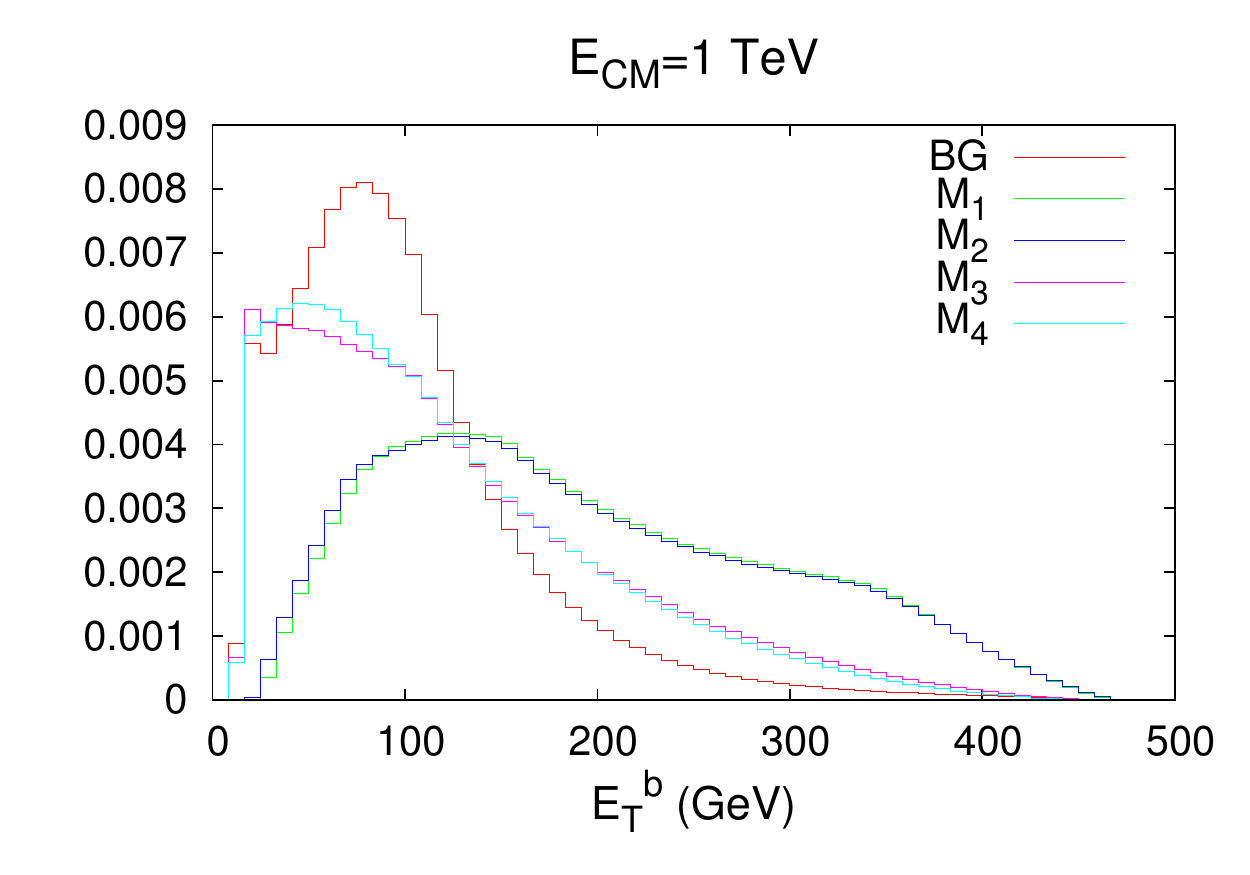}\\
 \includegraphics[width=0.33\textwidth]{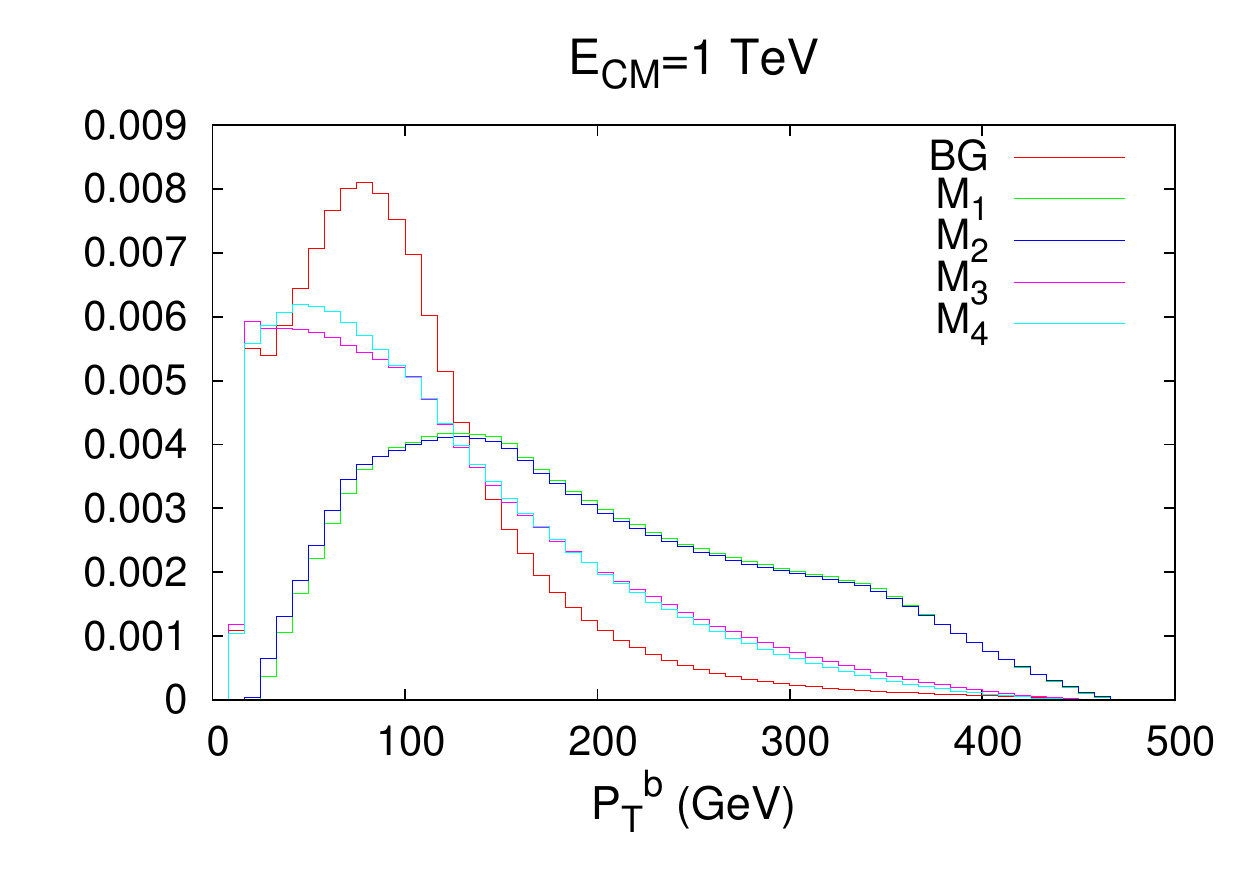}~\includegraphics[width=0.33\textwidth]{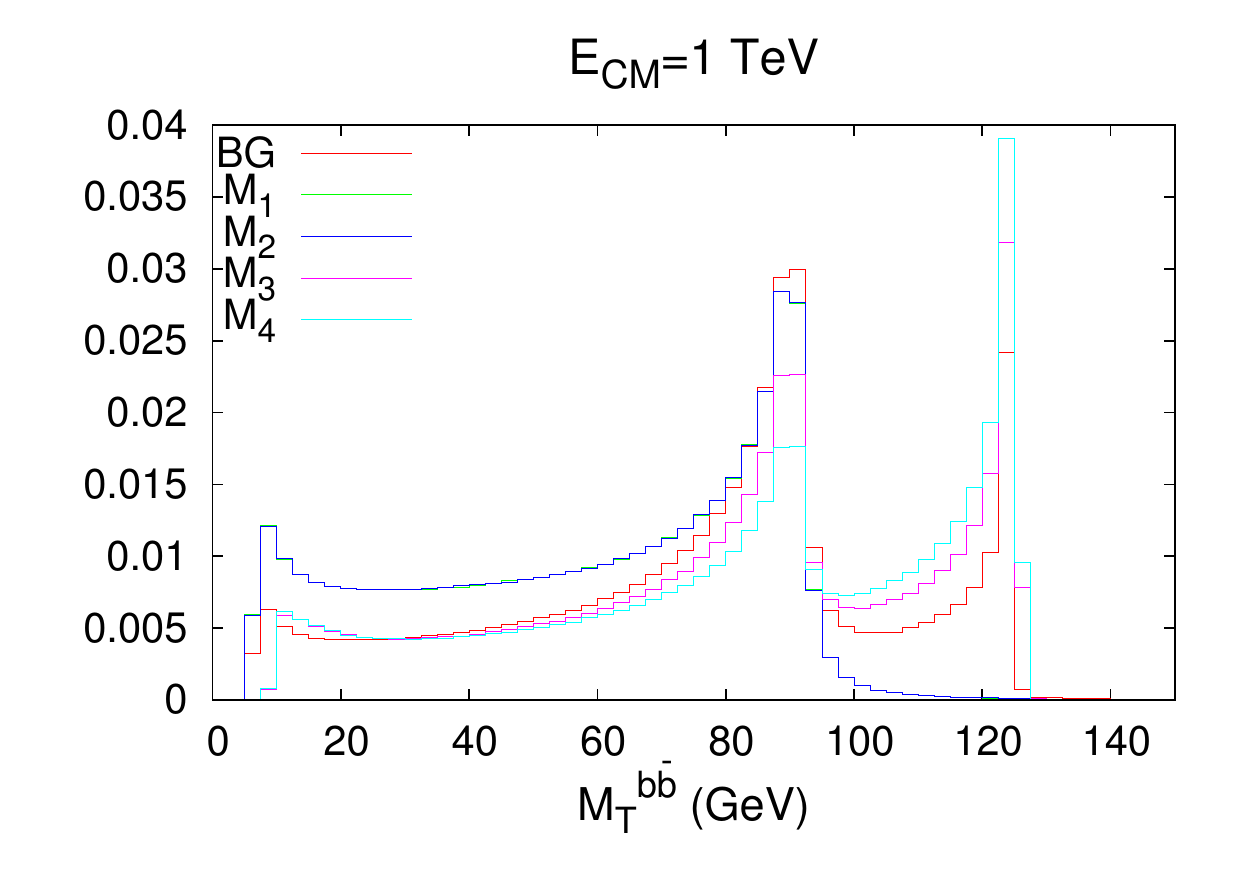}~\includegraphics[width=0.33\textwidth]{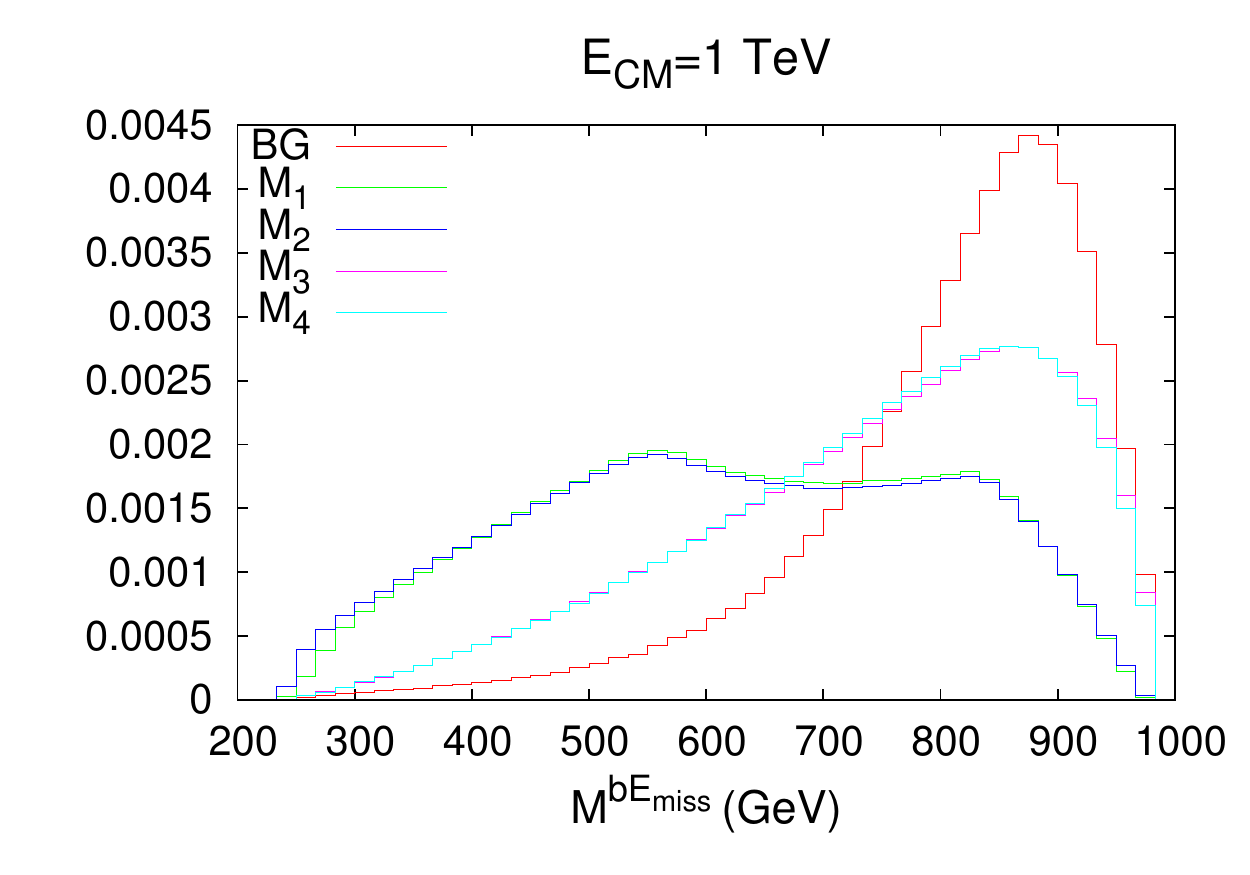}\\
 \includegraphics[width=0.33\textwidth]{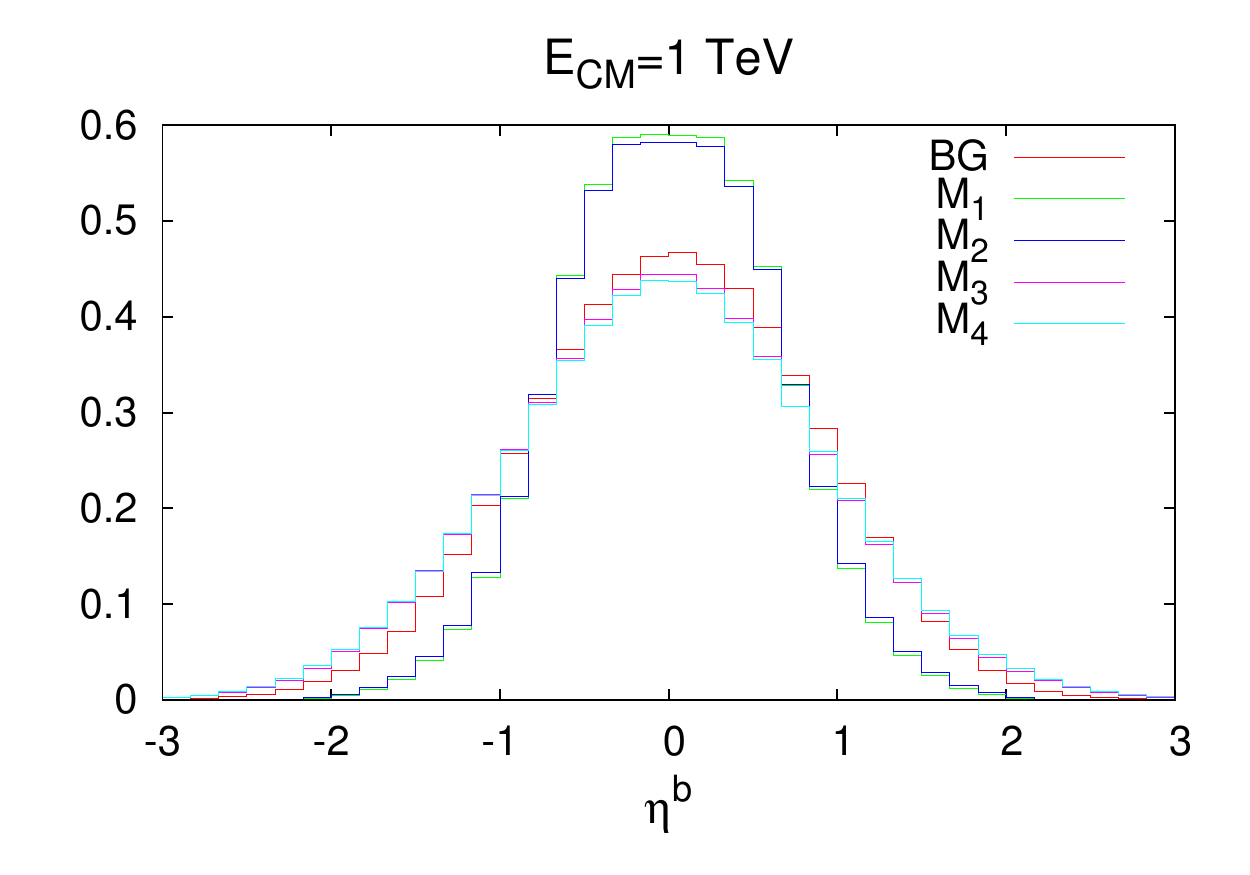}~\includegraphics[width=0.33\textwidth]{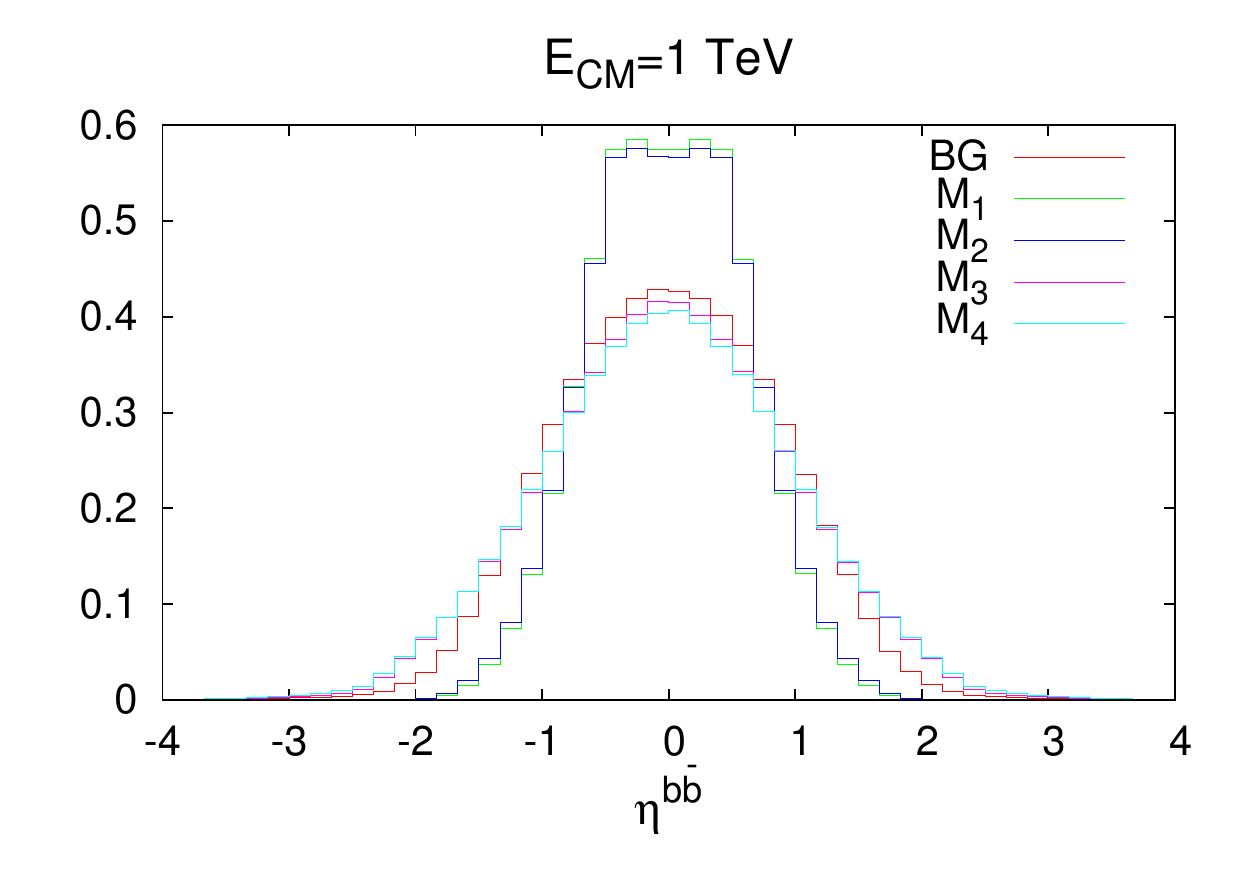}~\includegraphics[width=0.33\textwidth]{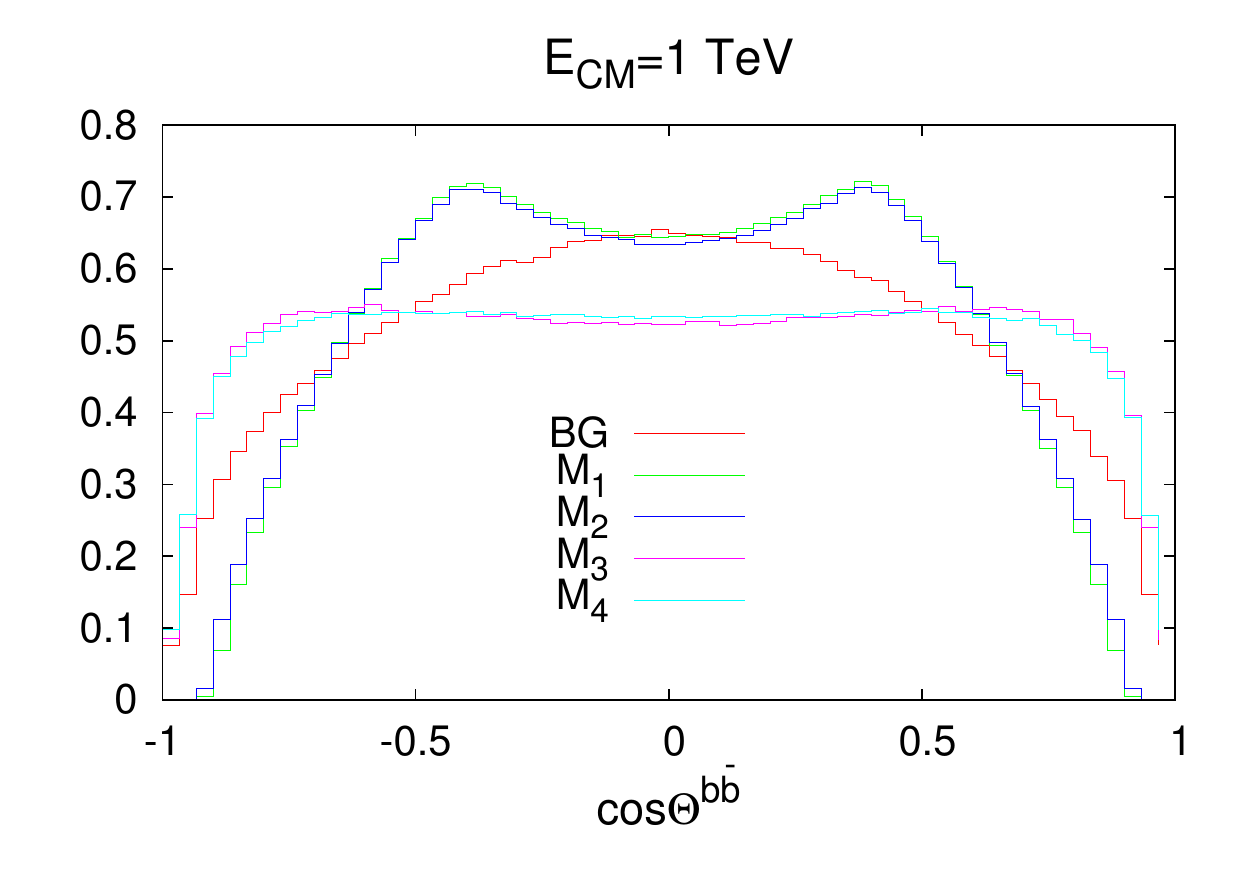}\\
 \caption{The relevant normalized distributions of the process $e^{-}e^{+}\rightarrow b\bar{b}+\slashed{E}_{T}$
at $E_{c.m.}=1~\mathrm{TeV}$.}

\label{dis.cuts1TeV} 
\end{figure}

In case of large luminosity values that allow the signal to be seen,
we show relevant normalized distributions in Figs.~\ref{dis.cuts500GeV}
and ~\ref{dis.cuts1TeV}, for $E_{c.m.}=500~\mathrm{GeV}$ and $1~\mathrm{TeV}$,
respectively. The relevant distributions here are the polar angle
between bottom-antibottom jets $cos(\theta^{b,\bar{b}})$, the jet
energy $E^{b}$, the jet transverse energy $E_{T}^{b}=\sqrt{m_{b}^{2}+\overrightarrow{p}_{T}^{2}}$,
the jet transverse momentum $p_{T}^{b}$, the transverse mass of the
bottom-antibottom jets $M_{T}^{b,\bar{b}}=\sqrt{\left(E_{T}^{b}+E_{T}^{\bar{b}}\right)^{2}-\left(\overrightarrow{p}_{T}^{b}+\overrightarrow{p}_{T}^{\bar{b}}\right)^{2}}$,
the invariant mass of the missing energy with a jet $M^{b,\slashed{E}_{T}}$,
the jet pseudorapidity $\eta^{b}$, the two-jet pseudo rapidity
$\eta^{b,\bar{b}}$, and the polar angle between the two jets in the
boost direction $cos(\Theta^{b,\bar{b}})$.

At $E_{c.m.}=500~\mathrm{GeV}$ (Fig.~\ref{dis.cuts500GeV}), for scalar
DM ($M_{1,2}$), the normalized distributions have different shapes
with respect to both the background and the fermionic DM case ($M_{3,4}$),
especially for the distributions $cos(\theta^{b,\bar{b}})$, $E^{b}$,
$E_{T}^{b}$, $p_{T}^{b}$, $M^{b,\slashed{E}_{T}}$, $\eta^{b,\bar{b}}$,
and $cos(\Theta^{b,\bar{b}})$. For the fermionic DM case ($M_{3,4}$),
the normalized distributions have the same shape with respect to the
background with a remarkable shift. For instance, if the DM is a scalar,
the normalized distributions of $cos(\theta^{b,\bar{b}})$ and $\eta^{b,\bar{b}}$
get maximized for $0.4<cos(\theta^{b,\bar{b}})<0.8$ and $0.3<|\eta^{b,\bar{b}}|<0.8$,
respectively. However, at $E_{c.m.}=1~\mathrm{TeV}$ (Fig.~\ref{dis.cuts1TeV}),
the two cases of scalar and fermionic DM could be easily distinguished
due to the different normalized distributions shapes.

\subsection{Analysis using polarized beams}

In search of new physics, the use of polarized beams at future electron/positron
colliders such as the ILC and CLIC could reduce the background and/or
enhance the signal~\cite{ILC4,Adolphsen:2013kya,Baer:2013c.m.a}. The
electron or positron polarization is defined as 
\begin{equation}
P\left(f\right)=(N_{f_{R}}-N_{f_{L}})/(N_{f_{R}}+N_{f_{L}}),
\end{equation}
where $N_{f_{R}}$ ($N_{f_{L}}$) is the number of right- (left-)
handed fermions. At the ILC, the polarization degree of the electron
(positron) beams could reach $80\%$ ($30\%$), i.e., $|P(e^{-})|<0.80$
($|P(e^{+})|)<0.30$)~\cite{Adolphsen:2013kya}. The positron polarization
could be improved up to $60\%$ at the CLIC~\cite{Battaglia:2004mw,CLICP2}.

Here, we reanalyze the same process at the same c.m. energy values within
the polarization $P\left(e^{-},e^{+}\right)=[+0.8,-0.3]$, while keeping
the same full set of cuts given in Table \ref{cut}. We present the
results compared to the case without polarization in Table \ref{pola}.

\begin{table}[ht]
\begin{adjustbox}{max width=\textwidth} \centering %
\begin{tabular}{|c|c|c|c|c|c|c|c|c|c|}
\cline{2-10} 
\multicolumn{1}{c|}{} & \multicolumn{1}{c}{} & \multicolumn{1}{c}{$P\left(e^{-},e^{+}\right)$ } & \multicolumn{1}{c}{$=$ } & \multicolumn{1}{c}{$[0,0]$ } & & \multicolumn{1}{c}{$P\left(e^{-},e^{+}\right)$ } & \multicolumn{1}{c}{$=$ } & \multicolumn{1}{c}{$[+0.8,-0.3]$ } & \multicolumn{1}{c|}{ }\tabularnewline
\hline 
$E_{c.m.}$ $\left(\mathrm{GeV}\right)$ & $\sigma^{BG}$$\left(fb\right)$ & Models & $\sigma^{S}$ $\left(fb\right)$ & $\mathcal{S}_{100}$ & $\mathcal{S}_{500}$ & $\sigma^{BG}$ $\left(fb\right)$ & $\sigma^{S}$ $\left(fb\right)$ & $\mathcal{S}_{100}$ & $\mathcal{S}_{500}$\tabularnewline
\hline 
 & & $M1$ & $0.520$ & $0.9808$ & $2.1936$ & & $0.558$ & $1.9488$ & $4.3584$ \tabularnewline
\cline{3-6} \cline{8-10} 
$500$ & $17.804$ & $M_{2}$ & $0.638$ & $1.2024$ & $2.6888$ & $5.061$ & $0.685$ & $2.3832$ & $5.3304$ \tabularnewline
\cline{3-6} \cline{8-10} 
 & & $M_{3}$ & $0.956$ & $1.7960$ & $4.0168$ & & $2.166$ & $7.2328$ & $16.1736$ \tabularnewline
\cline{3-6} \cline{8-10} 
 & & $M_{4}$ & $1.070$ & $2.0088$ & $4.4912$ & & $2.570$ & $8.4944$ & $18.9944$ \tabularnewline
\hline 
 & & $M_{1}$ & $0.282$ & $0.3216$ & $0.7192$ & & $0.303$ & $0.7640$ & $1.7096$ \tabularnewline
\cline{3-6} \cline{8-10} 
 & & $M_{2}$ & $0.292$ & $0.3328$ & $0.7448$ & & $0.313$ & $0.7896$ & $1.7656$ \tabularnewline
\cline{3-6} \cline{8-10} 
$1000$ & $49.072$ & $M_{3}$ & $0.942$ & $1.0720$ & $2.3976$ & $9.950$ & $5.472$ & $12.8312$ & $28.6912$ \tabularnewline
\cline{3-6} \cline{8-10} 
 & & $M_{4}$ & $0.760$ & $0.8656$ & $1.9352$ & & $4.219$ & $10.0520$ & $22.4784$ \tabularnewline
\hline 
\end{tabular}\end{adjustbox} \caption{The cross section values for the background $\sigma^{BG}$ and the
signal $\sigma^{S}$ estimated for the considered energies within
the full set of cuts given in Table \ref{cut}, without and with polarized
beams at both c.m. energies $E_{c.m.}=500~\mathrm{GeV}$ and $1~\mathrm{TeV}$.
The significances $\mathcal{S}_{100}$ and $\mathcal{S}_{500}$ correspond
to the two integrated luminosity values $L=100~fb^{-1}$ and $500~fb^{-1}$,
respectively.}

\label{pola} 
\end{table}

From Table \ref{pola}, by comparing the cases with and without polarization,
one remarks that the cross section value for the background $\sigma^{BG}$
is reduced by about $72\%$ ($80\%$) at $E_{c.m.}=500~\mathrm{GeV}$
($E_{c.m.}=1~\mathrm{TeV}$). On the contrary, the signal cross section
value $\sigma^{S}$ gets increased by about $7\%$ for both $M_{1,2}$
at both c.m. energies $E_{c.m.}=500~\mathrm{GeV}$ and $1~\mathrm{TeV}$.
One can also see the cross section value for $M_{3}$ ($M_{4}$) gets
raised by about $127\%$ ($140\%$) and by $481\%$ ($455\%$) for
c.m. energies $500~\mathrm{GeV}$ and $~1~\mathrm{TeV}$ respectively. Consequently,
the signal significance gets enhanced by $303\%$ ($323\%$) and by
$1097\%$ ($1061\%$) for $M_{3}$ ($M_{4}$) at $500~\mathrm{GeV}$ and $~1~\mathrm{TeV}$,
respectively. When considering the polarization $P\left(e^{-},e^{+}\right)=[+0.8,-0.3]$,
the background cross section gets decreased sharply due to the vertices
suppression of the electron-positron with gauge bosons unlike the
vertices of the charged scalar-Majorana fermion-charged lepton (for $M_{3,4}$),
which enhances the cross section.

For luminosity $L=100~fb^{-1}$, one discovers for $M_{3,4}$
at both $E_{c.m.}=500~\mathrm{GeV}$ and $1~\mathrm{TeV}$; however, for $L=500~fb^{-1}$,
one can see also a discovery for all models at $E_{c.m.}=500~\mathrm{GeV}$
except Model 1. At $E_{c.m.}=1~\mathrm{TeV}$ within
the same luminosity, we could not even see a deviation from the SM
for $M_{1,2}$, unlike $M_{3,4}$ in which one can clearly see a discovery.
Therefore, we require a large luminosity value (1 $ab^{-1}$ or more)
in order to see such a signal for two models $M_{1,2}$ in which DM is
a scalar.

\begin{figure}[ht]
\includegraphics[width=0.33\textwidth]{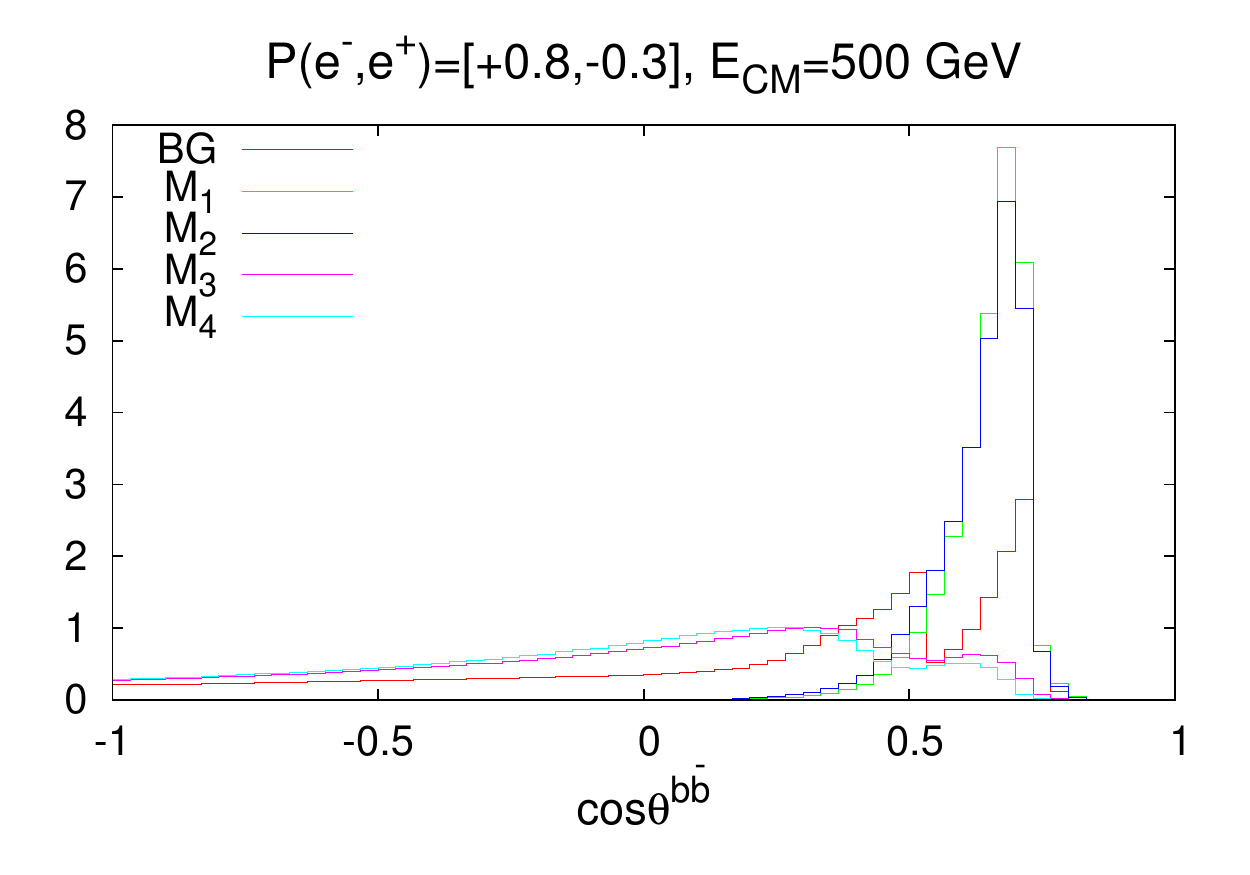}~\includegraphics[width=0.33\textwidth]{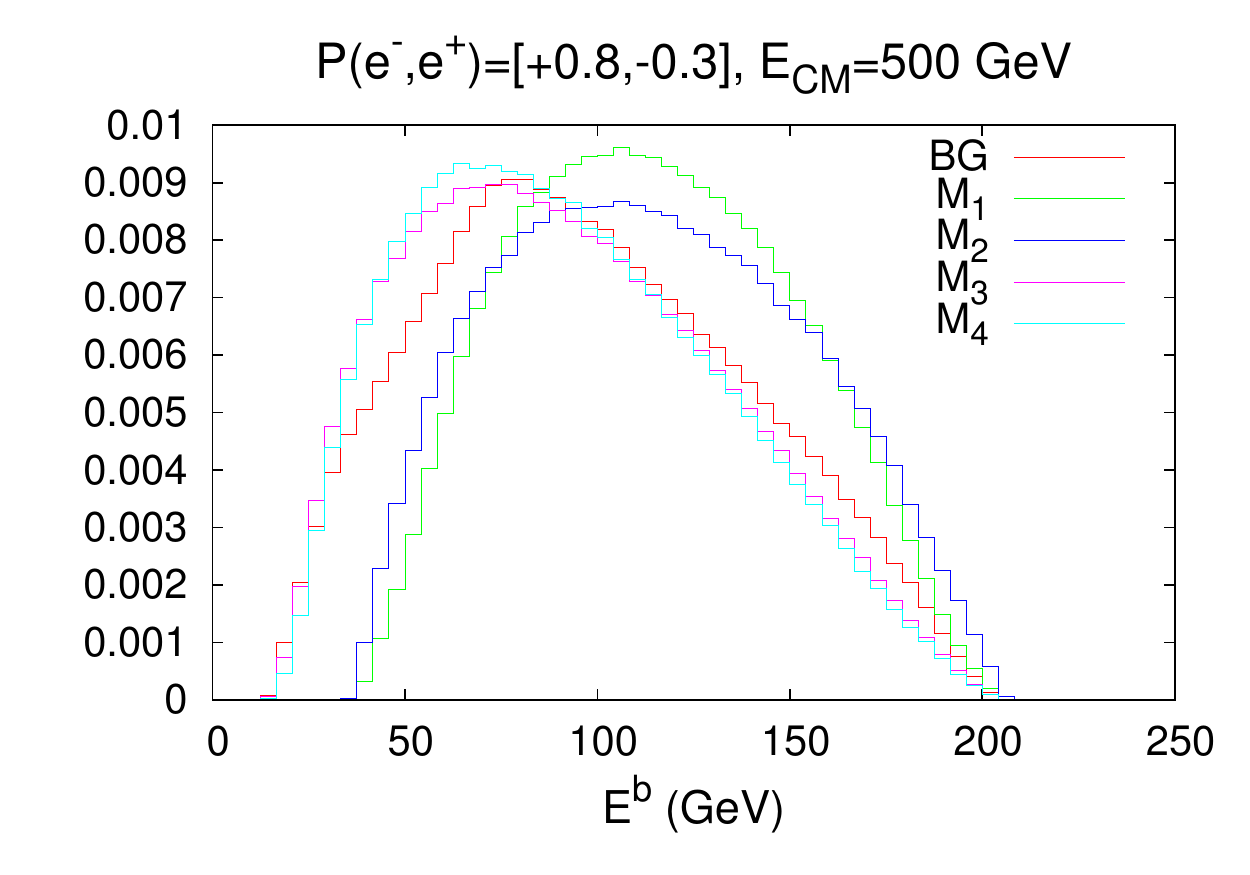}~\includegraphics[width=0.33\textwidth]{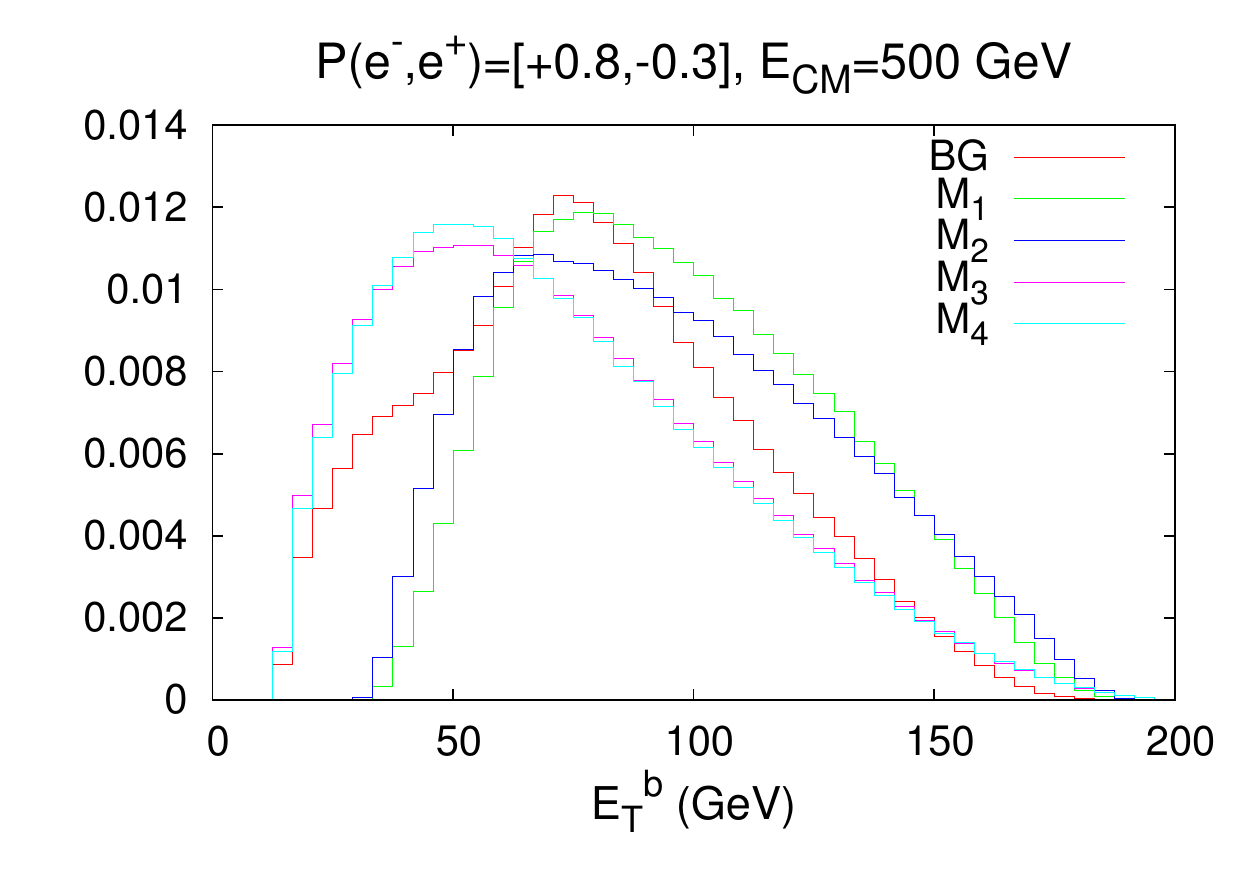}\\
 \includegraphics[width=0.33\textwidth]{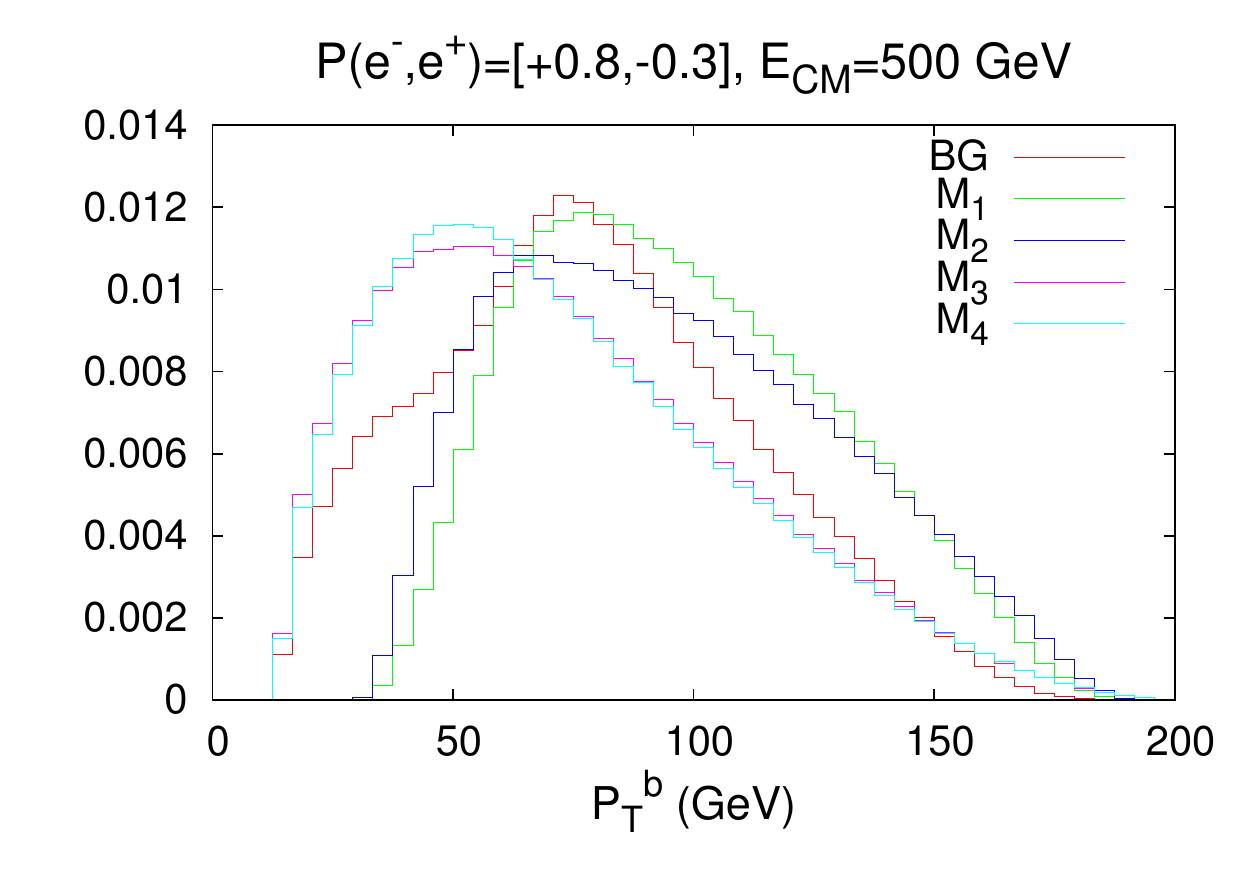}~\includegraphics[width=0.33\textwidth]{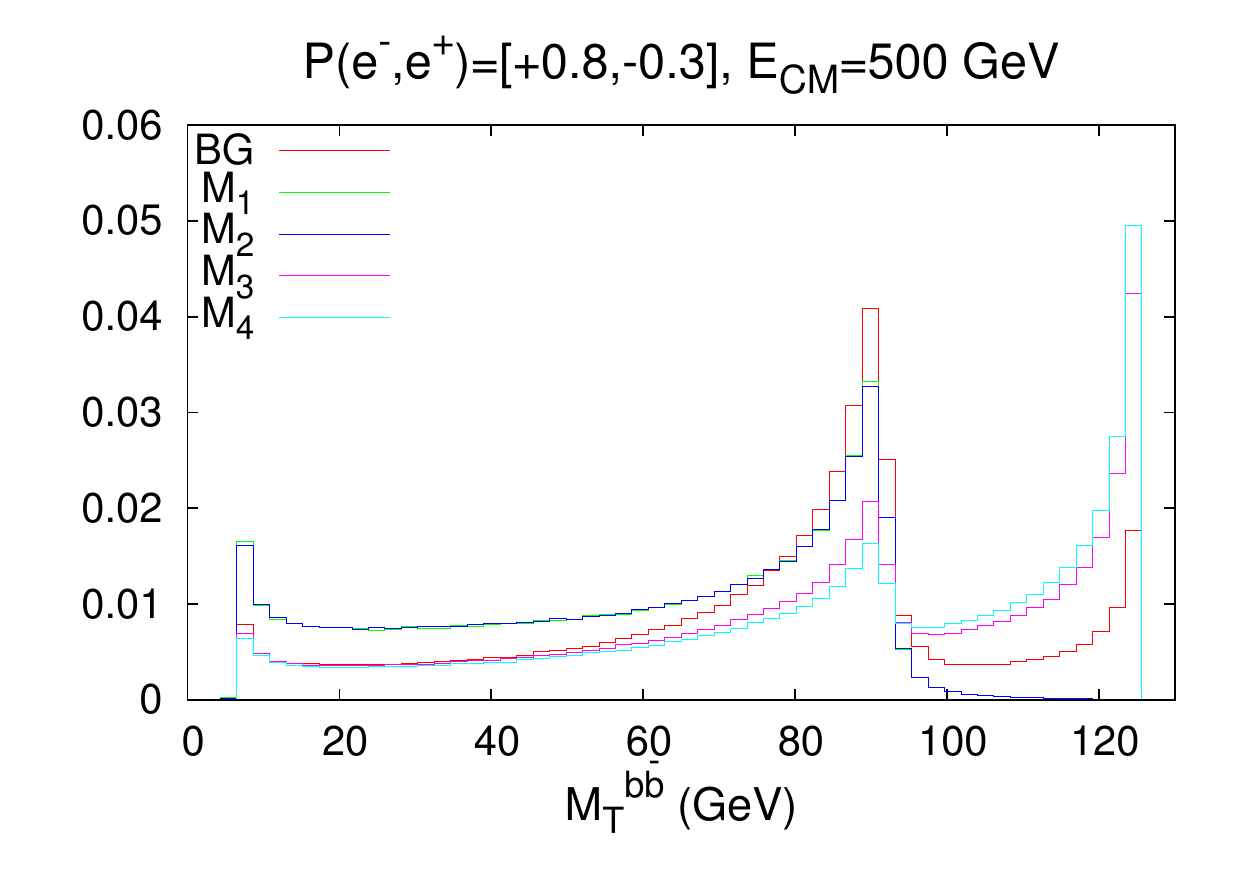}~\includegraphics[width=0.33\textwidth]{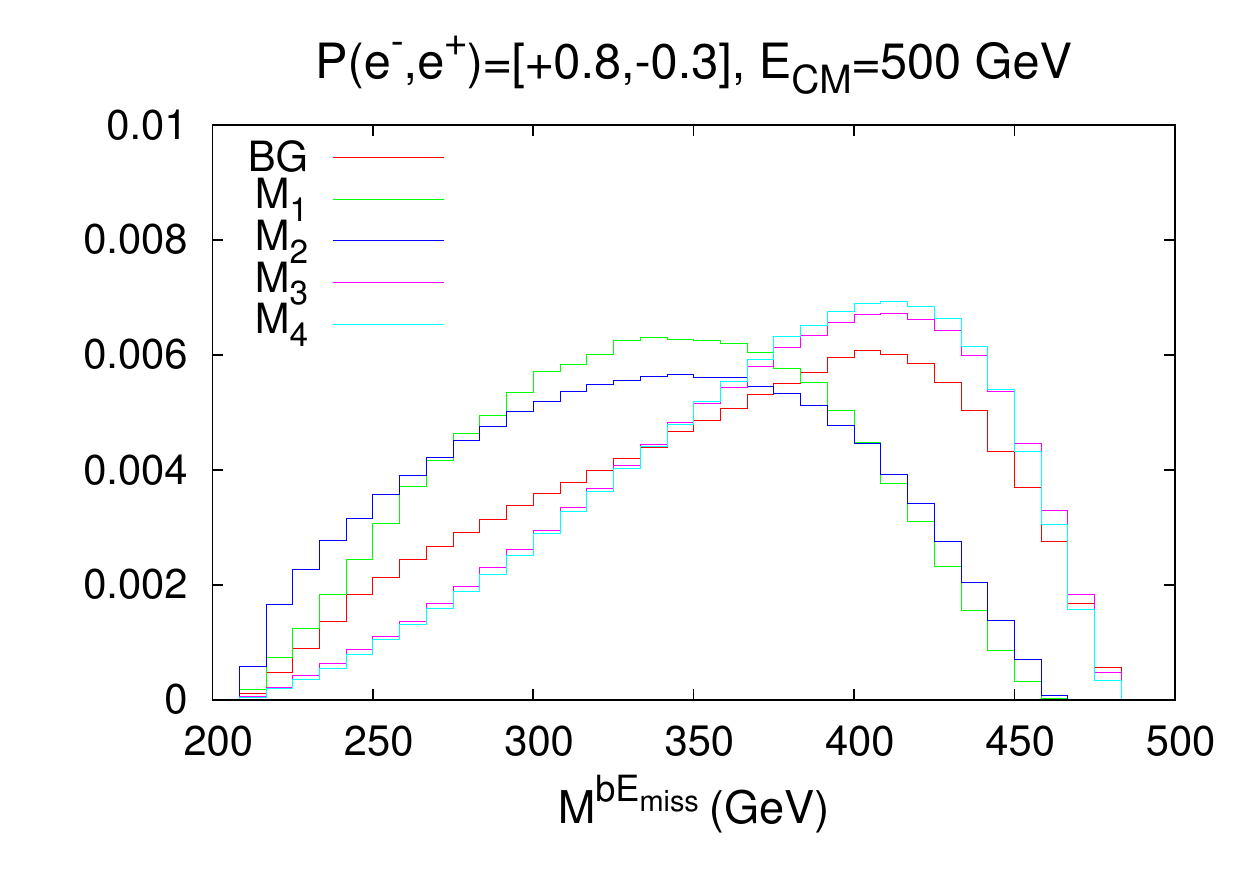}\\
 \includegraphics[width=0.33\textwidth]{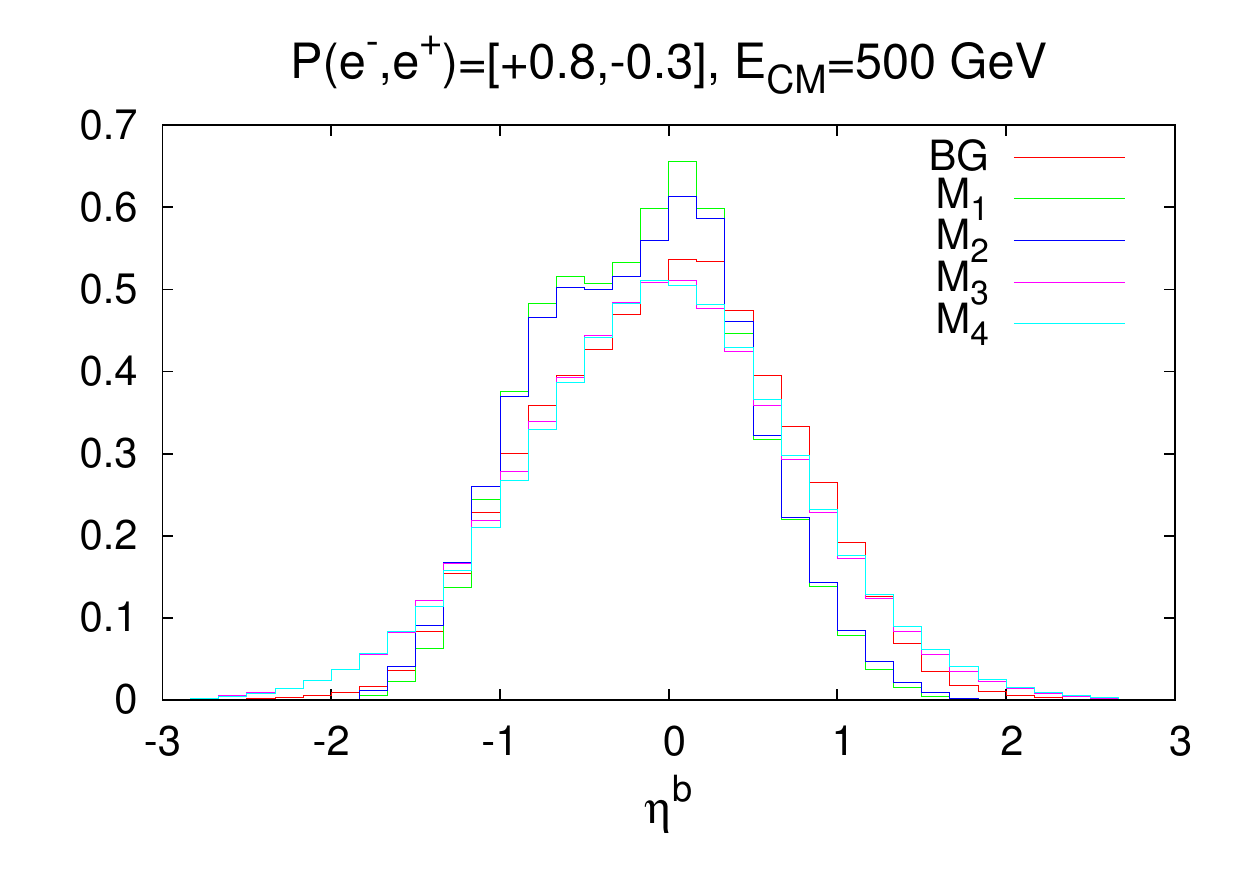}~\includegraphics[width=0.33\textwidth]{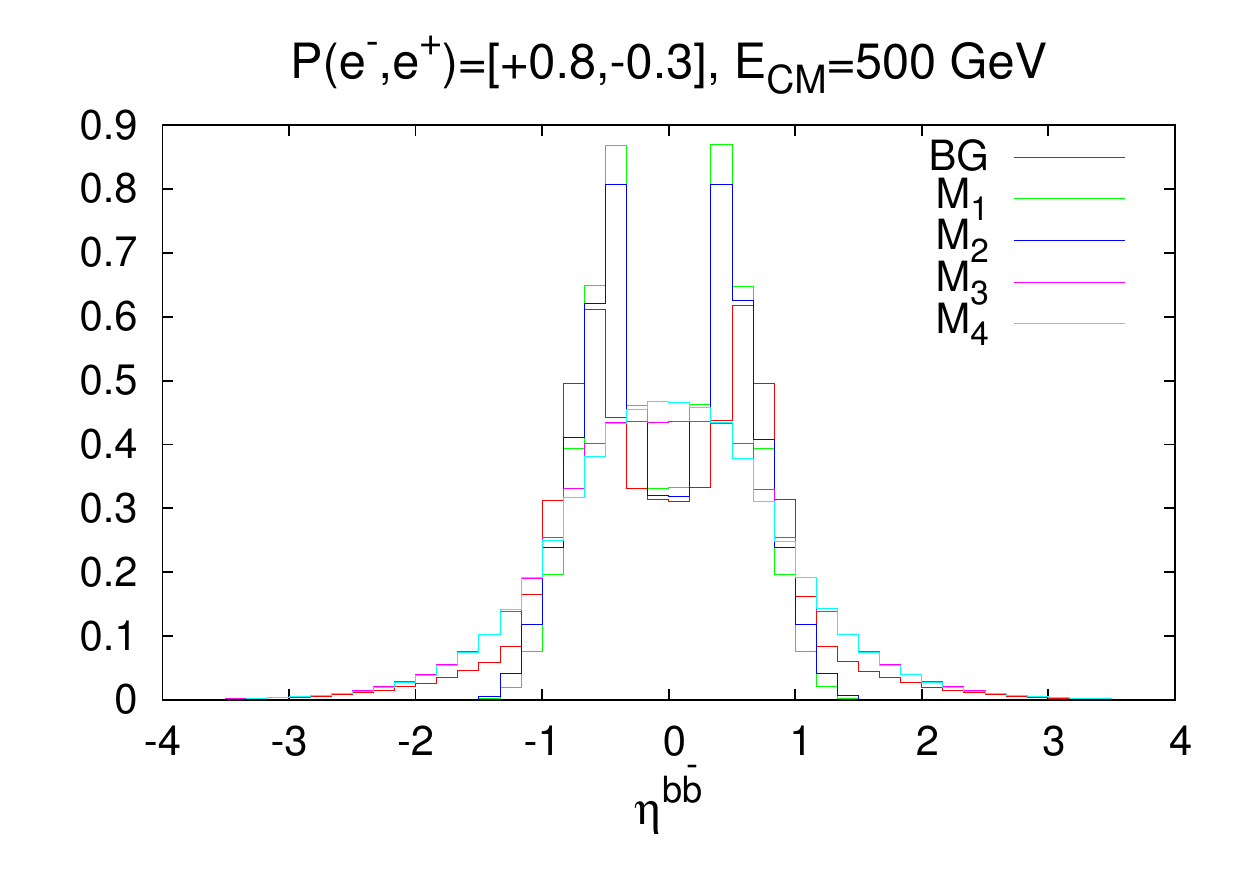}~\includegraphics[width=0.33\textwidth]{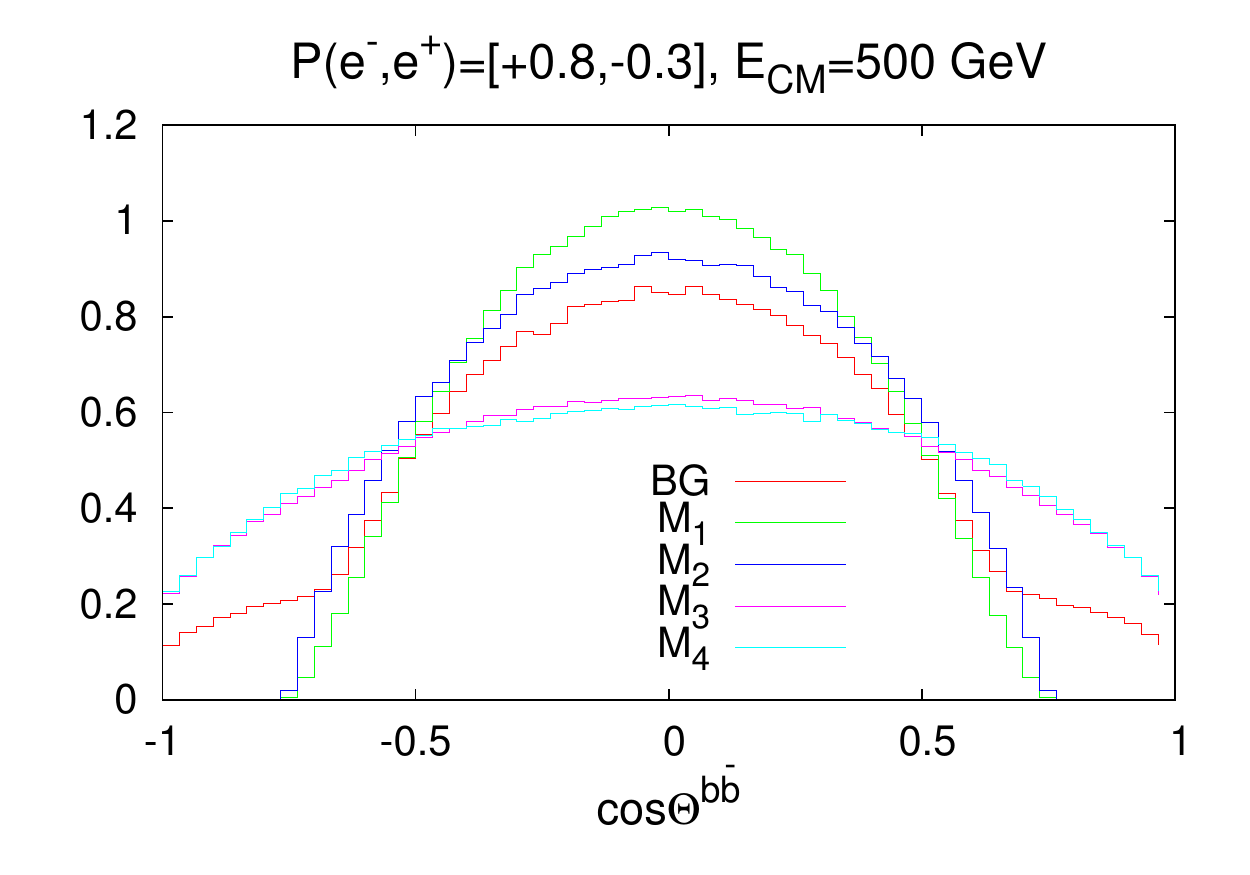}\\
 \caption{The relevant normalized distributions of the process $e^{-}e^{+}\rightarrow b\bar{b}+\slashed{E}_{T}$
at $E_{c.m.}=500~\mathrm{GeV}$ with polarized beams $P\left(e^{-},e^{+}\right)=[+0.8,-0.3]$.}
\label{dis.cutspola500GeV} 
\end{figure}

\begin{figure}[ht]
\includegraphics[width=0.33\textwidth]{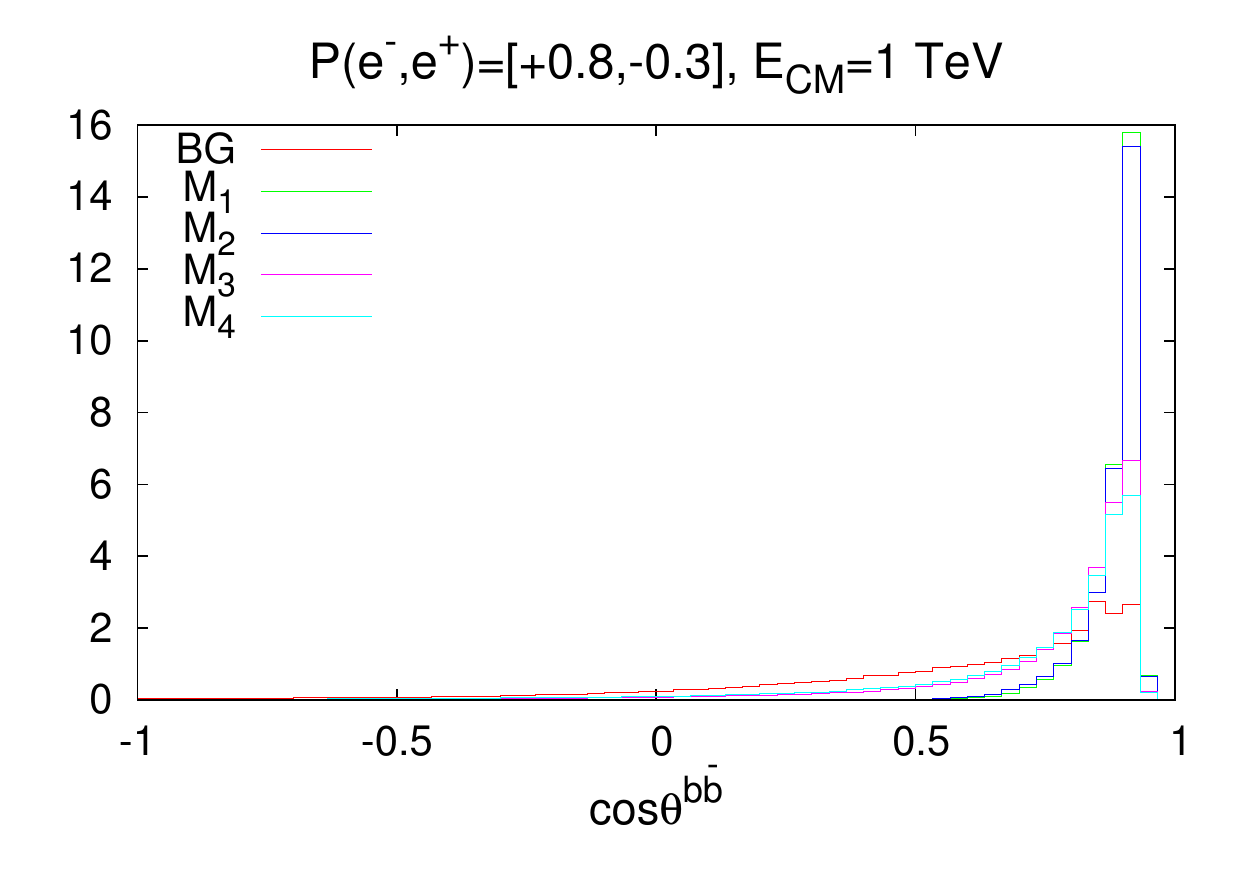}~\includegraphics[width=0.33\textwidth]{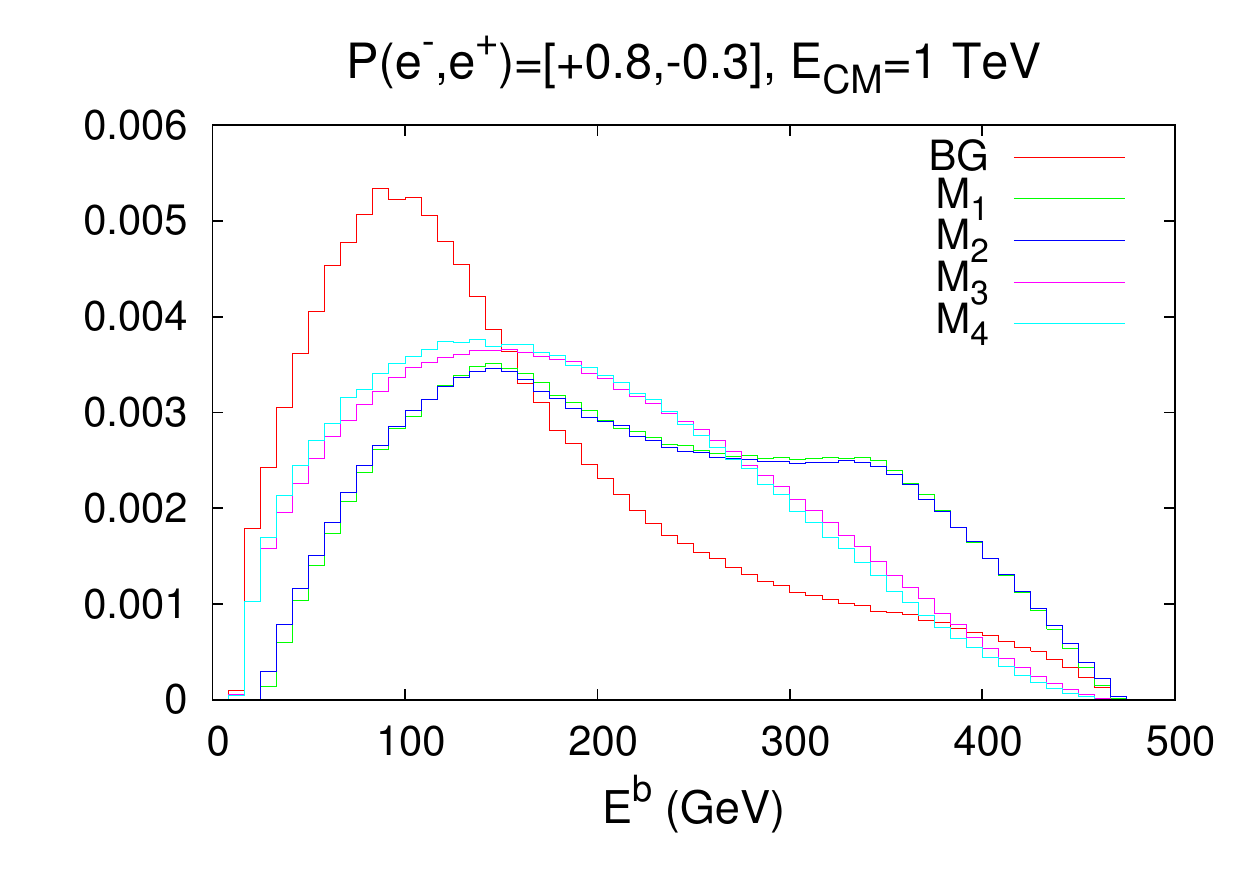}~\includegraphics[width=0.33\textwidth]{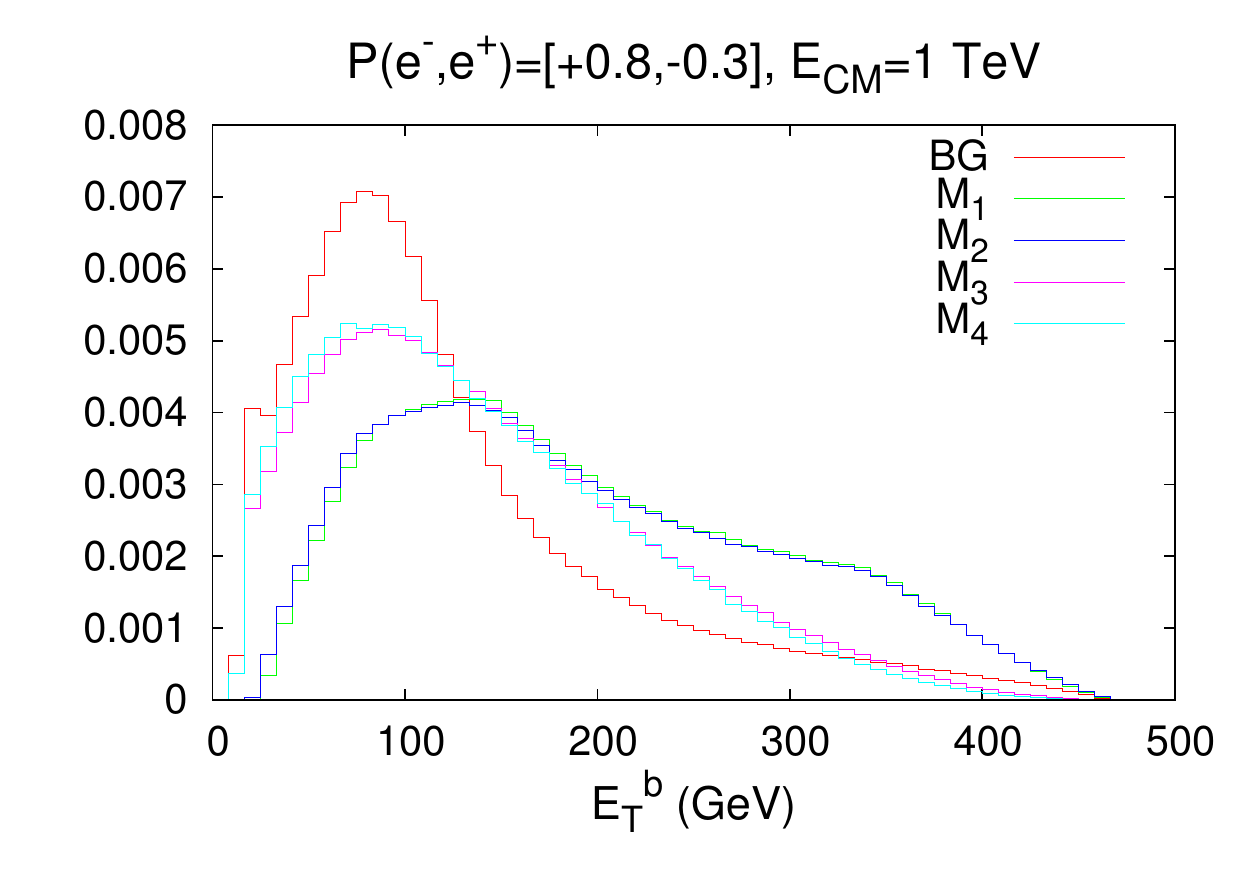}\\
 \includegraphics[width=0.33\textwidth]{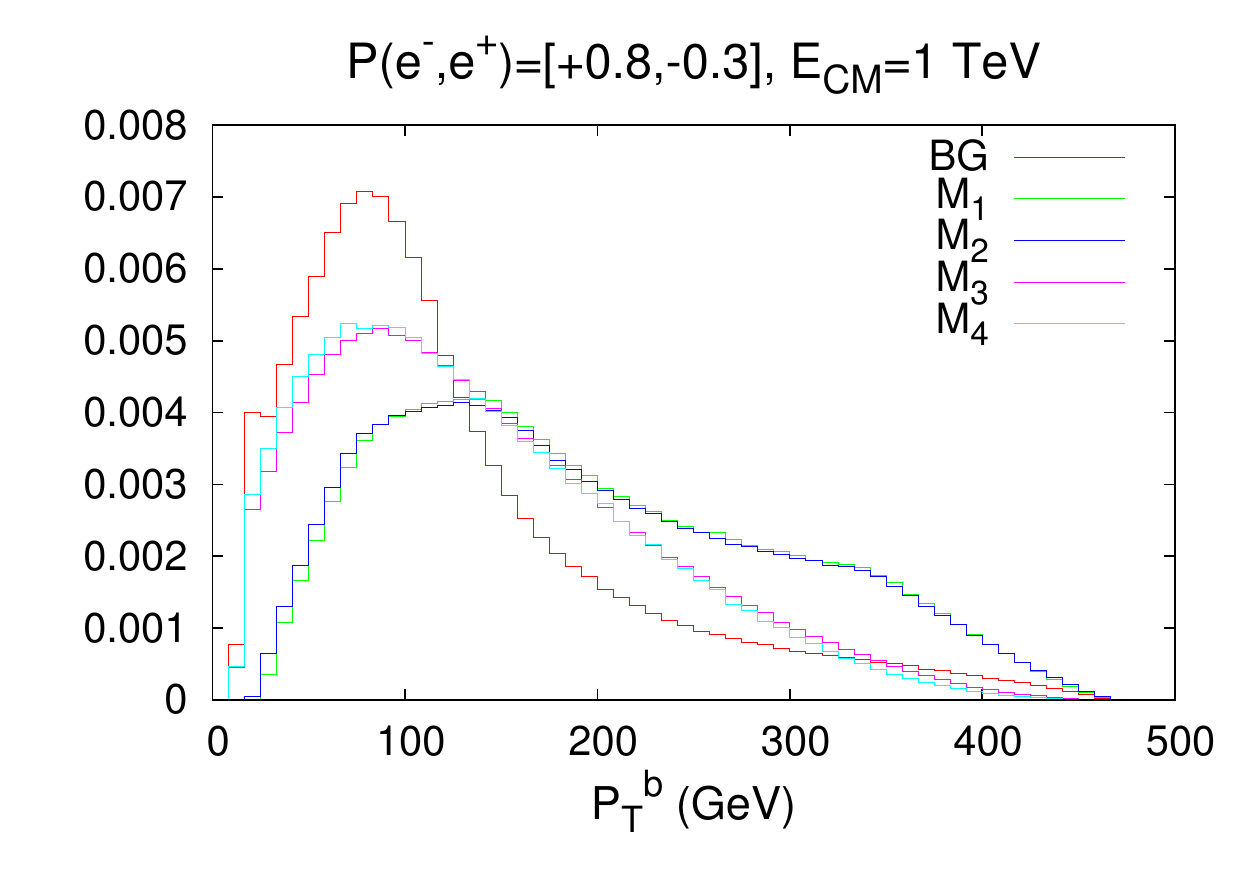}~\includegraphics[width=0.33\textwidth]{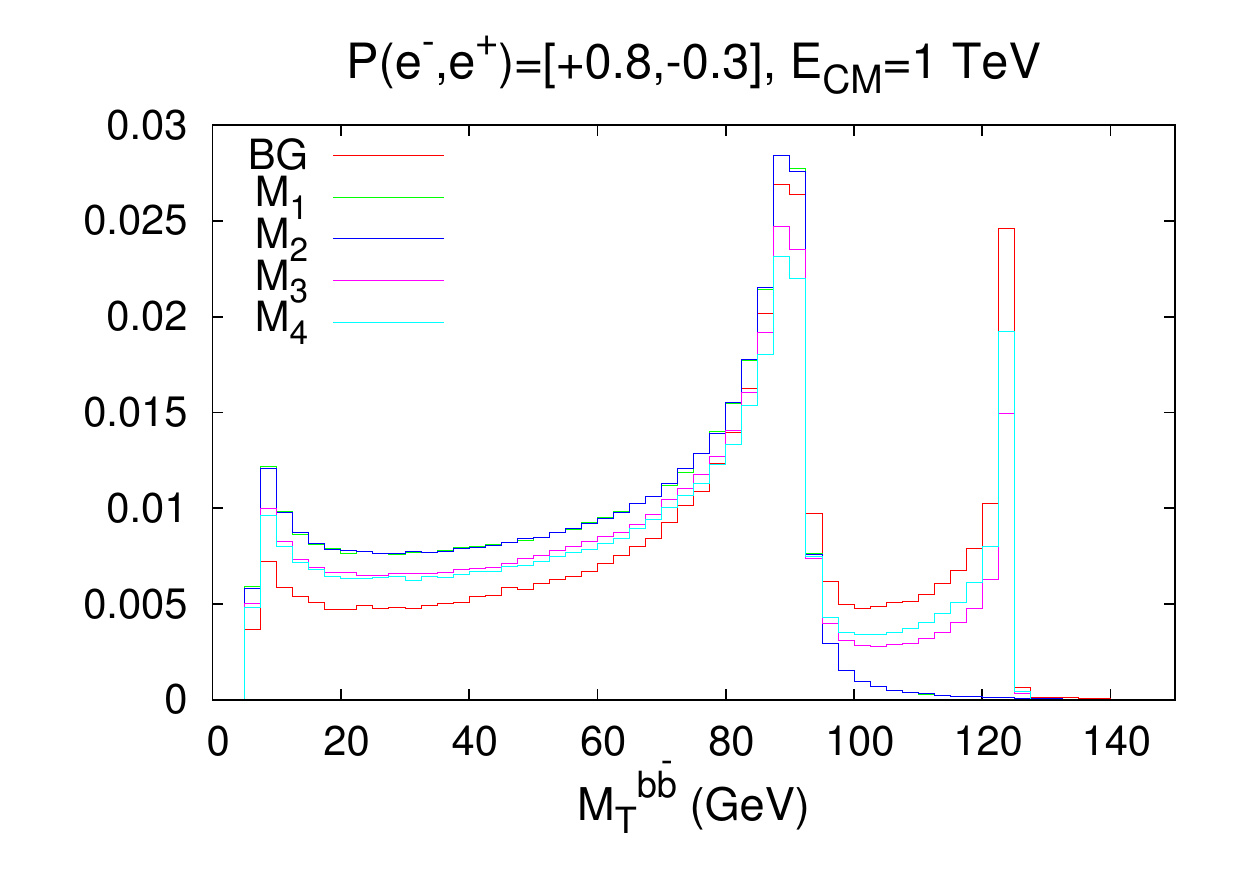}~\includegraphics[width=0.33\textwidth]{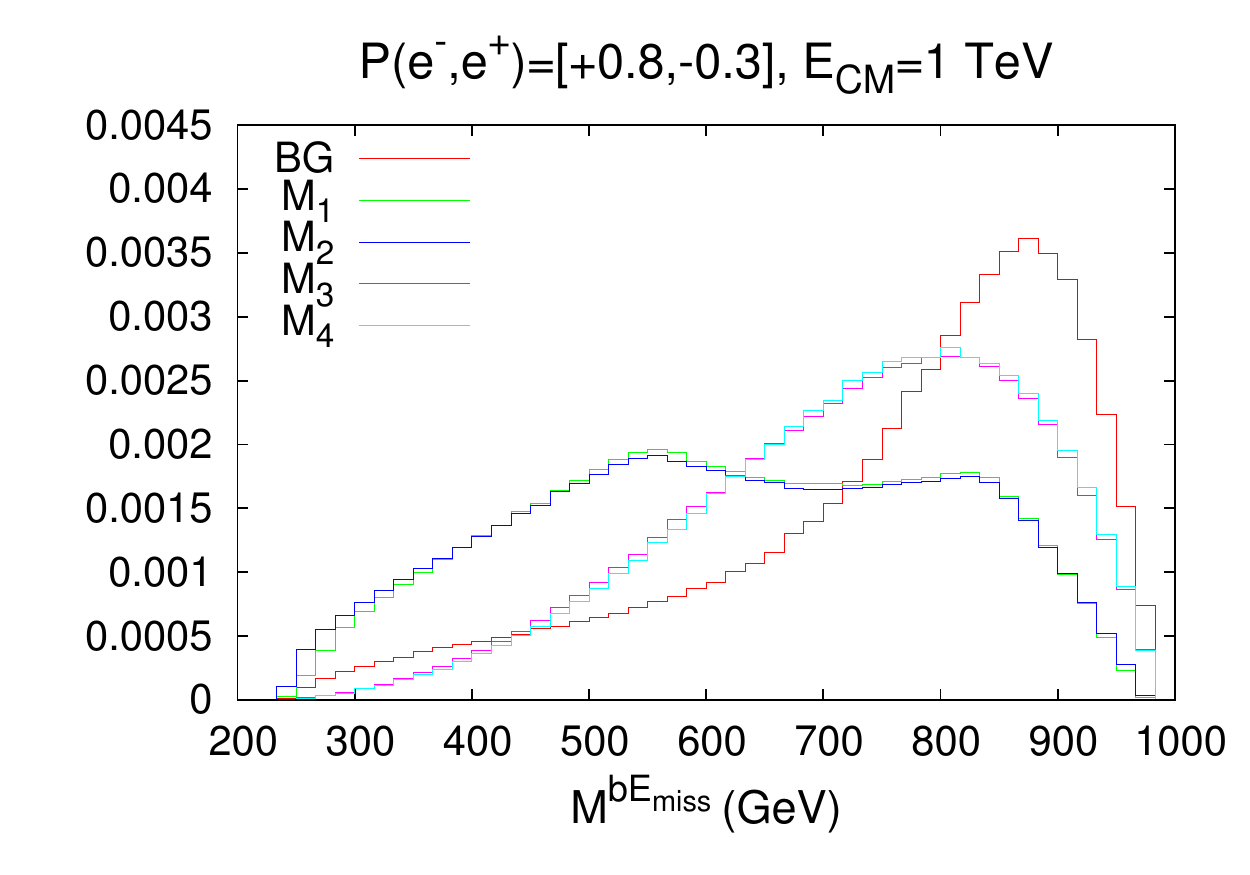}\\
 \includegraphics[width=0.33\textwidth]{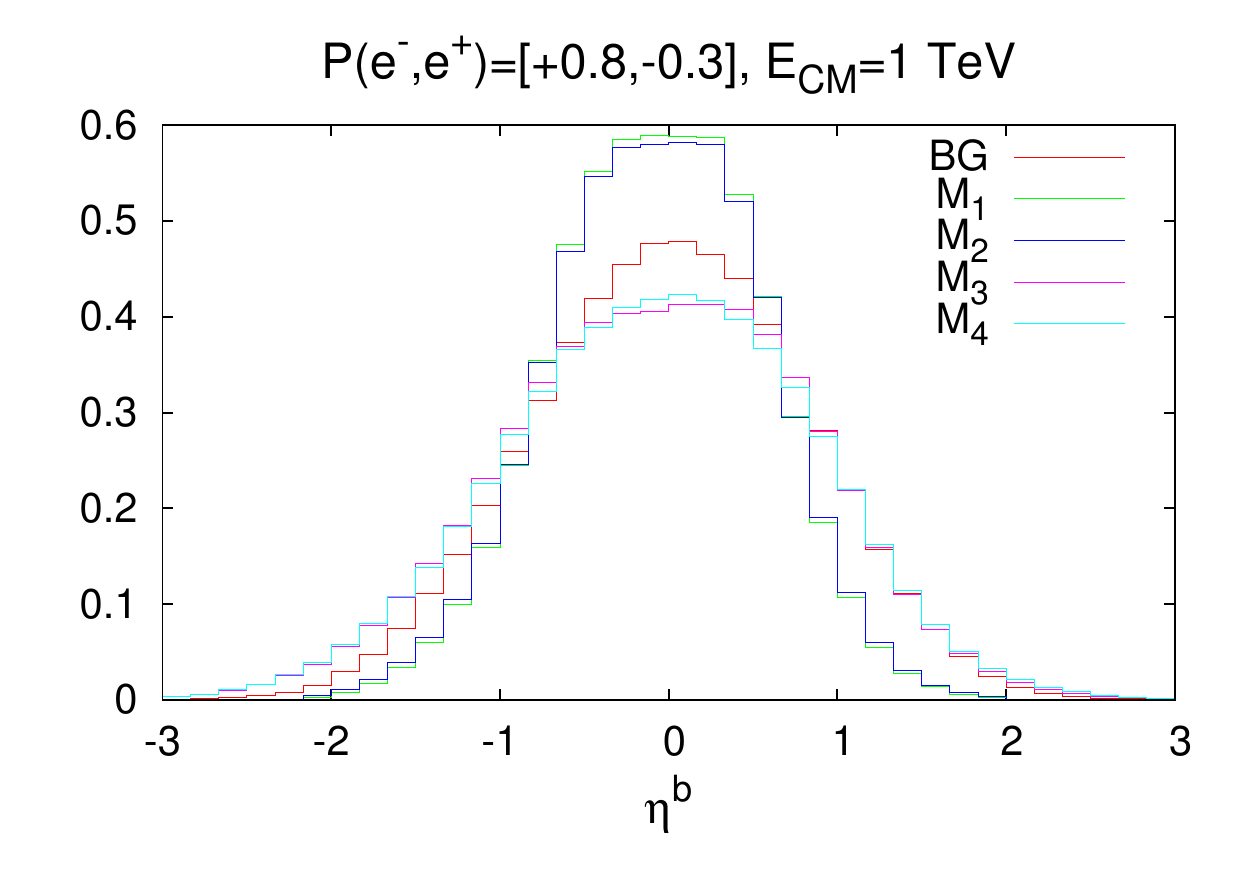}~\includegraphics[width=0.33\textwidth]{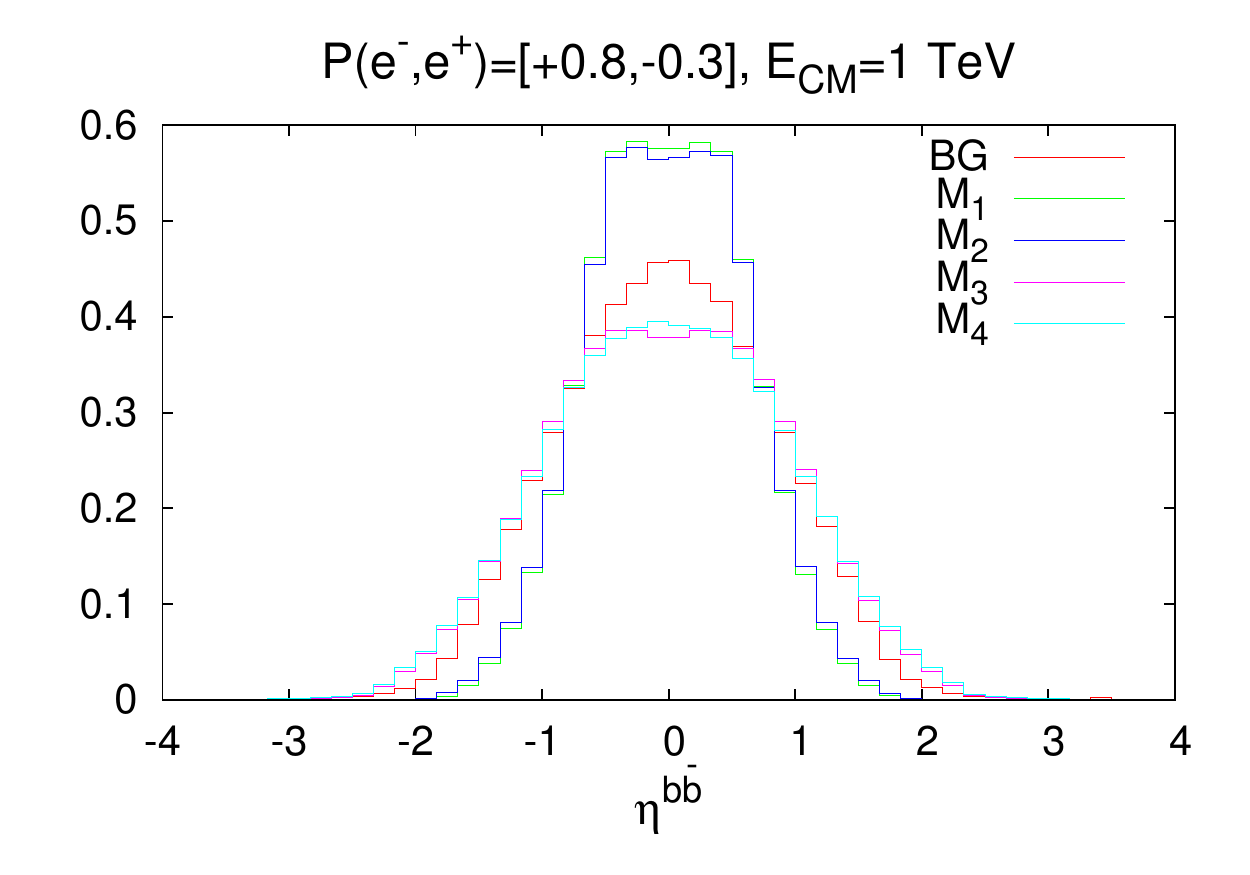}~\includegraphics[width=0.33\textwidth]{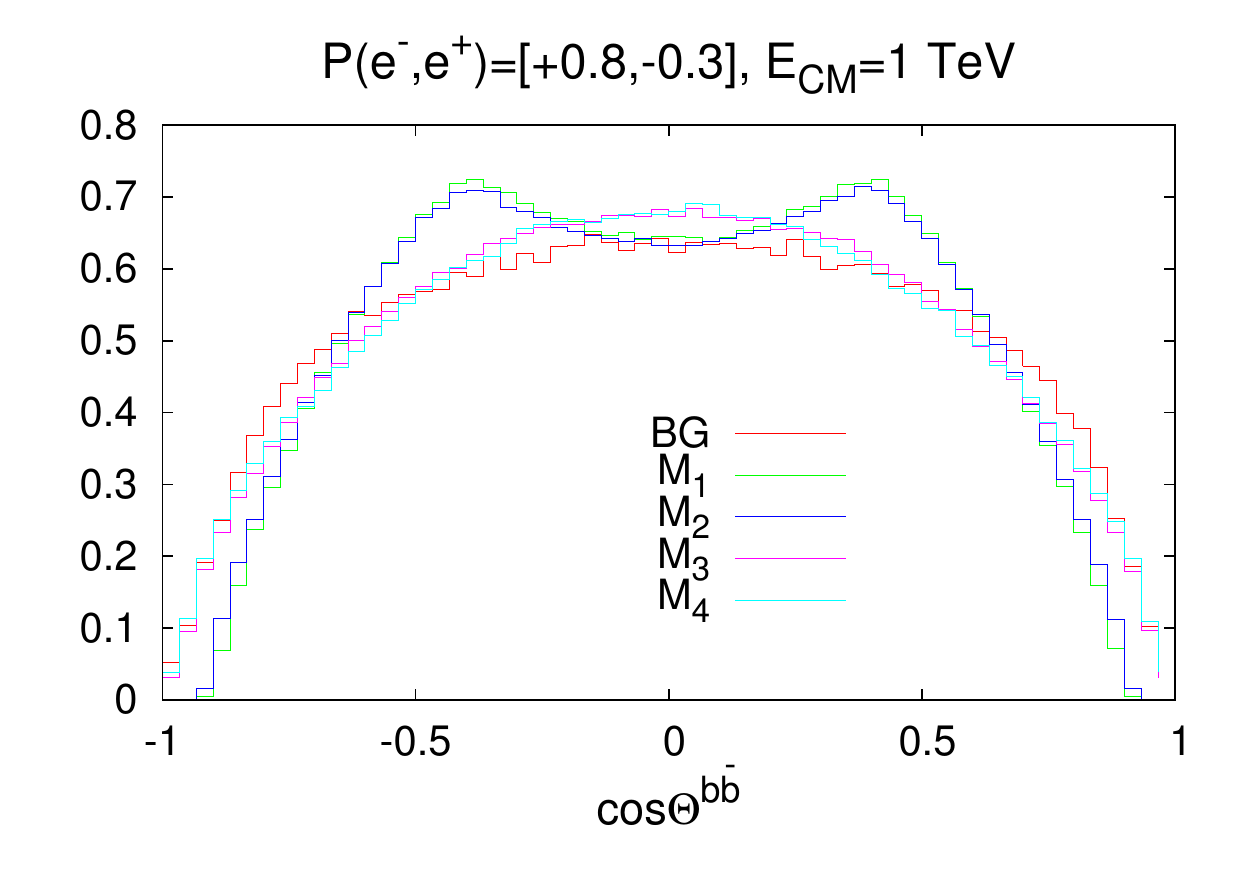}\\
 \caption{The relevant normalized distributions of the process $e^{-}e^{+}\rightarrow b\bar{b}+\slashed{E}_{T}$
at $E_{c.m.}=1~\mathrm{TeV}$ with polarized beams $P\left(e^{-},e^{+}\right)=[+0.8,-0.3]$.}
\label{dis.cutspola1TeV} 
\end{figure}

In Figs.~\ref{dis.cutspola500GeV} and ~\ref{dis.cutspola1TeV},
we show the relevant normalized distributions at $E_{c.m.}=500~\mathrm{GeV}$ and $1~\mathrm{TeV}$,
respectively, using the polarized beams $P\left(e^{-},e^{+}\right)=[+0.8,-0.3]$.

From Fig.~\ref{dis.cutspola500GeV}, for scalar DM ($M_{1,2}$),
the normalized distributions have different shapes with respect to
the background, especially for the distributions: $E_{T}^{b}$, $p_{T}^{b}$,
$M^{b,\slashed{E}_{T}}$, $\eta^{b}$, and $cos(\Theta^{b,\bar{b}})$.
However, for fermionic DM ($M_{3,4}$), the distributions shape is
different for $cos(\theta^{b,\bar{b}}),$$E^{b}$,$E_{T}^{b}$,
$p_{T}^{b}$, $M^{b,\slashed{E}_{T}}$, $\eta^{b,\bar{b}}$, and $cos(\Theta^{b,\bar{b}})$.
From Fig.~\ref{dis.cutspola1TeV}, one notices that the normalized
distributions $E^{b}$, $E_{T}^{b}$, $p_{T}^{b}$, $M^{b,\slashed{E}_{T}}$,
and $cos(\Theta^{b,\bar{b}})$ have different shapes between the
background, the scalar DM ($M_{1,2}$), and the fermionic DM cases ($M_{3,4}$).
One remarks also that for the background and the fermionic DM case
($M_{3,4}$) the normalized distributions have the same shape especially
for $cos(\theta^{b,\bar{b}}),~\eta^{b}$, and $\eta^{b,\bar{b}}$.

By comparing the results produced at $E_{c.m.}=500~\mathrm{GeV}$ using
polarized beams (Fig.~\ref{dis.cutspola500GeV}) with those without
polarization (Fig.~\ref{dis.cuts500GeV}), one can notice a clear difference.
For instance, if the DM is a scalar, the maximum of the normalized
distributions of $E^{b}$, $p_{T}^{b}$, and $\eta^{b}$ get shifted
into $30~\mathrm{GeV}<E^{b}<70~\mathrm{GeV}$, $15~\mathrm{GeV}<p_{T}^{b}<65~\mathrm{GeV}$,
and $-1<\eta^{b}<0.2$, respectively, with respect to the case without
polarization. At $E_{c.m.}=1~\mathrm{TeV}$, for the fermionic DM case
($M_{3,4}$), the maximum of the normalized distributions of $M^{b,\slashed{E}_{T}}$
and $cos(\Theta^{b,\bar{b}})$ get shifted also into $450~\mathrm{GeV}<M^{b,\slashed{E}_{T}}<780~\mathrm{GeV}$
and $|cos(\Theta^{b,\bar{b}})|<0.4$, respectively.

\begin{figure}[ht]
\includegraphics[width=0.48\textwidth]{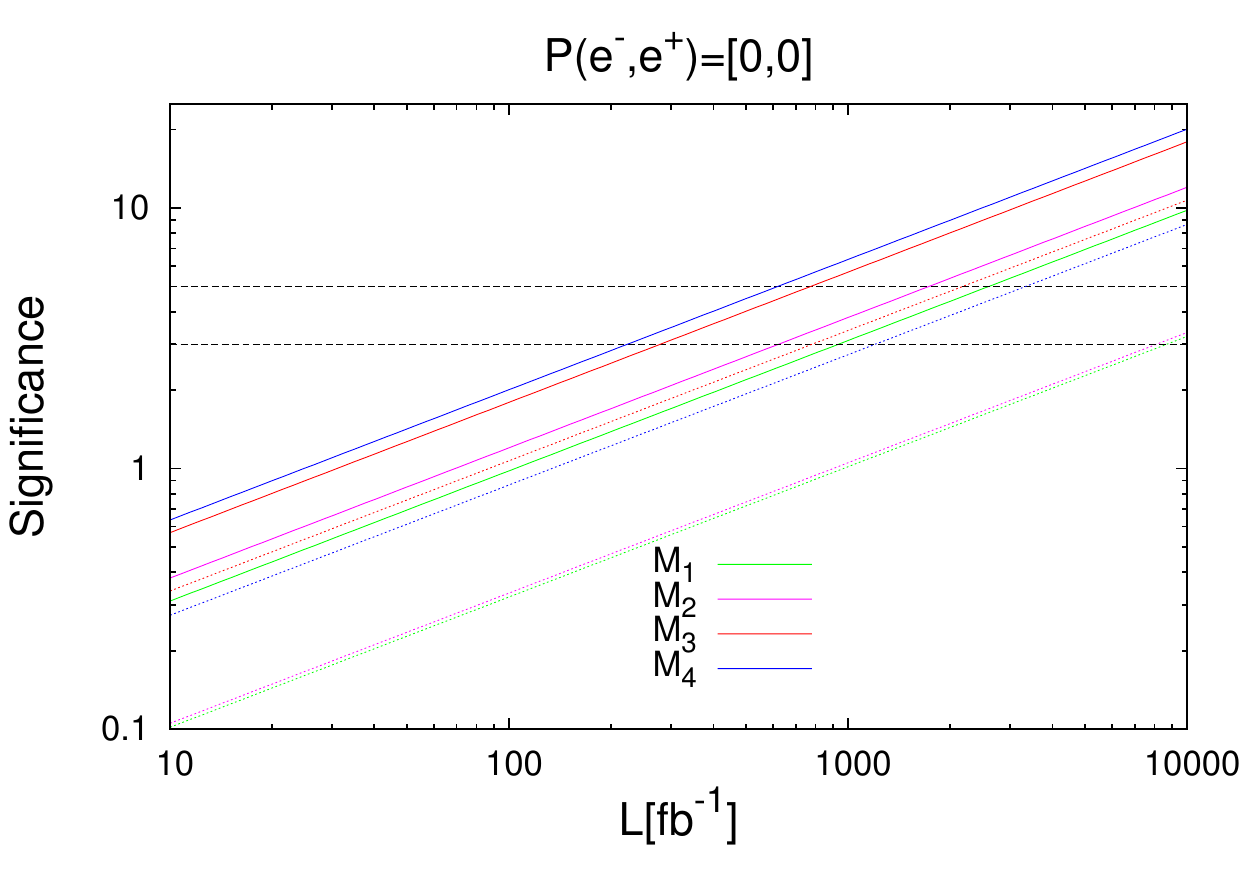}~\includegraphics[width=0.48\textwidth]{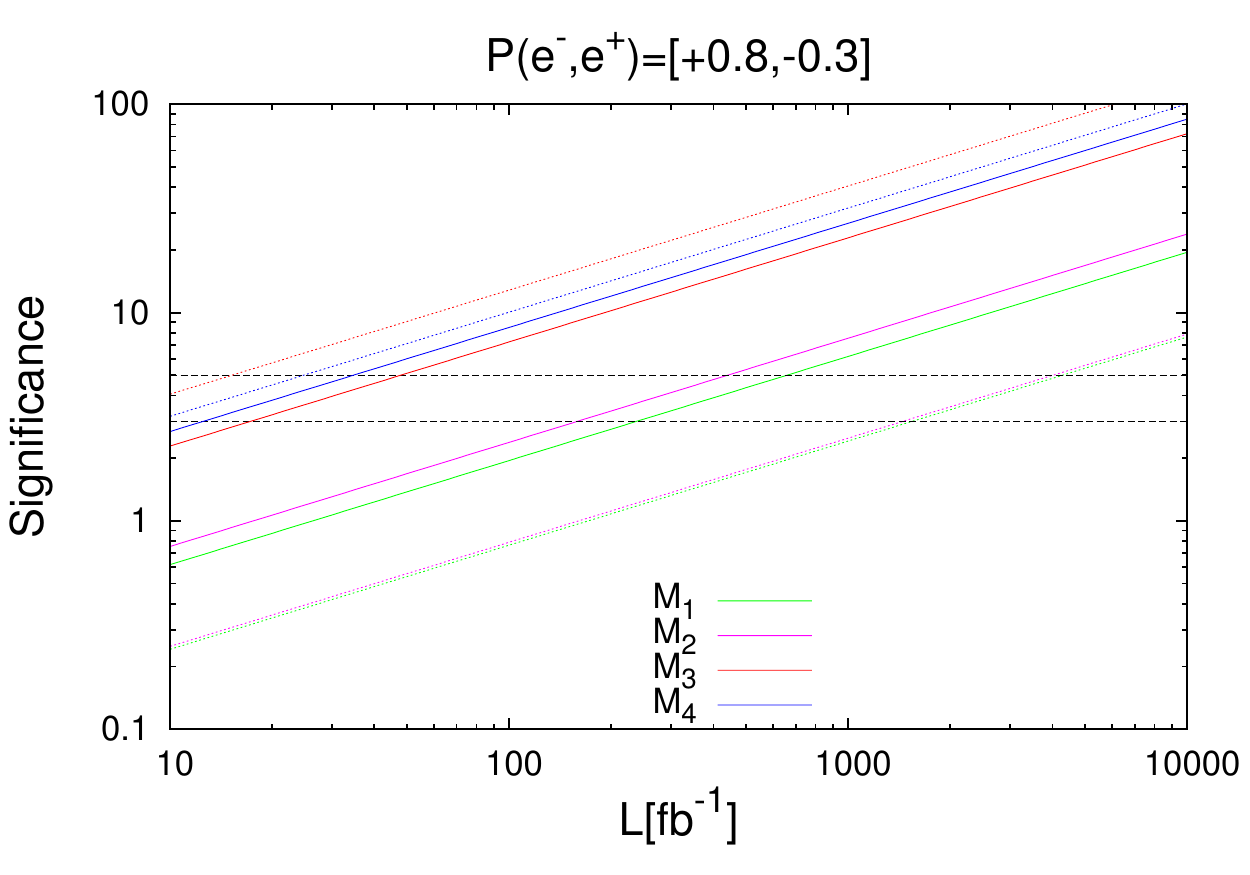}
\caption{The significance $\mathcal{S}$ vs luminosity $L$ at different
c.m. energies (solid lines at $500~\mathrm{GeV}$, the dashed lines
at $1~\mathrm{TeV}$) within the full set of cuts given in Table \ref{cut},
without (left) and with (right) polarized beam. The two horizontal
dashed lines represent $\mathcal{S}=3$ and $\mathcal{S}=5$, respectively.}
\label{SvsL} 
\end{figure}

To get idea about the required values for the luminosity
to observe such a deviation or a discovery, we estimate the signal
significance by varying integrated luminosity values $L$. We show
in Fig.~\ref{SvsL} the signal significance for different models,
using polarized and unpolarized beams

From Fig.~\ref{SvsL}, one remarks easily that the use of polarized
beams (with the polarization $P\left(e^{-},e^{+}\right)=[+0.8,-0.3]$)
makes the signal detected with smaller integrated luminosity as compared
to the case with unpolarized beams for each model and at $E_{c.m.}=500~\mathrm{GeV}$,$~1~\mathrm{TeV}$.
For example, at $E_{c.m.}=500~\mathrm{GeV}$, a $5\sigma$ significance
requires a minimal luminosity value $750~\mathrm{fb^{-1}}$ ($600~\mathrm{fb^{-1}}$
) for $M_{3}$ ($M_{4}$) using unpolarized beams. Using polarized
beams, this minimal luminosity value becomes $45~\mathrm{fb^{-1}}$
($30~\mathrm{fb^{-1}}$) for $M_{3}$ ($M_{4}$). Similar remarks
hold for the models $M_{1,2}$, where the required luminosity gets
decreased form 2500 (1700) to $650~\mathrm{fb^{-1}}$ ($430~\mathrm{fb^{-1}}$) for $M_{1}$
($M_{2}$).

In Table \ref{pola2}, we summarize the events number for the background
and the signal for the different models using polarized and unpolarized
beams at $E_{c.m.}=500~\mathrm{GeV}$ and $1~\mathrm{TeV}$.

\begin{table}[ht]
\begin{adjustbox}{max width=\textwidth} \centering %
\begin{tabular}{|c|c|c|c|c|c|c|c|c|c|}
\cline{2-10} 
\multicolumn{1}{c|}{} & \multicolumn{1}{c}{} & \multicolumn{1}{c}{$P\left(e^{-},e^{+}\right)$ } & \multicolumn{1}{c}{$=$ } & \multicolumn{1}{c}{$[0,0]$ } & & \multicolumn{1}{c}{$P\left(e^{-},e^{+}\right)$ } & \multicolumn{1}{c}{$=$ } & \multicolumn{1}{c}{$[+0.8,-0.3]$ } & \multicolumn{1}{c|}{ }\tabularnewline
\hline 
$E_{c.m.}~(\mathrm{GeV})$ & $N_{BG}$ & Models & $N_{S}$ & $\mathcal{S}_{100}$ & $\mathcal{S}_{500}$ & $N_{BG}$ & $N_{S}$ & \multicolumn{1}{c}{$\mathcal{S}_{100}$ } & $\mathcal{S}_{500}$\tabularnewline
\hline 
\hline 
 & & $M1$ & $33.2864$ & $0.9808$ & $2.1936$ & & $35.7120$ & $1.9488$ & $4.3584$ \tabularnewline
\cline{3-6} \cline{8-10} 
$500$ & $1139.456$ & $M_{2}$ & $40.8320$ & $1.2024$ & $2.6888$ & $323.904$ & $43.8400$ & $2.3832$ & $5.3304$ \tabularnewline
\cline{3-6} \cline{8-10} 
 & & $M_{3}$ & $61.1840$ & $1.7960$ & $4.0168$ & & $138.6368$ & $7.2328$ & $16.1736$ \tabularnewline
\cline{3-6} \cline{8-10} 
 & & $M_{4}$ & $68.4800$ & $2.0088$ & $4.4912$ & & $164.4864$ & $8.4944$ & $18.9944$ \tabularnewline
\hline 
 & & $M_{1}$ & $18.0608$ & $0.3216$ & $0.7192$ & & $19.4048$ & $0.7648$ & $1.7104$ \tabularnewline
\cline{3-6} \cline{8-10} 
 & & $M_{2}$ & $18.6944$ & $0.3328$ & $0.7448$ & & $20.0320$ & $0.7896$ & $1.7656$ \tabularnewline
\cline{3-6} \cline{8-10} 
$1000$ & $3140.608$ & $M_{3}$ & $60.2880$ & $1.0720$ & $2.3976$ & $636.8064$ & $350.2080$ & $12.8312$ & $28.6912$ \tabularnewline
\cline{3-6} \cline{8-10} 
 & & $M_{4}$ & $48.6528$ & $0.8656$ & $1.9360$ & & $270.0224$ & $10.0528$ & $22.4784$ \tabularnewline
\hline 
\end{tabular}\end{adjustbox} \caption{The background and signal events number $N_{BG}$, $N_{S}$ estimated
for the considered energies within the full set of cuts given in Table \ref{cut},
without and with polarized beams at both c.m. energies $E_{c.m.}=500~\mathrm{GeV}$
and $1~\mathrm{TeV}$. The significance $\mathcal{S}_{100}$ and $\mathcal{S}_{500}$
correspond to the two integrated luminosity values $L=100$
and $500~fb^{-1}$, respectively.}

\label{pola2} 
\end{table}

The results presented in Table \ref{pola2} give evidence that with
the polarization $P\left(e^{-},e^{+}\right)=[+0.8,-0.3]$ suppresses
the background $N_{BG}$ events number by $72\%$ and by $80\%$ for
$E_{c.m.}=500~\mathrm{GeV}$ and $1~\mathrm{TeV}$, respectively. Simultaneously,
the signal $N_{S}$ number of events for $M_{3}$ ($M_{4}$) gets
improved by $127\%$ ($140\%$) and by $481\%$ ($455\%$) for c.m.
energies $500~\mathrm{GeV}$ and $1~\mathrm{TeV}$, respectively.
This (significant) excess of events numbers could be an indication
of the nature of DM; i.e., if the DM is a heavy RHN, the excess could
be about five times.

\section{Conclusions}

\label{CON}

In this work, we have investigated the possibility of detecting the
signal significance of DM and identifying its nature. In our setup,
the DM could be either a real scalar or heavy RHN produced at future
electron-positron linear colliders such as the ILC and CLIC. To realize
this task, we considered the process $e^{-}e^{+}\rightarrow b\bar{b}+\slashed{E}_{T}$
at two different c.m. energies: $E_{c.m.}=500~\mathrm{GeV}$ and $1~\mathrm{TeV}$.
Here, we considered two parameter values sets for both scalar and
RHN cases, four models, and we defined and investigated different experimental
constraints for each case, such as the Higgs invisible decay, the muon
anomalous magnetic moment, lepton flavor violation, DM relic
density, and possible constraints from LEP-II. The latter constraint
comes from the negative search of the monophoton at LEP-II, i.e., from
the process $e^{-}e^{+}\rightarrow\gamma+\slashed{E}_{T}$, which
is translated into bounds on the DM and charged scalar masses and
the Yukawa coupling $|g_{1e}|$.

We found that when using appropriate cuts (in Table \ref{cut}), the
background gets significantly decreased and the signal significance
gets lifted especially for the heavy RHN DM case. Using unpolarized beams
at $E_{c.m.}=500~\mathrm{GeV}$, the DM nature can be distinguished
using the normalized distributions: $E_{T}^{b}$, $p_{T}^{b}$, $M^{b,\slashed{E}_{T}}$,
$\eta^{b,\bar{b}}$, and $cos(\Theta^{b,\bar{b}})$. However, a remarkable
shift can be observed in most of the distributions for the fermioinc DM
case. At $E_{c.m.}=1~TeV$, the DM nature can be also distinguished
whether it is scalar or fermioinc using the different distributions.

Using polarized beams, the shape difference with respect to the background
for most of the distributions is more clear, and smaller values of
luminosity with respect of the case without polarized beams are required.
Although, using the polarization $P\left(e^{-},e^{+}\right)=[+0.8,-0.3]$,
the background cross section gets suppressed by about 80\%, and/or
the signal one gets enhanced. This leads to a significant enhancement
on the statistical significance by double if the DM is a scalar and
by five times if the DM is a heavy RHN.

\subsection*{Acknowledgements}

This work is supported by the Algerian Ministry of Higher Education
and Scientific Research under the CNEPRU Project No. \textit{B00L02UN180120140040}.
N. Baouche thanks the ICTP where part of this project was realized for the warm hospitality.
We want to thank Junping Tian for his useful
comments and clarifications. We would like to thank Salah Nasri, Luigi
Delle Rose, and Rikard Enberg for reading the manuscript and for their
useful comments.

\end{document}